# Heat Capacity−A Powerful Tool for Studying Exotic States of Matter

K. Ramesh Kumar,[1,⊥] Xudong Huai,[1,⊥] Michał J. Winiarski,[2]
Allen O. Scheie,[3,†] and Thao T. Tran[1,4]*

[1] Department of Chemistry, Clemson University, Clemson, SC, 29634, USA

[2] Faculty of Applied Physics and Mathematics and Advanced Materials Center, Gdańsk University of Technology, ul. Narutowicza 11/12, 80-233 Gdansk, Poland

[3] MPA-Q, Los Alamos National Laboratory, Los Alamos, NM 87545, USA

[4] Department of Physics and Astronomy, Clemson University, Clemson, SC, 29634, USA

[⊥] These authors contributed equally

[†] scheie@lanl.gov

* thao@clemson.edu

**Abstract**

Heat capacity measurements are a powerful tool that researchers rely on when studying the relationship between microscopic degrees of freedom and macroscopic behavior in condensed matter. This uniqueness stems from heat capacity capturing contributions from lattice, electronic, and magnetic components, as well as energy-level populations, enabling an effective approach to studying phase transitions and excitations across different classes of materials. However, analyzing heat capacity data presents a common, appreciable challenge for new researchers. Although comprehensive theoretical aspects of heat capacity are presented in several elegant textbooks, practical application remains a daunting task. To overcome this challenge, this technical review guides researchers in collecting, analyzing, and interpreting heat capacity data in contemporary quantum materials. We outline the connections between thermodynamics, heat capacity, and entropy, as well as measurement methodology and data analysis for representative examples, including phonon dynamics, spin waves, superconductors, magnetic skyrmions, proximate quantum spin liquids, and heavy-fermion materials. Our goal is to provide a concise, accessible guide that enables new researchers to utilize heat capacity as a quantitative lens for understanding exotic states of matter.

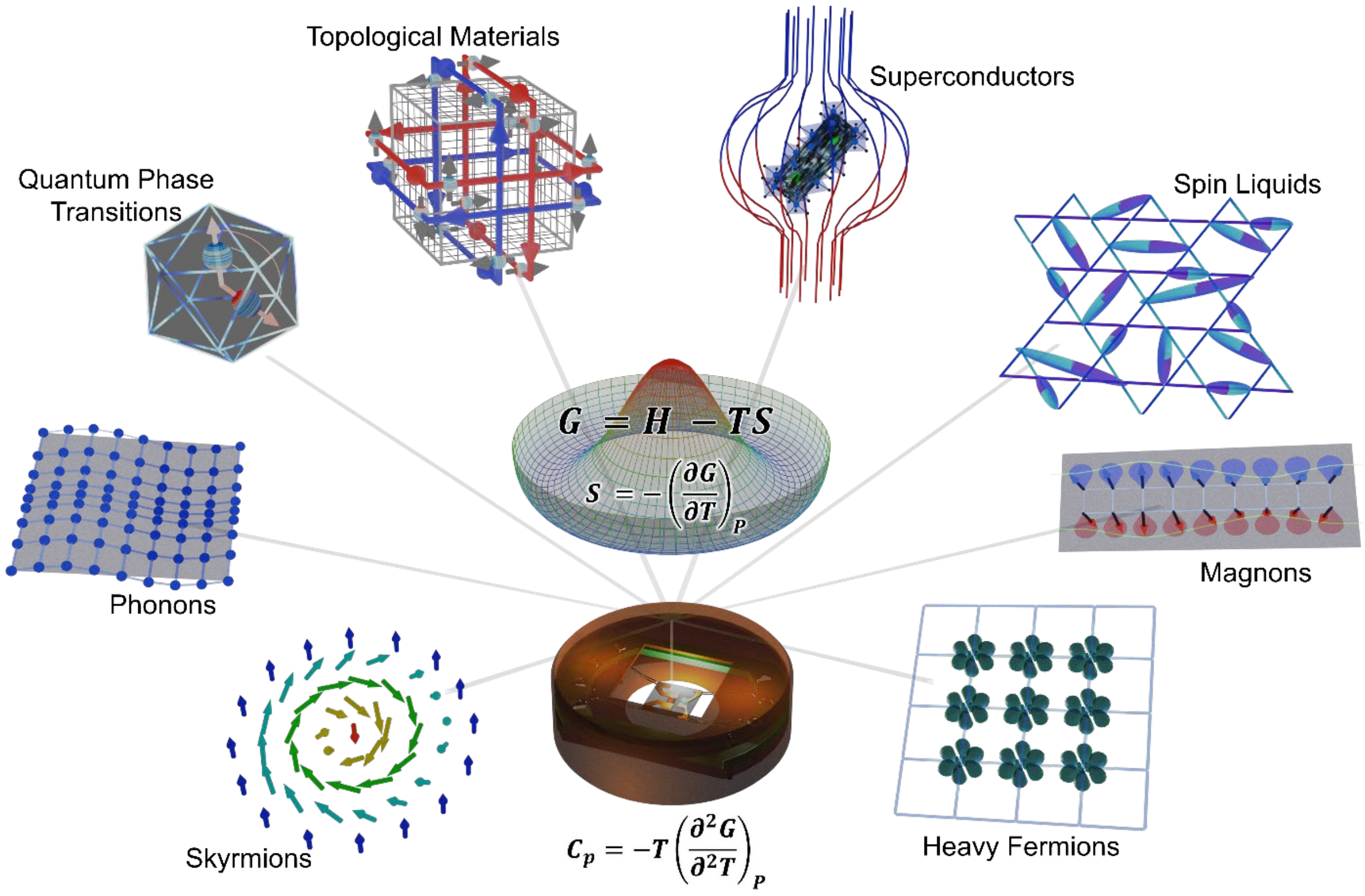

**Figure 1.** Heat capacity: a thermodynamic "microscope". A schematic diagram showing how heat capacity provides insight into entropy and Gibbs energy, revealing phase transitions, degrees of freedom, and low-energy excitations.

## 1. Introduction

Heat capacity is a powerful experimental tool for understanding the intrinsic and emergent properties of quantum materials (Figure 1). As one of the fundamental thermodynamic quantities, it is closely connected to other state functions—internal energy ($U$), entropy ($S$), enthalpy ($H$), and Gibbs energy ($G$)—all of which describe how a material exchanges heat with its surroundings.[1-4] Heat capacity is proportional to the first derivative of entropy with respect to temperature:

$$C = T\left(\frac{\partial S}{\partial T}\right) , \quad (1)$$

Anomalies in the entropy change are often associated with physical, magnetic, or electronic phase transitions.[2-8] One of the most critical applications of heat capacity measurements is identifying and characterizing symmetry-breaking phase transitions, which are accompanied by characteristic anomalies in heat capacity curves. For example, a first-order phase transition (associated with a discontinuity in the first derivative of free energy with respect to temperature) is characterized by a steep change in entropy, and the heat capacity curve shows a sharp symmetrical peak (Figure 2a). This peak is associated with latent heat and indicates an abrupt change in the order parameter. In a second-order phase transition (associated with a discontinuity in the second derivative of free energy with respect to temperature), the entropy varies continuously, but a discontinuity in the heat capacity is observed (Figure 2b). Such second-order transitions manifest as a lambda-shaped peak in heat capacity curves.[9-11] The peak position reveals the transition temperature while the peak shape indicates whether the transition is structural, magnetic, or superconducting. For example, a structural transition is associated with a sharp rise in heat capacity, whereas second-order magnetic or superconducting transitions often show a lambda-shaped cusp or asymmetrical peak. The magnetic heat capacity below $T_c$ is governed by the excitations of the ordered state, which are discussed in detail in section 3.

In addition, heat capacity is also expressed as the second derivative of Gibbs energy,

$$C_p = -T\left(\frac{\partial^2 G}{\partial T^2}\right)_p \qquad (2)$$

directly linking to microscopic degrees of freedom. Thus, heat capacity provides critical insights into changes in lattice, electronic, magnetic, or nuclear contributions. Mapping phase transitions as a function of external parameters, such as temperature, pressure, and magnetic or electric fields results in the construction of phase diagrams.[12, 13] Phase diagrams represent a thermodynamic landscape that visually depicts how different phases exist and evolve under varying external conditions, such as temperature, pressure, and field. Let us take the water phase diagram as an example for phase transitions (Figure 2c). The water phase diagram illustrates three distinct phases: gas (water vapor), liquid (water), and solid (ice), separated by phase boundaries and their evolution as a function of temperature and pressure.[14] A similar thermodynamic framework applies to structural and electronic phase transitions that occur in condensed matter. For instance, Figure 2d illustrates the magnetic landscape of $LiCrTe_2$, showing paramagnetic (PM), ferromagnetic (FM), and antiferromagnetic (AFM) phases.[15] Similar to the distinct phases in the water *P-T* phase diagram, different magnetic regions in an *H-T* phase diagram are characterized by macroscopic state variables, such as (*M*, *H*, *T*, and *S*). Heat capacity gives direct access to these quantities and the collective behavior of the system. As such, low-temperature heat capacity measurements can probe a wide range of emergent phenomena, including superconductivity, heavy-fermion behavior,

electron–quasiparticle scattering, crystal field effects, and exotic states such as spin liquids, skyrmions, and topological phases.

The theoretical definition of heat capacity—the heat required to raise the temperature of a substance by a unit amount—is relatively simple and intuitive.

$$C_p = \left(\frac{dQ}{dT}\right)_p \quad (3)$$

but in practice, heat capacity measurements are more complex and involve precise control of heat flow, accurate temperature measurement, and careful data analysis and interpretation. For beginners entering the fields of contemporary magnetic and quantum materials, understanding theoretical aspects, performing measurements, and modeling heat capacity data can be particularly challenging, especially in systems where many-body interactions are at play.[5, 16-19] In this technical review, we provide a comprehensive guide to the measurement and interpretation of heat capacity data, with an emphasis on quantum materials. Our goal is for this tutorial to complement textbooks and literature examples on understanding microscopic degrees of freedom through heat capacity—a powerful fundamental probe.

## 2. Measurement Techniques

Heat capacity measurements differ fundamentally from other physical property measurements in that the desired quantity cannot be obtained directly from a single reading. For example, a single magnetization measurement can already provide valuable information, whereas heat capacity requires multiple steps of background subtraction and data processing before reliable values can be extracted (Figure 3a). This is because heat capacity measurements do not measure the heat capacity of a sample alone. Instead, it measures the heat capacity of an assumed isothermal system consisting of the sample platform, the sample, and typically a medium that provides thermal contact between the platform and the sample (Figure 3b, 3c). There are a variety of instruments and techniques that can measure temperature and field-dependent heat capacity. One of the most accessible and widespread is a Physical Property Measurement System (PPMS) by Quantum Design; [20] we here make some comments about using this system.

## 2.1 Historical Development and Theoretical Aspects

Calorimetry techniques are broadly classified into two categories: adiabatic and non-adiabatic methods. In adiabatic calorimetry, heat flow and temperature changes are measured under high-vacuum conditions to minimize heat exchange with the environment. In contrast, in non-adiabatic methods, heat flow is controlled and accounted for by a continuous flow of exchange gas, such as argon or nitrogen. The adiabatic method was developed by Eucken and Nernst in 1911.[21, 22] It is a highly accurate technique based on a straightforward principle. However, this method requires relatively large sample masses and long measurement times. For small-sized samples, several modifications have been proposed, including the AC calorimetry technique, but high accuracy is observed only limited temperature range.[23] In 1972, Bachmann and co-workers introduced the relaxation method, which combined features of the heat-pulse method and relaxation analysis.[24] Later, Quantum Design (San Diego) adopted this method and fully automated the measurement, implementing the Hwang curve-fitting approach to simultaneously fit both the heating and cooling segments of the temperature curve.[25-27] The heat capacity measurement in PPMS is based on the following theoretical framework, which is briefly discussed here. The detailed description can be found in the PPMS Heat Capacity User Manual.[20]

The heat balance equation describes how the platform temperature changes with respect to time. In the simple model, the platform and the sample are in good thermal contact, and both have the same temperature during the measurement ( $T_P \approx T_s$). The total heat balance can be written as:

$$C_{\text{total}} \frac{dT_P}{dt} = -K_w (T - T_b) + P(t) \quad (4)$$

where $C_{\text{total}}$ is the total heat capacity of the platform and the sample, $K_w$is the thermal conductance between the platform and the heat sink at $T_b$, and $P(t)$ is the applied heating power. The solution to equation (4) gives the platform temperature as a function of time.

$$T_P(t) = T_b + \Delta T \exp\left(-\frac{t}{\tau}\right); \ \tau = \frac{C_{\text{total}}}{K_w} \qquad (5)$$

where the $\tau$ is the thermal time constant, and this is known as the single-tau method.

In the two-tau relaxation method, both the platform and sample temperature relaxations are described by two coupled differential equations:

$$C_p \frac{dT_p}{dt} = -K_w(T_p - T_b) - K_g(T_p - T_s) + P(t), \qquad C_s \frac{dT_s}{dt} = -K_g(T_p - T_s) \qquad (6)$$

where $C_p$ and $C_s$ are the heat capacities of the platform (addenda) and the sample, and $K_g$ is the thermal conductance between the platform and the sample. In the limit $K_w \ll K_g$, the solution for heating or cooling can be approximated as a sum of two exponentials toward a steady state $\bar{T}$:

$$T_p(t) = \bar{T} + A_1{}^{e^{-t/\tau_1}} + A_2{}^{e^{-t/\tau_2}} \qquad (7)$$

where $\tau_1$ and $\tau_2$ are two characteristic time constants. $\tau_1$ is referred to as the long time constant and is related to the thermal relaxation between the entire puck assembly and the thermal bath. The short time constant $\tau_2$ is related to the thermal relaxation between the sample and the platform. The amplitudes $A_1$ and $A_2$ are set by initial conditions.

However, this two-exponential solution is not valid near a first-order phase transition, as the latent heat associated with the transition is not captured by the exponential curve described in equations (6-7). One of the simplest methods, suggested by Lashley et al., avoids nonlinear curve fitting and instead uses real-time temperature changes to estimate heat capacity from the slope change. [25]

If we rewrite the heat balance equation (4) by dividing both sides by $\frac{dT_P}{dt}$,

$$C_{\text{total}} = \frac{-K_W\,(T - T_b) + P_0}{dT_P/dt} \qquad (8)$$

with known values for $K_w$, $P_0$, and $\frac{dT_P}{dt}$, we can estimate $C_{\text{total}}$ for every temperature. By carefully subtracting the addenda heat capacity, we can then obtain the intrinsic heat capacity of the sample. If the sample undergoes a first-order transition, the associated heat capacity will show either a positive or a negative peak, depending on whether heat is absorbed or released. [25, 27, 28] To obtain accurate heat capacity near a first-order transition, Suzuki and co-workers suggested an alternative approach that combines both relaxation and scanning methods.[28] Initially, the normal heat capacity measurements are carried out for both cooling and heating processes, and a correction factor is estimated for $K_w$ and $K_g$ until both curves merge.[28] Once the correction factor is found, we can rearrange the heat balance equation at equilibrium and estimate the sample temperature correctly as $K_w$ and $K_g$ are accurate enough to account for the heat leak error. The difference between $T_p$ and $T_s$ is crucial, hence using the set of derived parameters, such as $K_w$, $K_g$, $P_0$ and $T_b$ and $T_p$, we can estimate the sample temperature and sample heat capacity. The standard PPMS measurement technique uses Eq. 7; and the rest of this section focuses on the use of this instrument.

## 2.2 Sample Preparation

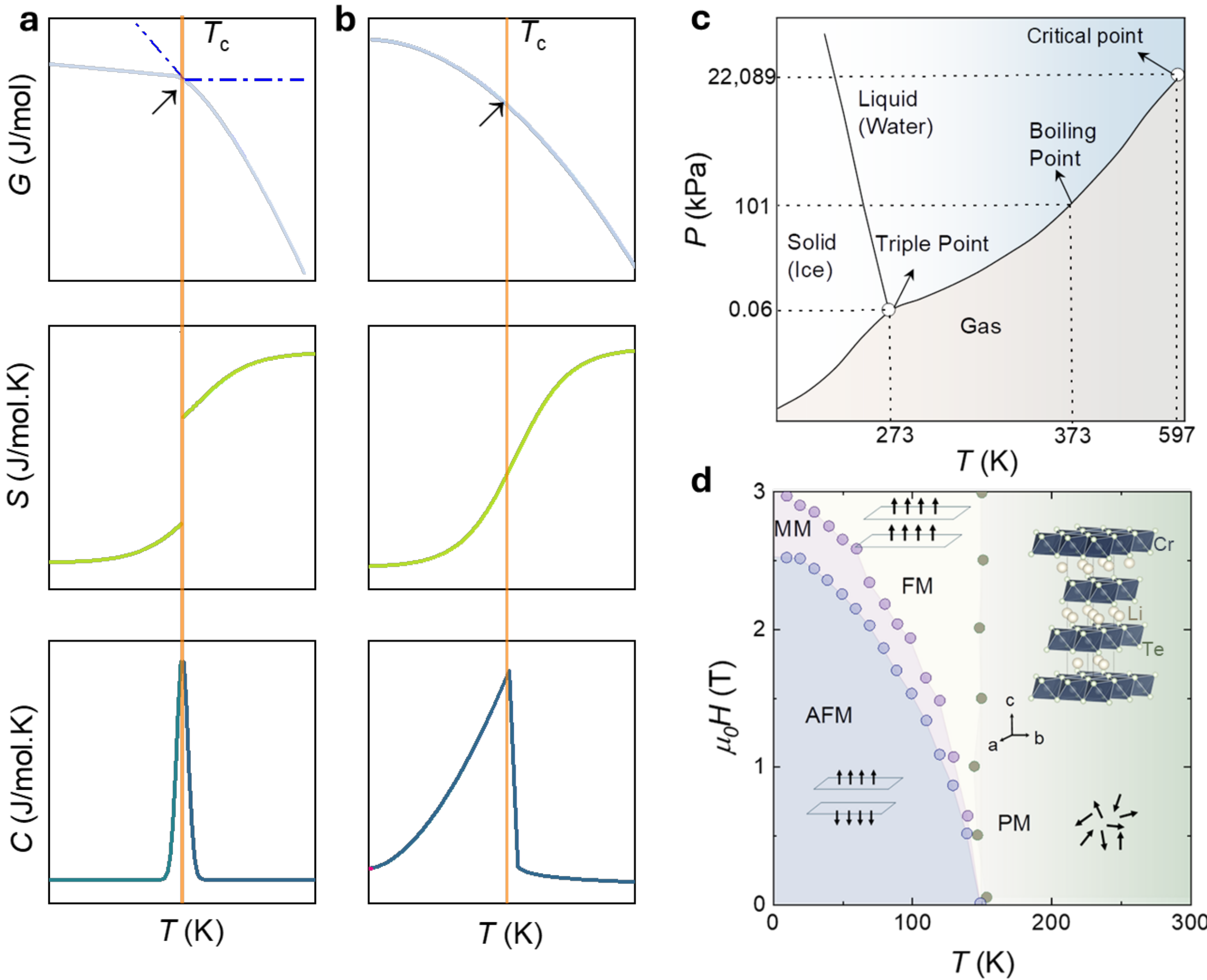

**Figure 2.** Temperature-dependence of Gibbs energy, entropy, and heat capacity for (a) first- and (b) second-order phase transitions, respectively. (c) *P-T* phase diagram for water showing liquid, solid, and gas phases. (d) *H-T* phase diagram for $LiCrTe_2$ showing different magnetic regions such as PM: paramagnetic; FM: ferromagnetic MM: metamagnetic; and AFM: antiferromagnetic phases (Adapted from Ref[15]: open access).

In the PPMS setup, each sample holder ("puck") contains an assembly of a heater and a thermometer (Figure 3b-d), which needs to be calibrated periodically. In addition, addenda (background specific heat of the sample holder and grease) must be measured before each experiment, as it will later be subtracted from the total signal.  Care is required when applying grease: (i) grease should cover only the area corresponding to the sample geometry and thickness, and (ii) its amount must be adjusted according to the sample type and mass. For example, metals, which conduct heat efficiently, require only a minimal amount of grease to avoid overwhelming the sample signal. In contrast, pressed pellets or insulating samples, which conduct heat poorly, require more grease and generally benefit from using a decent sample mass in a thin sample habit.

Once the addenda measurement is completed, the sample can be mounted by gently placing it onto the pre-applied grease, without removing any existing grease or introducing any extraneous material.

The overall quality of heat capacity data depends on several factors, including sample mass, geometry, thermal conductance, and density. There is no universal procedure that applies equally well to metals, oxides, and insulators. However, with careful preparation and measurement, it is viable to develop a methodology tailored to each specific case.

### 2.3 Heat Flow and Equilibrium

The PPMS experimental design features a cylindrical copper puck with a sapphire platform suspended at its center by platinum or manganin wires, schematically shown in Figure 3b. The platform is connected to the heat sink through thin manganin wires with thermal conductance $K_W$, while the sample is coupled to the platform through a thin layer of Apiezon N grease (Figure 3b). This grease exhibits two transitions at 215 K and 285 K, which can be accounted for by careful background subtraction. The thermal conductance of the grease is represented by $K_g$. For measurements below 300 K, Apiezon N grease is generally preferred. However, above 300 K, Quantum Design recommends Apiezon H grease, as it provides better thermal coupling at elevated temperatures.[20]

During heat capacity measurements (Figure 3a), the PPMS records both the platform and bath temperatures. The system assumes that the sample temperature equals the platform temperature, $T_s(t) = T_p(t)$, due to thermal coupling through the grease (Figure 3b). However, the sample and platform are linked by a finite conductance $K_g$, and as a result, $T_s(t)$ can deviate slightly from $T_p(t)$.

To ensure reliable data acquisition, a plate-shaped sample is generally preferred, with a large flat surface to improve thermal contact and a mass of about 2–20 mg. The optimal mass, however, depends on the material. For materials with good thermal conductivity (e.g., metals, semiconductors and semimetals), as little as 2–10 mg may suffice because their electronic and phononic contributions are large compared to the addenda (Figure S1a and S1b). For polycrystalline samples, especially intermetallics, it is good practice to polish one side of the sample that touches the grease and sapphire platform to increase the contact area and improve thermal coupling. For poorly thermal-conducting materials (e.g., oxides and insulators), a larger sample mass of 10–20 mg is sometimes necessary to ensure that the measured signal remains significantly above the background (grease and platform) (Figure S1c and S1d).

Several sample preparation techniques have been developed for measuring the heat capacity of polycrystalline, insulating samples.[5] For poor heat-conducting samples (e.g., glasses) or powder samples that are difficult to press into a pellet, a known amount of high-purity Cu powder can be mixed with the sample and then pressed into a pellet with a known total mass for measurements. Cu is diamagnetic and exhibits no phase transition down to 2 K; its heat capacity is well characterized. The heat capacity of Cu can then be subtracted from the measured total heat capacity. For samples that are highly toxic, radioactive, or can decompose under high vacuum, stycast resin can be used instead of Cu as a matrix material.[29]

**2.4 Heat Curve and Fitting**

To measure heat capacity, the PPMS applies a quasi-adiabatic measurement technique, in which temperature gradients are small relative to the sample temperature. A square-pulse heater power *P(t)* for a finite duration produces a small temperature rise on the sapphire platform ($T_p(t)$) relative to the bath $T_b$ (Figure 3e).[20, 26, 28] The system then records the subsequent cooling as heat flows back through the support wires. The platform and sample temperatures are $T_p(t)$ and $T_s(t)$, respectively, which are coupled by the grease conductance $K_g$. The platform is weakly linked to the bath by the wire conductance $K_w$.[20, 28, 30]

As described in equations (4) and (5), $C_p$ and $C_s$ are the effective heat capacities of the platform (addenda) and the sample. The solution (in the limit that $K_w \ll K_g$) for heating or cooling can be approximated as a sum of two exponentials (eq. 7) toward a steady state: $\overline{T}$ with time constants $\tau_1 \sim C_{total}/K_w$, and $\tau_2 \sim C_{eff}/K_g$. In the ideal strong-coupling limit, $A_2$ and $\tau_2$ approach 0. Typical $\tau_1$ values are ~ 10–150 s, $\tau_2$ should be small (tens to hundreds of ms) when the sample coupling is good. Large $\tau_2$ or sizable $A_2$ flags poor coupling.[20, 26]

Two practical cautions are as follows. First, a visually good exponential fit does not guarantee correctness of the value if the sample is porous, pelletized, or internally gradient-limited; always check the reported coupling fraction and compare the relative magnitudes of $C_{total}$, $C_s$ and $C_{addenda}$. As a rule of thumb, the coupling should exceed ~90% so that $T_s \approx T_p$ during the pulse, and $C_{total}$ should remain at least ~50 % higher than $C_{addenda}$ throughout the measurement temperature window. Second, the standard relaxation fit is reliable for second-order transitions (smooth $T(t)$) but can fail near first-order transitions where the heat curve shows inflection points in both the heating and cooling phases. This behavior is due to the latent heat associated with first-order transitions, such as melting, structural phase transitions, martensitic transformations, or field-induced metamagnetic transitions (Figure 3e).[26 25, 28]

In such cases, the extracted $C$ is broadened and biased; alternative methods such as scanning approaches (e.g., slope-based methods and hybrid protocols) and long-pulse methods are necessary to sufficiently capture heat capacity around the transition (Figure 3a).[24, 25, 28, 31, 32] For example, in the long-pulse method, the platform is heated for an extended duration of approximately 20–30 minutes, and the platform temperature and bath temperature are continuously monitored. Instead of fitting an exponential, the specific heat is computed directly from a time derivative of the sample temperature.[33, 34] Overall, the measurement parameters: $P_0$, $t_0$, $T_b$, $\bar{T}$, $\tau_1$, $\tau_2$, coupling %, and grease quantity should be carefully monitored, and care should be taken when subtracting the background. For ease of review, common issues and rapid diagnostics are given below:

- Low sample coupling (< 90%): Improve thermal contact by increasing the contact area. For metals, reduce the grease thickness; for pelletized samples, increase the mass and sinter the pellet to improve thermal coupling. Watch $\tau_2$ and $A_2$.
- Grease artifacts (215 K, 285 K): Confirm the addenda across these temperatures; avoid over-interpreting anomalies that track the addenda contribution.[35]
- First-order transitions (inflection in heating and cooling curves): Do not just rely on the automatic two-tau fit; instead, use a slope/scanning method or a hybrid protocol near transition temperature when appropriate.
- Addenda dominate ( $\frac{C_{total}-C_{addenda}}{C_{total}} < 0.5$): Increase sample mass to 10-20 mg or lower $P_0$ to keep within linear response.
- Addenda stray measurement error: It is a good practice to check the addenda measurement every time for potential stray measurement points. When the sample and addenda measurement points do not coincide, data interpolation by the MultiVu software may cause the stray measurement point to appear as a broad peak or downward hump. This should not be taken as an intrinsic feature.

- Cryo-pump regeneration: In the PPMS system the high vacuum environment necessary for semi-adiabatic heat pulse measurements is achieved by evacuating the sample space using first a scroll pump (down to a few Torr) and then by opening a valve of the cryopump, which is a liquid He cooled tube filled with pyrolytic graphite that can absorb the remaining gas (mostly He) from the chamber. If the cryopump is not regenerated regularly, or a leak occurs, the final vacuum may be insufficient, or the system may lose the high vacuum during the measurement, resulting in an anomalous heat curve due to heat leaking from the sample into the sample chamber. Therefore, it is important to check the chamber pressure and regenerate the cryo-pump frequently.
- Internal gradients / porous pellets: These can result in artificially large $\tau_2$ and hysteresis. It is recommended to re-press and/or sinter the pellet to improve its thermal contact.
- Baseline drift / nonlinearity: Verify that the platform temperature rise is only a few percent of $T_b$ ; if not, reduce $P_0$ or allow more time for the system to reach equilibrium.
- Non-exponential decay: sample may be non-ergodic or have internal relaxation (e.g., between nuclear and electronic degrees of freedom). In this case, the two-tau model is insufficient, and a more sophisticated fit is necessary.[36]
- Caution: if the sample contains large magnetic moment elements, such as Eu, Gd, Dy, Tb, Ho, Er, and has a thin-plate habit, applying a strong magnetic field can produce a large force acting on the sample. This may pull the sample out of the puck, bend the wires or even damage the platform. A small to moderate sample mass, prepared in a more "block-like" shape, and limiting the applied field can help reduce the force and avoid puck damage.

## 2.5 Measurement Units

Heat capacity, specific heat, and molar heat capacity are generally expressed in J K$^{-1}$, J Kg$^{-1}$ K$^{-1}$ and J mol$^{-1}$ K$^{-1}$, respectively. Molar heat capacity is preferred in certain cases as it directly connects to thermodynamic principles, such as the equipartition theorem, and to lattice, electronic and magnetic degrees of freedom. For instance, according to the Dulong–Petit law, the molar heat capacity approaches 3$R$ at high temperatures, reflecting the three vibrational degrees of freedom per atom. Further, at low temperature, electronic specific coefficients ($\gamma$) are expressed in J mol$^{-1}$ K$^{-2}$, which is a convenient way of understanding correlated electron behavior.

Specific heat units are preferred in certain cases where a material's performance is quantified by the mass or density rather than by the number of moles. Magnetocaloric effects are defined as the isothermal entropy ($\Delta S_m$) change or adiabatic temperature change ($\Delta T_{ad}$) of a material when a magnetic field is applied or removed. The performance metrics are governed by $\Delta S_m$ and $\Delta T_{ad}$ values, and to compare these values across different classes of materials and benchmark the cooling efficiency, mass-normalized units (J Kg$^{-1}$ K$^{-1}$) are preferred for $\Delta S_m$ and $C_p$. Finally, both molar heat capacity and specific heat are commonly used in literature, provided that the units are applied appropriately. In PPMS measurements, however, the raw data are typically reported in μJ $K^{-1}$. These values provide immediate insight into the relative magnitudes of the sample and the addenda contributions.

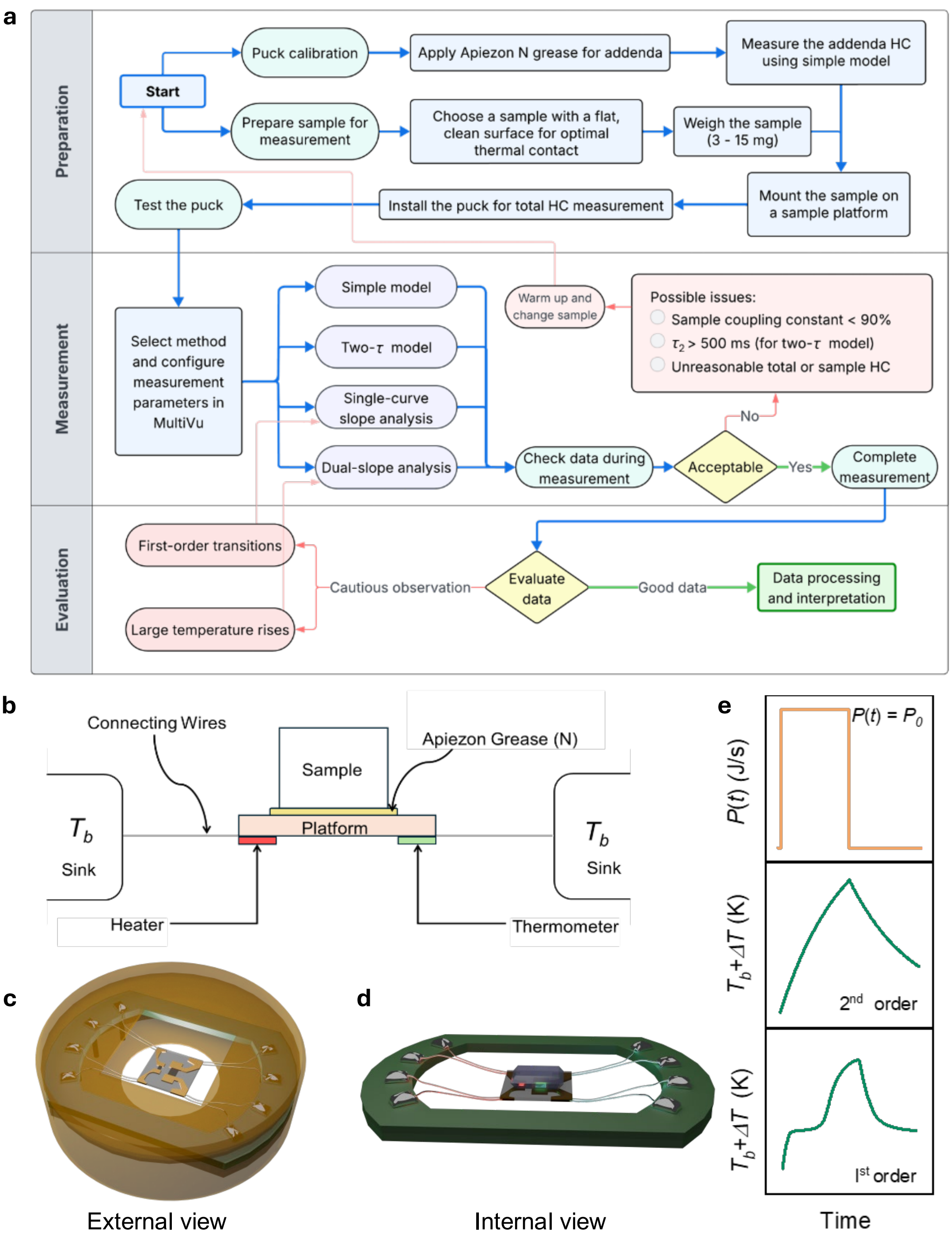

**Figure 3.** (a) Workflow for PPMS heat capacity measurements and data evaluation. (b) Schematic sketch of the PPMS heat capacity setup. (c) A heat capacity puck assembly includes the platform and the frame. (d) An isolated puck platform showing the heater, thermometer, and wire assembly. (e) Time-dependence of heater power *P(t)*. Platform temperature as a function of time for first and second order transitions.[37]

## 2.6 Other Measurement Methods

We note that the PPMS is just one of many instruments that can measure heat capacity, and there are, in fact, a plethora of different designs and methodologies, including methods for systems with internal relaxation, frequency-dependent calorimetry for non-ergodic systems, and so on (see the Supplemental materials for an extended discussion).[36, 38, 39] However, this tutorial focuses on one of the most common systems for studying quantum materials.

## 3. Theoretical Aspects of Heat Capacity in Solids

Understanding the heat capacity of solids is essential for linking microscopic degrees of freedom to macroscopic thermodynamic behavior. As a fundamental probe, heat capacity not only reveals how materials store and exchange thermal energy but also encodes critical information about lattice vibrations, electronic excitations, magnetic ordering, and discrete energy levels. In addition, heat capacity is sensitive to all types of excitations (lattice, electronic, and magnetic) and thus serves as a universal probe of a system's density of states. These insights are indispensable for interpreting emergent phases in quantum materials, superconductors, magnets, and correlated electron systems. In this sense, heat capacity serves as a unifying quantity that ties together thermodynamics, statistical mechanics, materials chemistry, and condensed-matter physics, and provides the theoretical foundation for constructing and analyzing phase diagrams.

Gibbs free energy ($G$), entropy ($S$), and heat capacity ($C_v$ *or* $C_p$) are three fundamental thermodynamic quantities that govern phase stability and phase transitions in materials. Together, they provide the foundation for constructing ($P$, $T$) phase diagrams. Gibbs free energy is defined as:

$$G \equiv E - TS + PV \qquad (9)$$

and its differential form is written as

$$dG = -SdT + VdP \qquad (10)$$

Within this framework, we now turn to heat capacity in the context of thermodynamic analysis.[3, 4, 7] Heat capacity is formally expressed as:

$$C_{x,y} = \lim_{dT\to 0} \frac{dQ}{dT} \qquad (11)$$

where $dQ$ and $dT$ represent infinitesimal changes in heat and temperature, and the subscripts $x$ and $y$ specify the thermodynamic constraints (e.g., constant pressure or constant volume). This general definition applies to gases, liquids, and solids alike.

For solids, however, the difference between heat capacity at constant pressure ($C_p$) and heat capacity at constant volume ($C_v$) is usually negligible. This can be seen from the relation:

$$C_p - C_v = TV \frac{\beta^2}{K_T} \qquad (12)$$

where $\beta$ is the volume expansion coefficient, and $K_T$ is the isothermal compressibility. Since $\beta^2/K_T$ is very small at low temperatures in solids, the difference between $C_p$ and $C_v$ can be negligible.[2, 7]

In practice, the heat capacity of a solid reflects contributions from multiple microscopic degrees of freedom. The total heat capacity can be expressed as:[2, 3]

$$C_p = C_{\mathrm{ph}} + C_{\mathrm{e}} + C_{\mathrm{mag}} + C_{\mathrm{Sch}} \qquad (13)$$

where $C_{\mathrm{ph}}$ is the phononic contribution from lattice vibrations, $C_{\mathrm{e}}$ is the electronic contribution from itinerant electrons, $C_{\mathrm{mag}}$ is the magnetic contribution from spin and orbital magnetic interactions, and $C_{\mathrm{Sch}}$ is the Schottky contribution from discrete energy level populations.

### 3.1 Lattice Heat Capacity

The dominant room-temperature contribution to the heat capacity of solids typically arises from lattice vibrations—the oscillations of atoms and molecules about their equilibrium positions. At high temperatures, this contribution approaches the classical limit given by the Dulong–Petit law,

$$C_v = 3nR \qquad (14)$$

where $n$ is the number of atoms per formula unit and $R$ is the universal gas constant.[40-42] This limiting value is independent of atomic mass, crystal structure, or bonding, making it a rapid, useful check of stoichiometry and atom count per formula unit.

At lower temperatures, however, heat capacity decreases smoothly and deviates from the Dulong–Petit prediction.[40, 42] To capture this behavior, Einstein adapted Planck's quantization concept—originally introduced for electromagnetic radiation—to atomic vibrations in solids.[40] In Einstein's model, atoms are treated as independent harmonic oscillators, each vibrating at a characteristic frequency ($v_E$). The excitation probability of these oscillators follows the Boltzmann distribution and is proportional to $exp(-hv_E / k_B T)$. Heat capacity is then expressed as:

$$C_v = 3rR \frac{\left(\frac{\Theta_E}{T}\right)^2 e^{\frac{\Theta_E}{T}}}{\left(e^{\frac{\Theta_E}{T}} - 1\right)^2} \quad (15)$$

where $\Theta_E$ is the Einstein temperature, $r$ is the number of atoms in the unit cell, and $R$ is the gas constant.[40, 42-44] While Einstein's model successfully introduces quantization, its assumption that all atoms vibrate at a single frequency has some limitations. In particular, it does not sufficiently capture the experimentally observed low-temperature behavior where $C_v \propto T^3$, rather than the exponential dependence predicted by the model.

To address this limitation, Debye proposed that vibrations in solids span a continuous spectrum of frequencies, from zero up to a maximum cutoff (the Debye frequency), as though the solid behaves like an isotropic elastic medium. The corresponding heat capacity is given by:

$$C_V = 9rR \left(\frac{T}{\Theta_D}\right)^3 \int_0^{\frac{\Theta_D}{T}} \frac{\left(\frac{\Theta_D}{T}\right)^4 e^{\frac{\Theta_D}{T}}}{\left(e^{\frac{\Theta_D}{T}} - 1\right)^2} d\left(\frac{\Theta_D}{T}\right) \quad (16)$$

where $\Theta_D$ is the Debye temperature. The Debye model is highly successful in explaining phonon contributions across a wide range of solids. At very low temperatures ($T < \Theta_D/10$), the expression simplifies to:

$$C_V = 1944\, r \left(\frac{T}{\Theta_D}\right)^3 \quad (17)$$

which is the well-known Debye $T^3$ law.[40]

Despite its success, the Debye model also has limitations. Born and von Kármán showed that the Debye temperature itself is not constant but varies with temperature.[40] As a result, modeling lattice heat capacity with a single $\Theta_D$ can oversimplify the behavior of real materials. An Einstein term describes an optical (gapped) phonon mode, while a Debye term corresponds to an acoustic (gapless) phonon mode. Real materials typically exhibit both terms. Lattice heat capacity can often be modeled by a sum of several Debye and Einstein terms.[40]

While the Einstein and Debye models provide a general description of lattice vibrations, practical analysis of experimental data requires different strategies depending on the temperature regime. At low temperatures, simplified fitting relations are particularly effective for extracting key parameters, whereas at higher temperatures, combinations of Debye and Einstein terms are often necessary. The following sections outline these complementary approaches.

**High-Temperature Heat Capacity Analysis**

At higher temperatures, where phonons dominate, the heat capacity can be modeled using the Debye integral or a combination of Debye and Einstein functions. The Debye model accounts for the entire lattice contribution by treating the crystal as an elastic medium with a continuous spectrum of vibrational modes. However, real solids often contain atoms or ligands that are loosely bound to the lattice. These may vibrate at distinct frequencies (optical phonon modes) that cannot be captured by a single Debye function. In such cases, Einstein oscillators are introduced to represent these localized vibrations.

A useful way to identify these additional contributions is through a $C_p/T^3$ vs. $T$ plot (with a logarithmic $T$-axis). In this representation, an out-of-phase vibration appears as a peak around $\frac{\Theta_E}{5}$. For example, in our simulation with $\Theta_E$ = 100 K, the peak appears near 20 K (inset of Figure S3a). The prefactor (1/5) stemming from the extreme condition of $C_p/T^3$ is discussed in Supplemental Information. Even when a peak is not obvious, it is common practice to include one or two Einstein terms to improve the fit over the full temperature range. The total heat capacity can then be written as a weighted sum of Einstein and Debye contributions:

$$C_p(T) = mC_E(T) + n\,C_D(T) \quad (18)$$

where $m$ and $n$ are weight factors for Einstein and Debye modes, respectively. The fitting yields $m$, $n$, the Einstein temperature ($\Theta_E$) and the Debye temperature ($\Theta_D$). Note that the sum of $m$ and $n$ must be equal or close to the number of atoms per formula unit. The Einstein temperature represents the characteristic frequency of vibrational modes of localized atomic/polyatomic units. Low values typically correspond to heavy atoms or covalently bonded units, while high values are associated with lighter atoms or stiffer bonded units (more details about $\frac{C_p}{T^3}$ are available in Supplemental Information). Since the Debye temperature is related to the cut-off frequency of the phonon density of states, the nonlinear fitting of the high-temperature heat capacity can be used to model the phonon density of states. In the Debye model, the density of states $g(\omega) \propto \omega^2$ and the fitting parameters, including oscillator strength (weight parameter) and Debye temperature, can be used to simulate the approximate phonon density of states. This exercise allows us to assess the appropriateness and accuracy of the fitting results.[45] A useful Python script for automated heat capacity fitting with the combination of Debye and Einstein terms is available in the following reference.[46]

It is often important to isolate the magnetic heat capacity from the total heat-capacity data. In such cases, the phononic contribution must be carefully determined.

Two straightforward methods are commonly applied. The first involves high-temperature modeling using the Debye and Einstein expressions described above. From such modeling, the Debye and Einstein temperatures can be obtained and extrapolated to simulate the phononic contribution down to low temperature. For metallic systems, the $\gamma T$ electronic term must also be included. The magnetic heat capacity is then estimated by subtraction:

$$C_{mag} = C_p^{total}(T) - \gamma T - [m\, C_E(T) + n\, C_D(T)] \tag{19}$$

This method, although useful when no nonmagnetic analog is available, has limitations: the extracted parameters depend on the chosen fitting range, and the low-temperature phonon contribution cannot always be reliably mapped from high-temperature fits. Nevertheless, hybrid Debye–Einstein modeling was used to capture subtle features essential for identifying spin-liquid behavior in $\alpha$-$RuCl_3$.[47]

The second method relies on subtraction using a nonmagnetic analog that crystallizes in the same structure. This approach is valid, provided that the phonon density of states and dispersion remain nearly unchanged, apart from mass difference. In the case of 4f lanthanide compounds, La- or Y-based systems are commonly used as nonmagnetic analogs for lighter members, while Lu-based compounds are used for heavier lanthanides from Gd to Yb. For $Eu^{2+}$ compounds , $Sr^{2+}$ nonmagnetic analogu is a good option. A mass correction is applied by scaling the Debye temperature, since vibrational frequencies depend on the molar mass:

$$\Theta_D^{(\mathrm{mag})} = \Theta_D^{(\mathrm{ref})} \sqrt{\frac{M_{\mathrm{ref}}}{M_{\mathrm{mag}}}} ; \omega_D = \frac{k_B \Theta_D}{\hbar} \qquad (20)$$

$\Theta_D^{(\mathrm{mag})}$ $\Theta_D^{(\mathrm{ref})}$ are corresponding Debye temperatures for magnetic and nonmagnetic compounds. The above scaling shifts the temperature axis appropriately to account for mass difference.

### 3.2 Electronic Heat Capacity

In metals, the total heat capacity is a sum of contributions from lattice vibrations and conduction electrons. If conduction electrons were treated as classical particles, their contribution would add $3R/2$ per mole, leading to a total high-temperature heat capacity of $9R/2$. However, experiments show that the electronic contribution is much smaller, especially at low temperatures. This discrepancy is explained by Sommerfeld's theory of the free electron gas, which recognizes that only electrons near the Fermi level can be thermally excited.

For most metallic solids, the Fermi temperature $T_F$ is on the order of $10^4$ K, far above the range of experimental measurements. As a result, only a fraction ($T/T_F$) of electrons participate in thermal excitations at accessible temperatures. If the Fermi surface is approximately constant with energy (which holds for most metals), the electronic heat capacity in this low-temperature regime is linear with respect to $T$:

$$C_e = \frac{\pi^2}{3} \mathrm{k}_\mathrm{B}^2 N(E_F) T = \gamma T \quad (21)$$

Here, $N(E_F)$ is the electronic density of states at the Fermi level, and

$$\gamma = \frac{\pi^2}{3} \mathrm{k}_\mathrm{B}^2 N(E_F) \qquad (22)$$

is the Sommerfeld coefficient. Experimentally, $\gamma$ provides a direct measure of the density of states at the Fermi level and is often used to assess electronic contributions.

This linear approximation typically holds up to $T/T_F \sim 0.1\,T$, well above the temperature range of typical heat capacity experiments. At higher temperatures, the electronic contribution becomes nonlinear, but such conditions are rarely encountered in practice (Figure S3b).[48] It is worth noting that the linear approximation breaks down when the density of states is small or varies strongly near the Fermi level, as in Dirac or Weyl semimetals.[49] Thus, for most experimental studies, the Sommerfeld model provides an accurate and practical description of the electronic heat capacity.[48, 50-58] As mentioned in section 2.3, for polycrystalline metallic samples, 2-10 mg is sufficient; however, more sample mass is generally recommended when measuring samples with low Sommerfeld coefficients. Additional guidance on sample preparation can be found in Table 9.1 in the PPMS Heat Capacity User Manual.[59]

**Low-Temperature Heat Capacity Analysis**

For most quantum materials, standard practice is to measure heat capacity between base temperature and 300 K. As discussed in section 3.1, at high temperatures ($\Theta_D/20 < T < \Theta_D$), the phononic contribution dominates. In contrast, at low temperatures, typically between $\Theta_D/50 < T < \Theta_D/20$, both phonons and electrons contribute significantly. This temperature window is particularly valuable because it allows us to extract fundamental parameters that describe the phononic and electronic properties of a material. In this regime, the total heat capacity can be modeled as the sum of two simple terms:

$$C_p = \gamma T + \beta T^3 \qquad (23)$$

The first term represents the linear electronic contribution while the second term describes the lattice contribution from the Debye $T^3$ law. The equation can be rearranged as:

$$\frac{C_p}{T} = \gamma + \beta T^2 \qquad (24)$$

Plotting $C_p/T$ versus $T^2$ produces a straight line. The intercept gives the Sommerfeld coefficient $\gamma$, which is directly related to the electronic density of states at the Fermi level. The slope gives $\beta$, from which the Debye temperature $\Theta_D$ can be estimated using the relation:

$$\Theta_D = \left(\frac{12\pi^4 RN}{5\beta}\right)^{1/3} \qquad (25)$$

where $N$ is the number of atoms per formula unit.

This approach is often the first step in analyzing low-temperature data because it provides a straightforward way to separate the electronic and lattice contributions.[50, 60-62]

For example, we consider the ferromagnetic heavy-fermion compound $Ce_2Ru_3Ge_5$, which has a magnetic ordering temperature of 8 K and a Kondo temperature of approximately 18.9 K.[63] Figure S3c shows its low-temperature specific heat plotted as $C_p/T$ versus $T^2$. A linear fit between 10 K and 25 K (blue line in Figure S3c) yields a Sommerfeld coefficient of $\gamma$ = 85 mJ/mol K$^2$, indicating a moderate enhancement in the electronic density of states at the Fermi level relative to most metals. The phononic coefficient is $\beta$ = 0.94 mJ/mol K$^4$. Using $N$ =10 atoms per formula unit for $Ce_2Ru_3Ge_5$, the calculated Debye temperature is approximately 195 K. In some cases where the presence of anharmonic lattice vibrations or Einstein modes, $C_p/T$ versus $T^2$ may show non-linear behavior. In such cases, the equation (24) should include an additional term $\delta T^4$ to accurately estimate $\gamma$.[59] For disordered or glassy materials, it is worth noting that $C_p/T$ function can scale with $T^{\alpha}$ ($1 < \alpha < 2$). This can sometimes mimic electronic heat capacity when $\alpha$ is close to 1. Thus, additional care is needed when interpreting the data.

### 3.3 Magnetic Contribution

Magnetism in solids originates from both orbital and spin angular momentum of electrons. Core electrons produce a weak diamagnetic response due to their orbital motion, which is present in all materials but is typically negligible compared with contributions from unpaired valence electrons.[64] Unpaired valence electrons give rise to magnetism and, under certain conditions, long-range magnetic order.

In atoms or ions with partially filled d- or f-shells, both spin ($S$) and orbital ($L$) angular momentum contribute to the magnetic moment. For rare-earth elements (4f electrons) and certain transition metals, the orbital contribution is particularly important. In these cases, the total angular momentum $J$ is determined by vector addition of $L$ and $S$, following Hund's rules.[64, 65]

Magnetic ordering emerges when atomic moments align cooperatively below a characteristic temperature, driven by exchange interactions (Figure S3d). The most universal and simplest model for these interactions is the Heisenberg Hamiltonian:

$$H_{ex} = -\sum_{i>j} J_{ij}\ S_i \cdot S_j \qquad (26)$$

where $J_{ij}$ is the exchange integral representing the energy cost or gain of aligning two spins on sites $i$ and $j$. A positive $J_{ij}$ favors parallel spin alignment, leading to a ferromagnetic (FM) order, while a negative $J_{ij}$ favors antiparallel alignment, producing an antiferromagnetic (AFM) order. It is worth noting that strong orbital moments can make this exchange anisotropic, such that the interaction is no longer a simple dot product.

In the mean-field approximation, long-range ordering is accompanied by a discontinuous jump in the heat capacity at the transition temperature. The magnitude of this jump is given by:[66, 67]

$$\Delta C = \frac{5S(S+1)}{S^2+(S+1)^2} \cdot R \qquad (27)$$

This expression applies to both FM and AFM transitions, since a heat capacity anomaly reflects changes in free energy rather than net magnetization. For ferrimagnetic systems with two inequivalent sublattices, the discontinuity is reduced according to

$$\Delta C = Q(\lambda, \alpha) \cdot \Delta C_{\mathrm{FM/AFM}} \qquad (28)$$

where $Q(\lambda, \alpha)$ depends on the ratio of molecular-field coefficients and the relative population or moment strength of the two sublattices.[67] Equations (27) and (28) assume mean-field theory, and real materials can strongly diverge from this. In particular, low-dimensional or strongly frustrated systems can have widely varying amounts of entropy recovered across a magnetic transition. [68, 69]

Magnetic transitions appear as anomalies in the heat capacity curve. In second-order (continuous) transitions, the anomaly takes the form of a pronounced lambda-shaped peak, often referred to as a "lambda anomaly" because of its resemblance to the Greek letter $\lambda$.[70, 71] In first-order transitions, the feature is sharper and more discontinuous, reflecting the release or absorption of latent heat. The positions and shapes of heat capacity anomalies reveal transition temperatures and provide insight into the underlying magnetic ordering.[72-83]

Once the phononic contribution is subtracted and the magnetic heat capacity is isolated, the magnetic entropy can be determined from the relation:

$$\Delta S_{\mathrm{mag}}(T) = \int_0^T \frac{C_{\mathrm{mag}}(T')}{T'}\, dT' \qquad (29)$$

The temperature dependence of $\Delta S_{\mathrm{mag}}(T)$ typically follows a sigmoid-like function. For first- and second-order transitions, schematic examples are shown in Figure 2a and 2b. The entropy approaches $R\ ln(2S+1)$ for spin-only systems and $R\ ln(2J+1)$ for systems that have both spin and orbital contributions.[84-88] Deviations from complete entropy recovery are often linked to crystalline electric-field effects, strong correlations or a loss of ergodicity (as in spin glasses) .[60, 89-95] Consequently, accurate phonon modeling is a prerequisite for interpreting magnetic entropy in complex materials.

**Spin Waves**

At $T > 0$ K, thermal fluctuations disrupt the alignment of ideal FM and AFM spin configurations, and the resulting collective excitations, spin waves (also known as magnons, which are bosonic quasiparticles), propagate over many lattice sites. These quasiparticles dominate the low-temperature magnetic heat capacity.[5, 96, 97]

Although the full magnon spectrum of a material can be quite complex, in the $T \to 0$ limit, only the lowest-energy band has appreciable population and contributes to specific heat. Conveniently, this allows one to analyze low-$T$ $C_{\mathrm{mag}}(T)$ using very simple models to determine basic magnetic properties such as the dimensionality, sign, and gap of the magnon spectrum.

For a generic set of $s$ bosonic dispersions $\omega_{\mathrm{s}}(k)$ (where $k$ is wavevector), the heat capacity is computed by integrating over a Brillouin zone:

$$C_{\mathrm{mag}}(T) = \mathrm{k_B} \sum_{\mathrm{s}} \int \mathrm{d}k \left(\frac{\hbar\omega_{\mathrm{s}}(k)}{\mathrm{k_B}T}\right)^2 \frac{e^{\hbar\omega_s(k)/k_B T}}{(e^{\hbar\omega_s(k)/k_B T}-1)^2}. \quad (30)$$

In the low-temperature limit with Heisenberg interactions (isotropic exchange), ferromagnetic (FM) magnons have a quadratic dispersion $\omega_k \propto k^2$,[95] and equation (30) gives a low-$T$ magnetic heat capacity $C_{mag} \propto T^{n/2}$ where $n$ is the dimensionality of the system.[95] Meanwhile, Heisenberg antiferromagnetic (AFM) magnons have a linear dispersion $\omega_k \propto k$, which gives a low-$T$ magnetic heat capacity $C_{mag} \propto T^n$.[98] Materials exhibiting long-range order are three-dimensional at sufficiently low energies, and thus at low temperatures $n$ takes the value of 3. Then, the total low-temperature heat capacity can therefore be modeled with a spin-wave term plus electronic and phononic contributions:

$$C_p = \gamma T + \beta T^3 + \delta T^n \qquad (31)$$

where the first term represents the electronic contribution and the second term accounts for phonons,[5] $\delta$ is a spin-wave coefficient that depends on the exchange interaction strength and spin-wave stiffness, and $n$ is determined by the dimensionality and sign of the magnetic exchange (usually $n = 3/2$ for ferromagnets, and $n = 3$ for antiferromagnets).

In practice, fitting spin-wave heat capacity is often non-trivial due to the overlap of phononic, Schottky, and magnonic contributions (Figure S3e). Although equation (31) works well for simple ferromagnets (Figure S3f), linear fitting frequently fails in both metallic and insulating materials, and a scaling-plot approach can be more robust than direct least-squares fitting.

Take yttrium iron garnet, a canonical ferrimagnetic insulator, as an example. Although yttrium iron garnet is ferrimagnetic with two inequivalent $Fe^{3+}$ sublattices, its low-energy excitations behave as ferromagnetic magnons at sufficiently low $T$.[99, 100] Plotting $C_p/T^{3/2}$ as a function of $T^{3/2}$ in the temperature range between 0.1 K and 2.5 K yields a linear region from 0.7–2.5 K consistent with equation (31); the fitting coefficient is directly related to the exchange constant and spin-wave stiffness. In this temperature window, phonons are already suppressed, making the scaling analysis more reliable (Figure S3f inset).

In practice, for an antiferromagnet $C_{mag} \propto T^3$, it is more challenging to separate AFM magnon contributions from phonons in heat capacity (especially in metals, where an additional electronic $\gamma T$ term can be present). Usually, a nonmagnetic analog must be subtracted to sufficiently distinguish these contributions.

If magnetic anisotropy opens a spin-wave gap $\Delta$, the low-temperature behavior crosses over to an activated form. Although such specific heat is often modeled as $C_{\mathrm{mag}} = \delta T^3 e^{-\Delta/k_B T}$ for antiferromagnets, this form is inaccurate and yields an incorrect value for $\Delta$ (by orders of magnitude, see Supplemental Information). A more accurate equation for fitting gapped linear dispersive magnons is:

$$C_{\mathrm{mag}} = \delta T^3 e^{-\frac{5}{8}\left(\frac{\Delta}{\pi k_B T}\right)^2} \quad (32)$$

where $\delta$ is a coefficient related to spin stiffness and magnon velocity (see Supplemental Information). Alternatively, if the magnon dispersion is known—for example from neutron spectroscopy—one can numerically compute the specific heat from equation (30).[49, 101-103] This can be useful as an exercise in searching for non-magnon contributions to specific heat, e.g., from more exotic quasiparticles.[49]

### 3.4 Schottky Contribution

Another important contribution to the heat capacity arises from the finite thermal population of excited energy levels in a material. Consider a magnetic ion in a crystal or ligand-field environment. In the absence of external fields, an ion with total angular momentum $J$ has $2J+1$ degenerate states. Under a crystal electric field (CEF), these levels are split into a set of crystal-field states. At finite temperatures, these states can be thermally populated according to the Maxwell–Boltzmann distribution.[104, 105] Similarly, at much lower energy scales, an atomic nuclear spin is split by hyperfine interactions and is also thermally populated according to the Maxwell-Boltzmann distribution (see next section).

The Schottky contribution reflects these excitations. Once the phononic and electronic contributions are carefully subtracted, it appears as a bell-shaped anomaly. The peak position reflects the energy gap between levels, while the peak height depends on the relative degeneracies. This makes Schottky analysis a powerful tool for probing local electronic environments and energy level splitting.

#### 3.4.1 Two-Level Schottky Model

The simplest case is a two-level system with equal degeneracies. This model captures the essential features of Schottky behavior and is widely used as a starting point:

$$\frac{C_{\mathrm{Sch}}}{R} = \left(\frac{\Delta}{T}\right)^2 \cdot \frac{e^{\Delta/T}}{\left(1+e^{\Delta/T}\right)^2} \quad (33)$$

Here $\Delta$ is the energy separation between the two levels. At low temperatures ($T \ll \Delta$), the anomaly is approximately exponential, while at high temperatures ($T \gg \Delta$) it falls off as $1/T^2$. This asymmetry gives the Schottky peak its characteristic skewed shape, with the maximum generally occurring near $T \approx 0.4\,\Delta$.[106-113]

#### 3.4.2 Multi-Level Schottky Model

Real materials often exhibit more complex level schemes, especially for ions with larger $J$ or $S$ values and in low-symmetry crystal fields.[114-117] For example, a three-level system with ground-state energy $E_0$, excited-state energies $E_1$ and $E_2$, and degeneracies $g_0$, $g_1$, and $g_2$, produces more complex heat-capacity profiles (Figure S6a and S6b).[118-120]

The generic equation for a nuclear Schottky anomaly from a set of discrete energy levels $[E_i]$ is

$$C(T) = \frac{1}{Z k_B T^2}\left[\sum_i E_i^2 e^{\frac{-E_i}{k_B T}} - \frac{1}{Z}\left(\sum_i E_i e^{\frac{-E_i}{k_B T}}\right)^2\right] \quad (34)$$

where $Z = \sum_i e^{\frac{-E_i}{k_B T}}$ is the partition function.[98] Simulations of equation (34) show that the relative degeneracies strongly affect the anomaly's shape (Figure S6a). The high-temperature Schottky peak is largely insensitive to the degeneracy ratio, whereas the low-temperature peak is significantly influenced by it. If the higher excited states are more degenerate, the low-temperature peak intensity increases (Figure S6a). Similarly, when the energy-level splittings are asymmetric, the Schottky curve can exhibit a double-peak feature (Figure S6b).

For example, $CeCuGa_3$ (non-centrosymmetric tetragonal, space group *I*4*mm,* Figure S6c) exhibits long-range magnetic order below 2.5 K and a broad Schottky-like anomaly between 5 K and 80 K.[100, 121] The hump near 30 K is typical of many Ce-based intermetallics and originates from CEF splitting of $Ce^{3+}$ levels. For $Ce^{3+}$ (a Kramers ion with half-integer spin) in tetragonal symmetry, the six-fold ground-state multiplets split into three doublets (Figure S6d). The thermal population of these states produces an asymmetric anomaly in the heat capacity. This can be modeled using a three-level Schottky expression: [65]

$$\frac{C_{\mathrm{Sch}}}{R} = \beta^2 \frac{g_1\Delta_1^2 e^{\beta\Delta_1} + g_2\Delta_2^2 e^{\beta\Delta_2} + g_1 g_2(\Delta_1-\Delta_2)^2 e^{\beta(\Delta_1+\Delta_2)}}{\left(g_0+g_1 e^{\beta\Delta_1}+g_2 e^{\beta\Delta_2}\right)^2} \quad (35)$$

where $\Delta_1$ and $\Delta_2$ are the energy gaps, and $g_0$, $g_1$, $g_2$ are the corresponding degeneracies. For $CeCuGa_3$, fitting yields three doublet levels at 0 K, 70 K, and 240 K, in excellent agreement with the observed anomaly (Figure S6d).

In practice, however, fitting multi-level systems can be a challenging task. Highly degenerate systems, such as $Er^{3+}$ and $Dy^{3+}$, often require many parameters, which may lead to unphysical results[119]. In such cases, Schottky fitting should be treated as a guide to the energy level scheme and ideally corroborated by spectroscopic or inelastic neutron-scattering data.

### 3.4.3 Nuclear Schottky Anomalies

In addition to electronic Schottky anomalies, nuclear Schottky anomalies are often observed in at very low temperatures in materials with large nuclear spins or strong hyperfine coupling. .[122, 123] The energy scale of nuclear Schottky anomalies is typically much lower than that of electronic ones. The temperature at which the nuclear Schottky anomaly becomes detectable depends on the strength of the hyperfine interaction, which varies from ion to ion. In some cases, the nuclear Schottky anomaly can be visible up to 4 K, but in most cases it appears only at sub-Kelvin or millikelvin temperatures (~50-300 mK).[55, 124-127]

For example. $CaMnTeO_6$ exhibits a clear Schottky anomaly below 0.5 K (Figure S6e, S6f).[128] The $Mn^{4+}$ ion has electronic spin $S = 3/2$ and nuclear spin $I = 5/2$. The hyperfine interaction gives rise to an 18-level system according to hyperfine coupling selection rules. However, the small overall energy scale and the observed data only exhibit a shallow upturn, indicating that all levels are nearly thermally populated within the measurement window. Thus, the heat capacity anomaly can be approximated by a two-level model, as mentioned above (equation 33). Note however that this extracted gap corresponds to the total splitting between the lowest and highest levels, and the extracted two-level gap is less accurate as one fits to lower temperatures (see Supplemental Information). Fitting this low-temperature anomaly yields a Schottky gap of approximately 0.015 K at zero field. Below $T = 1$ K, the phonon contribution is essentially quenched, leaving only electronic terms. The electronic contribution can be modeled as:

$$C_e = A + BT^c \qquad (36)$$

where $A$, $B$, and $c$ are fitting parameters; the exponent '$c$' often reflects the dimensionality of spin-wave excitations (see section 3.3 and Figure S6f).[129]. The nuclear Schottky contribution is more prominent in the mK range.[130, 131] In these cases, the total non-Schottky heat capacity is modeled as:

$$C_p - C_{sch} = (A + BT^c) + \beta T^3 \qquad (37)$$

The first term represents the low-temperature electronic contribution and the second term is the simplified Debye model accounting for acoustic phonons (Einstein modes are negligible at low $T$).[132] This approach enables effective isolation of the nuclear Schottky anomaly, which, if the hyperfine coupling constants are known, can be used as a local measure of the root-mean-square static electronic magnetism.[133]

## 4. Exotic Phases

With the fundamentals and practical tools in place, we now demonstrate how heat capacity measurements and analysis provide insights into real quantum materials, such as superconductors, heavy-fermion metals, frustrated magnets and spin liquids, skyrmion hosts, and topological or multipolar systems.

### 4.1 Superconductivity

Superconductivity is a quantum phenomenon characterized by zero electrical resistance and the expulsion of magnetic fields below a critical temperature, $T_c$. Since its discovery in mercury by Kamerlingh Onnes in 1911, the field has grown with major advances in materials discovery, theoretical understanding, and applications.[134-144]

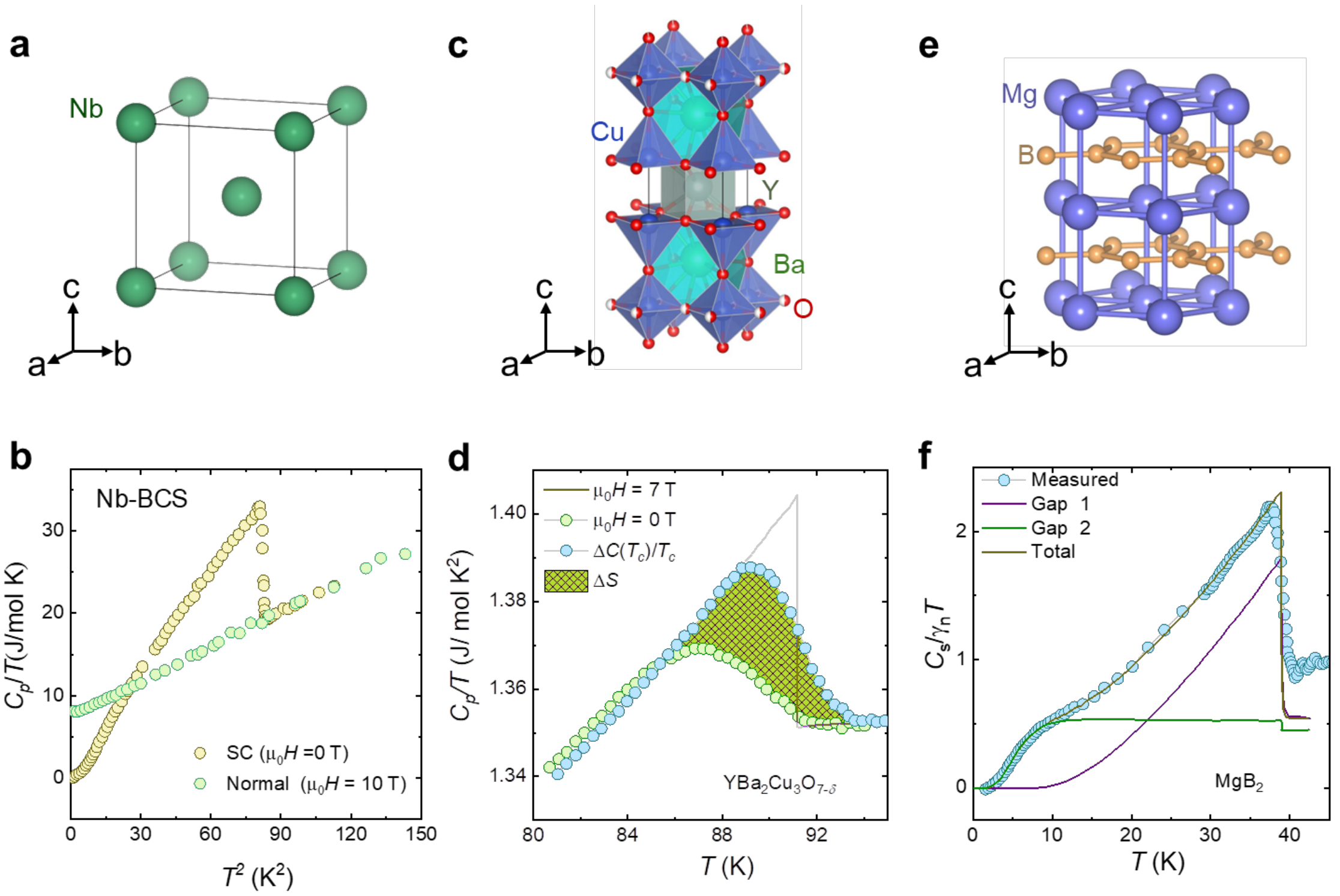

**Figure 4.** (a) Niobium body-centered cubic crystal structure. (b) $C_p/T$ vs $T^2$ plot for $\mu_0H$ = 0 and 1 T (Adapted with permission from Ref[145] © 1962 American Physical Society) (c). Orthorhombic crystal structure of YBCO showing the $Cu^{2+}$ square pyramid and octahedral network. (d) $C_p/T$ vs. $T$ at $\mu_0H$ = 0 and 1 T, the residual entropy is shown as the shaded area indicating the non-BCS type superconductivity (Adapted from Ref[146] : open access) . (e) Hexagonal $AlB_2$-type structure of $MgB_2$ showing that B atoms form a honeycomb layer that is sandwiched between two hexagonal Mg layers. (f) $C_p/\gamma_n T$ vs. $T$ showing a two-gap model fit. (Adapted with permission from Ref[147] © 2015 Elsevier B.V.).

To analyze the heat capacity of a conventional superconductor, one must first account for its normal state phononic and electronic contributions. This can be achieved by fitting the low-temperature normal-state data using equation (24). The ratio $\Delta C/\gamma_n T_c$ ($\Delta C$- electronic heat capacity jump at $T_c$) provides information about the coupling strength; for weak-coupling BCS superconductors, it is expected to be 1.43.[134, 136, 148, 149] It should be noted here that this ratio is valid only for fully gapped *s*-wave superconductors. For *d*-wave and nodal *s*-wave superconductors, the ratio is reduced due to the presence of nodes.

Further insights can be gained by applying a magnetic field to suppress superconductivity. In conventional cases, the thermodynamic critical field follows:

$$H_c(T) = H_0(1 - (T/T_c)^2) \qquad (38)$$

Above $H_c$, the system returns to the normal state. The behavior of heat capacity under applied field is noticeably different for type-I and type-II superconductors, as shown examples for pure Ta and its alloys.[150] From the data with applied fields, one can extract the modified $\gamma_s$ and analyze the superconducting electronic heat capacity $C_{es}$.[7, 151, 152] An exponential suppression of $C_{es}/\gamma_s T_c$ at low temperature is a hallmark of a fully gapped *s*-wave state, resulting in an estimate of the superconducting gap $\Delta$ (Figure S7).

Let us consider Niobium as an example. The material is a conventional type-II superconductor with $T_c$ = 9.2 K (Figure 4a).[153-155] The upper critical field of Nb is between 0.4 T and 0.8 T, depending on the sample purity. Figure 4b shows that the heat-capacity anomaly is suppressed at 1 T. The extracted normal-state parameters are $\gamma_n$ = 7.5 mJ/mol K$^2$ and $\Theta_D$ = 238 K. Plotting $C_{es}/\gamma T_c$ versus $T/T_c$ reveals a near-linear correlation consistent with a full superconducting gap (Figure S7).[145, 152] The superconducting gap $\Delta(T)$ decreases with increasing temperature, and the gap vanishes at the critical temperature as predicted by the BCS theory. While the linear fit of $C_{es}/\gamma T_c$ versus $T/T_c$ results in the zero-temperature gap $\Delta(0)$, the temperature-dependent superconducting gap $\Delta(T)$ can be extracted from measurements and fitting (Figure S8).[156] The temperature dependence of $\Delta(T)$ reflects the collective nature of charge carriers, the binding energy of Cooper pairs, and the phonon-assisted process. The BCS theory predicts that the binding energy weakens as temperature increases. Hence, the linear fit of $C_{es}/\gamma T_c$ versus $T/T_c$ data must be carried out only within the $T < T_c/6$ temperature window, where the gap value is saturated.

In contrast, unconventional superconductors such as $YBa_2Cu_3O_{7-\delta}$ (YBCO) do not follow the BCS theory (Figure 4c).[157, 158] While an *s*-wave superconductor exhibits an exponential decay of $C_{es}$ below $T_c$, YBCO retains a finite linear term $\gamma(0)$ even in the superconducting state. This residual term is often taken as evidence for *d*-wave symmetry, where nodes exist in the superconducting state. $C_p/T$ vs. $T$ curves at 0 and 7 T reveal residual entropy, highlighting the non-BCS character (Figure 4d). Another possible explanation for the residual entropy derives from $Cu^{2+}$ impurities. $\gamma(H)$ scales approximately with $\sqrt{H}$, and the heat capacity is sensitive to impurities such as $Cu^{2+}$. As impurity concentration increases, $\Delta C$ decreases without a change in $T_c$, suggesting a reduction in the superconducting volume fraction rather than a uniform gapless state.[146, 159]

These observations highlight the need for careful interpretation. Residual linear terms may result from impurities or structural inhomogeneities rather than intrinsic behavior. When analyzing heat capacity in unconventional superconductors, it is essential to distinguish between intrinsic properties and extrinsic contributions, especially in disordered systems.

Further complexity arises in multiband superconductors such as $MgB_2$, which possesses a relatively high $T_c = 39$ K (Figures 4e and 4f).[160, 161] Although initially classified as a conventional *s*-wave system, $MgB_2$ exhibits $\Delta C/\gamma T_c = 1.09$, smaller than the expected 1.43, and shows a second hump in $C_{es}(T)$ at low temperature.[147, 162] These features cannot be explained by a single-gap model. Instead, a two-gap model was proposed, where the total electronic heat capacity is modeled as a weighted sum:

$$C_{es}(T) = \omega_1 C_{\Delta_1}(T) + \omega_2 C_{\Delta_2}(T) \qquad (39)$$

Using this two-gap model, Bouquet and co-workers successfully reproduced the full temperature dependence of the heat capacity, resulting in $\Delta_1 = 6.8$ meV for the strongly coupled $\sigma$-band and $\Delta_2 = 1.8$ meV for the weaker $\pi$-band.[163] The heat-capacity analysis plot depicts the experimental data, individual gap contributions, and total fit (Figure 4f). This approach is now standard for identifying multiband superconductivity in complex materials, although it has to be noted that in some cases heat capacity measurements may not be sufficient to distinguish between multi-gap and anisotropic single gap system.

### 4.2 Magnetic Skyrmions

Magnetic skyrmions are topologically protected spin textures in which magnetic moments form swirling, vortex-like configurations.[164-166] Across the skyrmion core, the spin directions continuously rotate from "up" to "down," maintaining a fixed chirality. This unique arrangement cannot be removed or transformed without introducing a discontinuity, giving skyrmions exceptional stability against perturbations. Examples of magnetic skyrmions include MnSi[167, 168], FeGe[169, 170], $GaV_4S_8$[171], $Co_8Zn_8Mn_4$[172], EuPtSi[173], $Gd_3Ru_4Al_{12}$[174], $GdRu_2X_2$ (X = Si, Ge) and $Gd_2PdSi_3$.[175, 176]

Heat capacity, especially when paired with entropy mapping, offers a thermodynamic probe of skyrmion formation. The isothermal magnetic entropy change, $\Delta S(H, T)$, can be evaluated using the Maxwell relation:

$$\left(\frac{dS}{dH}\right)_T = \left(\frac{dM}{dT}\right)_H \qquad (40)$$

This implies that field-induced entropy changes can be estimated from magnetization isotherms. Entropy-driven phase transitions (such as skyrmion formation) typically produce positive entropy peaks, while conventional magnetic transitions (e.g., antiferromagnetic order) yield negative peaks.[134]

For example, $GeRu_2Ge_2$ displays two skyrmion regions at 0.9 T < $H$ < 1.2 T and 1.3 T < $H$ < 1.7 T within the temperature range of 2 K < $T$ < 30 K (Figure 5a).[177] To gain insight into its phononic contributions, the high-temperature heat capacity was analyzed using both the one-Debye-one-Einstein and the two-Debye models. Both models reproduce the experimental data between 50 K and 300 K; however, given the absence of a peak in $C_p/T^3$, the two Debye model is preferred. After extracting the two Debye temperatures and Sommerfeld coefficient, we extrapolate the model across the full range down to 2 K. This background subtraction yields the magnetic contribution:

$$C_{mag}(T) = C_p(T) - C_{ph}(T) - C_e(T) \qquad (41)$$

Several anomalies appear in the 2–33 K range. The peak at $T_N$ = 33 K gradually diminishes with increasing field, whereas the anomaly at $T_0$ = 20 K and $\mu_0 H$ = 1.2 T vanishes at higher fields (Figure 5b). To assess the entropy change associated with these transitions, we calculate the magnetic entropy:

$$\Delta S_{mag}^{ph,e} = \int_0^T \frac{C_{mag}}{T'} \, dT' \qquad (42)$$

We then obtain the isothermal magneto-entropy change by referencing the zero-field curve:

$$\Delta S_{mag}^{H} = \Delta S(H,T) - \Delta S(0,T) \qquad (43)$$

This analysis reveals a positive entropy peak at 20 K, associated with the skyrmion evolution, and a negative peak at 33 K, corresponding to AFM ordering (Figure 5c).[177]

These results underscore the importance of thermodynamic measurements for probing subtle phase transitions, such as the emergence of skyrmions. However, heat-capacity analysis should be complemented by other characterizations, such as Hall effect, magnetization, Lorentz TEM, and small-angle neutron scattering measurements, to achieve a comprehensive understanding of these topologically nontrivial spin states.

**4.3 Quantum Spin Liquids**

A quantum spin liquid (QSL) is an exotic magnetic phase in which no long-range magnetic order is present—even at very low temperatures—despite strong exchange interactions that would typically favor ordering. Instead, spins remain in a highly entangled quantum state, characterized by dynamic fluctuations and emergent excitations, such as spinons, visons, or Majorana fermions.[117, 178-186] QSL behavior is commonly associated with frustrated spin lattices. Examples include (i) triangular-lattice compounds, such as the organic salts κ-$(ET)_2Cu_2(CN)_3$ and $EtMe_3Sb[Pd(dmit)_2]$;[187-189] (ii) kagome-lattice systems, such as herbertsmithite ($ZnCu_3(OH)_6Cl_2$) and Zn-doped barlowite;[190-193] and (iii) honeycomb-lattice materials, such as α-$RuCl_3$ and $Na_2IrO_3$,[194-197] Heat capacity and entropy analysis can provide important thermodynamic evidence for exotic excitations.

As a case study, we examine α-$RuCl_3$, a field-induced QSL candidate (Figure 5d).[47, 198] At zero field, α-$RuCl_3$ undergoes zigzag AFM ordering around 6.5 K, but this order is suppressed under an external magnetic field, pushing the system toward a quantum spin liquid regime. Unlike sharp transitions in ordered magnets, QSLs may not exhibit sharp ordering transitions (depending on the type of QSL that forms). Therefore, careful subtraction of the phononic contribution is essential. Standard Debye or Einstein models may be insufficient, especially when Schottky-like anomalies or localized Einstein modes overlap with the signal. Subtraction using a nonmagnetic analog, scaled by mass-corrected Debye temperature, is generally more reliable.[197]

After phononic subtraction, the magnetic heat capacity $C_{mag}(T)$ of α-$RuCl_3$ shows two broad features: one near the suppressed AFM transition around 10 K, and another broader anomaly near $T_H \sim 100$ K (Figure 5e).[197, 199] These are interpreted as thermodynamic signatures of localized and itinerant Majorana fermions. The corresponding magnetic entropy is evaluated using equation 29.

This reveals a two-step increase in entropy that forms a plateau around $0.46Rln2$, consistent with spin-½ moment fractionalization.[47] Another hallmark of QSL behavior is observed at low temperatures: in α-$RuCl_3$, the magnetic heat capacity $C_{mag}$ exhibits a linear dependence on temperature, even though the material is an insulator. This is striking, because such $T$-linear behavior is normally a signature of Fermionic quasiparticles (like in metals), and is inconsistent with 2D antiferromagnetic magnons (equation (31)). Its appearance here points to fermionic fractional excitations in the QSL state (Figure 5f). [200]

These broad, unconventional signatures make heat capacity a vital probe for identifying QSL behavior—but they must be interpreted in conjunction with complementary data from neutron scattering, thermal transport, and magnetic measurements.

Before moving on, we note that QSL ground states generically do not involve residual entropy in their ground states. Although classical spin liquids like $Dy_2Ti_2O_7$ are famous for a residual $T = 0$ entropy, a quantum spin liquid is a well-defined eigenstate of the system (in the language of quantum mechanics, its ground state is diagonal in the correct basis). [94, 201] As such, it has no degeneracies and no residual entropy, despite the fact that its spins are in superposition in the local spin basis. This absence of zero-point entropy can also be demonstrated by induction: a spin singlet (two spins in a superposition of up-down and down-up) has no entropy despite its fluctuating ground state. Likewise, a spin quadruplet also has no residual entropy, and so on up to an arbitrary number of spins in a collective superposition state.

### 4.4 Heavy Fermions

In some intermetallic systems containing rare-earth (Ce, Yb) or actinide (U) elements, the partially filled 4*f*/5*f* states hybridize with conduction electrons at low temperatures. This hybridization renormalizes the carriers into heavy quasiparticles that behave as if they have an effective mass, $m_e$, much larger than that of free electrons. This results in a large Sommerfeld coefficient $\gamma$, enhanced magnetic susceptibility $\chi$, and a characteristic $\rho = \rho_0 + AT^2$ behavior for their conductivity.[202-204] At sufficiently low temperatures, many compounds exhibit Landau-Fermi-liquid behavior, characterized by an enhanced resistivity coefficient $A$.

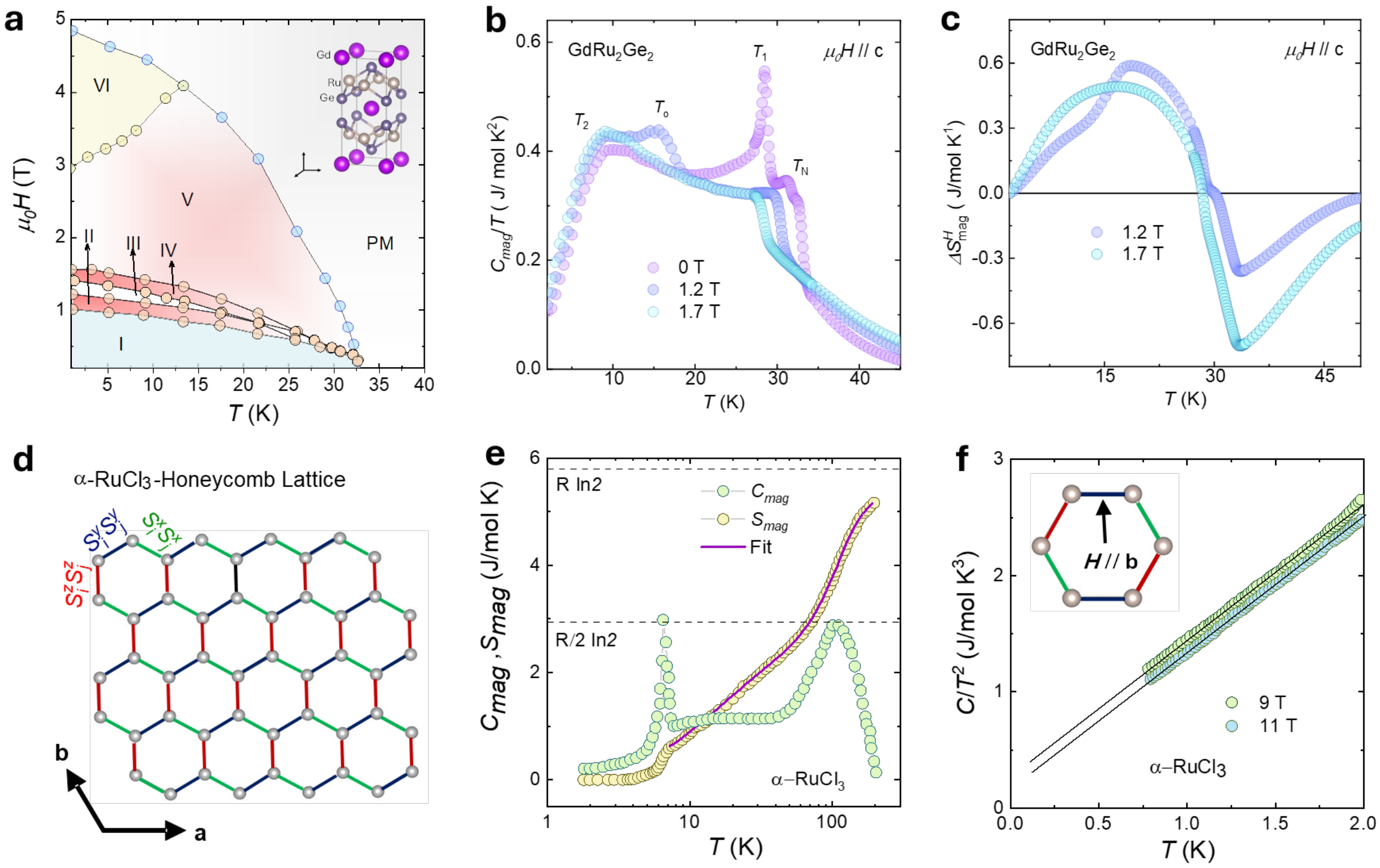

**Figure 5.** (a) Magnetic phase diagram of $GdRu_2Ge_2$ showing several magnetic phases, including including a screw spin (Phase *I*), elliptic skyrmion (Phase *II*), meron-antimeron pair (Phase *III*), circular skyrmion (Phase *IV*) and vortex lattice (Phase *V*). Phase II and IV are two nontrivial topological spin structures (Adapted with permission from Ref[205] © 2024 Springer Nature), with the inset showing $GdRu_2Ge_2$ tetragonal unit cell. (b) $C_{mag}/T$ as a function of temperature for $\mu_0H$ = 0, 1.2 and 1.7 T. (c) Magnetic entropy change $\left(\Delta S^H_{mag}\right)$ as a function of temperature showing a positive peak at 20 K for $\mu_0H$ = 1.2 and 1.7 T. (d) $Ru^{3+}$ ($J_{eff}$ = 1/2) honeycomb layer of the α-$RuCl_3$ compound. The bond-dependent Ising interactions are highlighted with different colors. (e) Temperature variation of $C_{mag}$ (green symbols) and $S_{mag}$ (yellow symbols). The pink solid line is the model fit (Adapted with permission from Ref[47] © 2017 Springer Nature). (f) $C/T^2$ vs. $T$ plot for $\mu_0H$ || **b** at 10 and 11 T (Adapted with permission from Ref[206] © 2025 American Physical Society).

Two empirical scaling approaches are particularly useful. First, the Kadowaki–Woods relation places a wide range of heavy-fermion compounds on a near-universal line when $A$ is plotted versus $\gamma^2$ in log scales, indicating that renormalized quasiparticles control charge transport. Second, the Wilson ratio, relating $\chi$ to $\gamma$, reveals a nearly linear correlation on log–log axes for many compounds, underscoring the coupling between localized moments and conduction-electron spin density (Figure 6a). [203, 207-210]

$Ce_2Ru_3Ge_5$ illustrates the ideal scenario: a linear window in $C_p/T$ vs $T^2$ allows straightforward extraction of $\gamma$ and $\beta$.[63] In practice, however, nonmagnetic impurities, spin fluctuations, or spin-glass-like behavior can masquerade as heavy-fermion signatures.[211] Caution is therefore essential when applying the simple $C_p = \gamma T + \beta T^3$ model, especially if multiple phases or anomalies coexist.

Another example is $CeCu_2Si_2$, a prototypical heavy-fermion superconductor (Figure 6b).[212, 213] Heat capacity reveals an extraordinarily large $\gamma \approx 1.3$ J mol$^{-1}$ K$^{-2}$, nearly three orders of magnitude above those of ordinary metals, signaling huge $m_e$. Yet the $C_p/T$ vs $T^2$ plot is non-linear (U-shaped) over 0.4-20 K (Figure 6c). The basic $\gamma T + \beta T^3$ form is reliable only in the mK range, and the superconducting transition near 0.4-0.6 K adds further curvature. Any "linear fit" over a finite window thus yields effective parameters rather than intrinsic $\gamma$.

$CeAl_3$ presents a classic heavy-fermion compound with $\gamma \approx 1.620$ J mol$^{-1}$ K$^{-2}$ (Figure 6d).[214] The $C_p/T$ vs $T^2$ plot remains curved even at the lowest temperatures, reflecting residual entropy from magnetic impurities, Schottky contributions, and Kondo-lattice effects. Reported $\gamma$ values are therefore extracted from a narrow temperature window (0.015-0.15 K) where curvature is minimal (Figure 6e).

Heavy fermions also provide canonical examples of non-Fermi-liquid behavior. $UPt_3$ exhibits a large $\gamma \approx 0.45$ J mol$^{-1}$ K$^{-2}$, deviating from Fermi-liquid scaling and unconventional superconductivity.[203, 215] When no clear linear region exists in $C_p/T$ vs $T^2$ plot, a common approach is to augment the low-$T$ expression with a logarithmic term:[203]

$$C_p(T) = \gamma T + \beta T^3 + \delta T^3 \ln T. \qquad (44)$$

Dividing by $T$ and rearranging Equation (44) leads to:

$$\frac{C_p/T - \gamma}{T^2} = \beta + \delta \ln T \qquad (45)$$

So, a plot of $C_p/T - \gamma/T^2$ vs. $\ln T$ should be linear, with a slope of $\delta$ and an intercept of $\beta$. This provides a practical diagnostic for non-Fermi-liquid behavior (Figure S9).

Overall, before attempting a linear fit, it is important to verify that superconducting, magnetic, or Schottky anomalies are not present in the chosen window; and select a small temperature range. Broader ranges often mix multiple contributions and bias $\gamma$ and $\beta$. As discussed in section 2.4, if the system possesses active nuclear and electronic degrees of freedom, the different relaxation time scales will introduce additional internal relaxation. We discuss an example in Supplemental Information (Figure S10).

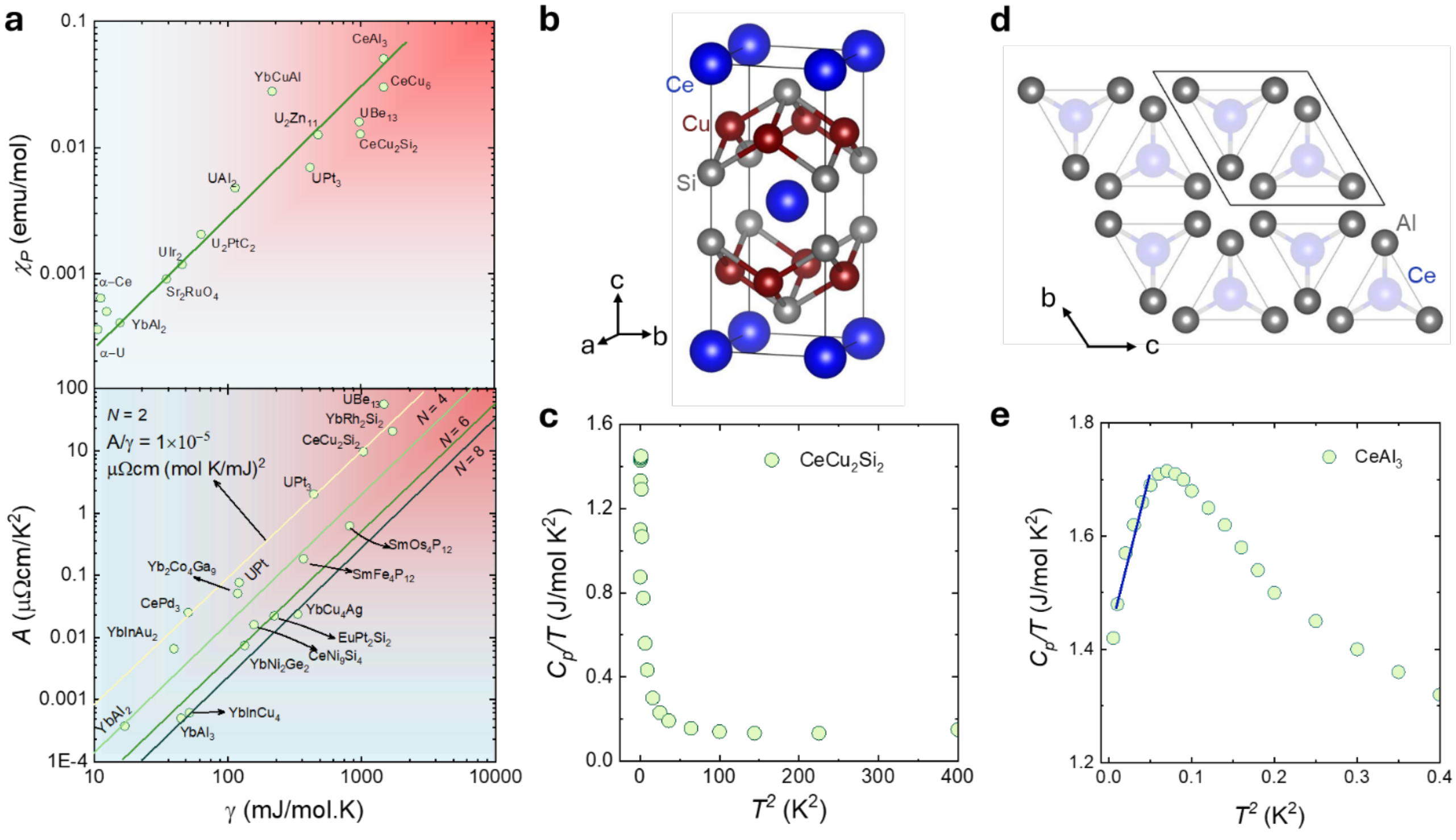

**Figure 6.** (a) Wilson ratio scaling plot between Sommerfeld electronic coefficient ($\gamma$) vs. Pauli spin susceptibility maximum (top)[216]. Kadowaki-Woods plot constructed using the $\gamma$ value and resistivity coefficient $A$ (bottom). The lines $N = 2, 4, 6, 8$ represent the f-orbital degeneracy (Adapted with permission from Ref[208] © 2005 American Physical Society). (b) $ThCr_2Si_2$-type tetragonal unit cell of $CeCu_2Si_2$. (c) $C_p/T$ as a function of $T^2$ for $CeCu_2Si_2$ (Adapted with permission from Ref[217] © 1983 American Physical Society), (d) Hexagonal $CeAl_3$ crystal structure showing Ce atoms in hexagonal layers surrounded by Al trigonal prisms and (e) $C_p/T$ vs. $T^2$ for $CeAl_3$ (Adapted with permission from Ref[203] © 1984 American Physical Society), respectively.

## 5. Thermopower and Heat Capacity

The thermoelectric effect is a non-equilibrium process: an applied temperature gradient drives charge carriers to diffuse, generating an open-circuit voltage set by the thermopower (Seebeck coefficient) $\alpha$.[218] When spin/orbital degeneracy is active near $E_F$, thermopower and heat capacity carry consistent entropy information: $C_p$ from equilibrium and $\alpha$ from non-equilibrium transport. Recently, spin-driven thermoelectric materials have attracted interest because spin entropy can enhance performance (effective $ZT$) in systems where magnetic degrees of freedom contribute to carrier transport.[219-221] Spin entropy originates from the degeneracy of magnetic ions, especially when multiple orbital and spin states are available, so that hopping processes can carry both charge and spin (Figures 7a-b).

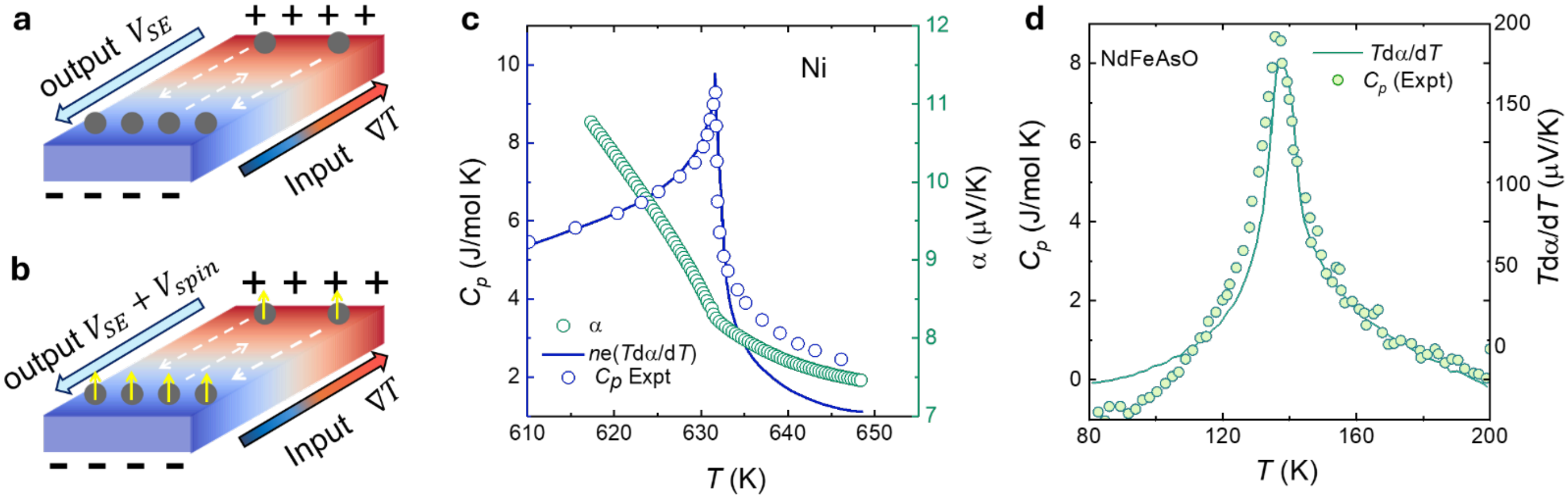

**Figure 7.** (a) Schematic diagram of carrier diffusion in a conventional thermoelectric measurement. (b) spin (magnetic) entropy contribution in a correlated magnetic system (c). Temperature dependence of heat capacity (blue symbols), Seebeck coefficient (green symbols) and the first derivative of Seebeck coefficient $ne(Td\alpha/dT)$ in heat capacity units (blue line) for nickel (Adapted with permission from Ref[222] © 1971 American Physical Society) (d). Temperature dependence of heat capacity (symbols) and $Td\alpha/dT$ (solid lines) for NdFeAsO (Adapted with permission from Ref[223] © 2009 American Physical Society)

In strongly correlated electron systems, heat capacity $C_p$ and thermopower $\alpha$ can both probe carrier entropy. They are, however, complementary: $C_p$ measures entropy in equilibrium, whereas $\alpha$ measures entropy transported per carrier under a temperature gradient. A thermodynamic link between them follows from the Kelvin relations:[223]

$$\left(\frac{\partial S}{\partial N}\right)_U = \frac{\mu}{T};\ \alpha = \frac{1}{e}\left(\frac{\partial S}{\partial N}\right)_U \quad (46)$$

Differentiating with respect to temperature yields:

$$T\,\frac{d\alpha}{dT} = \frac{1}{ne}C_e \quad (47)$$

where $C_e$ is the electronic heat capacity, $n$ is the carrier concentration (per formula unit), and $e$ is the elementary charge. Equation (47) is the practical bridge: the slope of $\alpha(T)$, scaled by $T$, mirrors $C_e$ once the carrier density is known.

Nickel exhibits ferromagnetic ordering below 631 K. Its magnetism derives from a partially filled 3d band that overlaps a broader s band, placing Ni on the border between itinerant and localized behavior. The thermopower shows a positive hump (Figure 7c). Using equation (47), the curve $T\,d\alpha/dT$ closely tracks the magnetic/electronic heat capacity plotted in the same units.[222] The unit conversion is delicate: one must use the itinerant carrier density rather than a multiband total. For Ni, there are ~9.4 d electrons and ~0.6 holes. Using the d-hole density $n$, $N_A$ and $e$, and $\mu$V K$^{-1}$ to J mol$^{-1}$ unit conversion, a quantitative superposition of $ne(T\,d\alpha/dT)$ and $C_p$ can be derived.[221]

NdFeAsO—a spin-density-wave material—displays similar features (Figure 7d). Both $C_p$ and $\alpha$ display a hump near ~140 K. For Ni, we demonstrate the comparison in common energy units; for NdFeAsO we show the correlation with $C_p$ and $\alpha$ in their native units. A scaling plot of $T\,d\alpha/dT$ versus $C_p$ is linear (Figure S11), implying that transport is governed by the spin degrees of freedom. From the slope ($\propto -1/ne$), we obtain an effective carrier count of −0.26 carriers per formula unit. This provides an effective method for estimating $n$ when Hall coefficients are ambiguous due to multiband or anomalous contributions.[223]

## 6. Summary and Outlook

Heat capacity is a uniquely versatile probe of quantum materials. Unlike transport, which primarily addresses metals, or magnetometry, which requires a magnetic response, heat capacity applies broadly—from metals to insulators and magnets to non-magnets—and captures phononic, electronic, magnetic, and energy-level excitations in a single thermodynamic observable. As such, it bridges the insight gap between charge and spin sectors and reveals physics that other characterization techniques cannot capture. This technical review has integrated essential background with practical methodology for new researchers. If the data are analyzed properly, one can gain valuable insights into spin, orbital, and lattice degrees of freedom, as well as energy-level populations. Integrating heat capacity analysis with complementary techniques, such as neutron scattering, Hall effect, thermal transport, magnetic susceptibility characterization, and first-principles calculations of phonon and electron band structures, provides a comprehensive understanding of quantum materials. Technological advances enabling measurements under high magnetic fields, sub-mK temperatures, and high pressures will further enhance the utility of heat capacity measurements. Heat capacity will therefore continue to be one of the most powerful and versatile tools for characterizing quantum materials.

## Acknowledgement

R.K., X.H., and T.T.T acknowledge support from the Arnold and Mabel Beckman Foundation, the NSF CAREER award NSF-DMR-2338014, and the Camille Henry Dreyfus Foundation. A.S. acknowledges support from the Laboratory Directed Research and Development program of Los Alamos National Laboratory under project number 20240083DR.

## References

(1) Pei, G.; Xiang, J.; Li, G.; Wu, S.; Pan, F.; Lv, X. A Literature Review of Heat Capacity Measurement Methods. In *10th International Symposium on High-Temperature Metallurgical Processing*, 2019; Springer: pp 569-577.
(2) Gopal, E. S. R. In *Specific Heats at Low Temperatures*, Plenum Press, 1966; pp 1-18.
(3) Tari, A. Basic Concepts and Definitions. In *The Specific Heat of Matter at Low Temperatures*, Tari, A. Ed.; Imperial College Press, 2003; pp 1-18.
(4) Tishin, A. M.; Spichkin, Y. I. *The Magnetocaloric Effect and Its Applications*; CRC Press, 2016.
(5) Rosen, P. F.; Woodfield, B. F. Standard Methods for Heat Capacity Measurements on a Quantum Design Physical Property Measurement System. *The Journal of Chemical Thermodynamics* **2020**, *141*, 105974.
(6) Phillips, N. E. Low-temperature Heat Capacity of Metals. *Critical Reviews in Solid State and Material Sciences* **1971**, *2*, 467-553.
(7) Sparks, L. L. Specific Heat. In *Materials at Low Temperatures*, Reed, R. P., Clark, A. F. Eds.; ASM International, 1984; pp 47-71.
(8) Aamlid, S. S.; Oudah, M.; Rottler, J.; Hallas, A. M. Understanding The Role of Entropy in High Entropy Oxides. *Journal of the American Chemical Society* **2023**, *145*, 5991-6006.
(9) Nishimori, H.; Ortiz, G. *Elements of Phase Transitions and Critical Phenomena*; Oup Oxford, 2010.
(10) Domb, C. *Phase Transitions and Critical Phenomena*; Elsevier, 2000.
(11) Karigerasi, M. H.; Kang, K.; Ramanathan, A.; Gray, D. L.; Frontzek, M. D.; Cao, H.; Schleife, A.; Shoemaker, D. P. In-plane Hexagonal Antiferromagnet in the Cu-Mn-As system $Cu_{0.82}Mn_{1.18}As$. *Physical Review Materials* **2019**, *3* (11), 111402.
(12) Stishov, S. M.; Petrova, A. E. Helical Itinerant MnSi Magnet: Magnetic Phase Transition. *Physics-Uspekhi* **2017**, *60*, 1268.

(13) Shibauchi, T.; Carrington, A.; Matsuda, Y. A Quantum Critical Point Lying Beneath the Superconducting Dome in Iron Pnictides. *Annual Review of Condensed Matter Physics* **2014**, *5*, 113-135.
(14) Schwartz, M. Lecture 9: Phase Transitions. *Harvard Univ* **2019**, *1*, 1-22.
(15) Witteveen, C.; Nocerino, E.; López-Paz, S. A.; Jeschke, H. O.; Pomjakushin, V. Y.; Månsson, M.; von Rohr, F. O. Synthesis and Anisotropic Magnetic Properties of $LiCrTe_2$ Single Crystals With a Triangular-lattice Antiferromagnetic Structure. *Journal of Physics: Materials* **2023**, *6*, 035001.
(16) Woodfield, B. F. *Specific Heat of High Temperature Superconductors: Apparatus and Measurements*; University of California, Berkeley, 1995.
(17) Dachs, E.; Bertoldi, C. Precision and Accuracy of The Heat-pulse Calorimetric Technique: Low-temperature Heat Capacities of Milligram-Sized Synthetic Mineral Samples. *European Journal of Mineralogy* **2005**, *17*, 251-261.
(18) Morin, F.; Maita, J. Specific Heats of Transition Metal Superconductors. *Physical Review* **1963**, *129*, 1115.
(19) Schwall, R.; Howard, R.; Stewart, G. Automated Small Sample Calorimeter. *Review of Scientific Instruments* **1975**, *46*, 1054-1059.
(20) *PPMS Heat Capacity Option User's Manual*; Quantum Design, San Diego, CA, 2010. PPMS Heat Capacity Option User's Manual, "1085-150, Rev. L3" Quantum Design.
(21) Nernst, W. Research on specific heat at low temperatures. II Sitzb. *Kgl. Preuss. Akad. Wiss* **1910**, *12*, 261.
(22) Eucken, A. Über die Bestimmung spezifischer Wärmen bei tiefen Temperaturen. *Physikalische Zeitschrift* **1909**, *10*, 586-589.
(23) Sullivan, P. F.; Seidel, G. Steady-state, ac-temperature calorimetry. *Physical Review* **1968**, *173* (3), 679.
(24) Bachmann, R.; DiSalvo Jr, F.; Geballe, T.; Greene, R.; Howard, R.; King, C.; Kirsch, H.; Lee, K.; Schwall, R.; Thomas, H.-U. Heat Capacity Measurements on Small Samples at Low Temperatures. *Review of Scientific Instruments* **1972**, *43*, 205-214.
(25) Lashley, J.; Hundley, M.; Migliori, A.; Sarrao, J.; Pagliuso, P.; Darling, T.; Jaime, M.; Cooley, J.; Hults, W.; Morales, L. Critical Examination of Heat Capacity Measurements Made on a Quantum Design Physical Property Measurement System. *Cryogenics* **2003**, *43*, 369-378.
(26) Tsui, S.; Sherman, S.; Averitt, R. Heat Capacity of $Fe_3O_4$ (Magnetite)(Heat Capacity Option).
(27) Kennedy, C. A.; Stancescu, M.; Marriott, R. A.; White, M. A. Recommendations for accurate heat capacity measurements using a Quantum Design physical property measurement system. *Cryogenics* **2007**, *47*, 107-112.
(28) Suzuki, H.; Inaba, A.; Meingast, C. Accurate Heat Capacity Data at Phase Transitions From Relaxation Calorimetry. *Cryogenics* **2010**, *50*, 693-699.
(29) Javorský, P.; Wastin, F.; Colineau, E.; Rebizant, J.; Boulet, P.; Stewart, G. Low-temperature heat capacity measurements on encapsulated transuranium samples. *Journal of Nuclear Materials* **2005**, *344* (1-3), 50-55.
(30) Hwang, J. S.; Lin, K. J.; Tien, C. Measurement of Heat Capacity by Fitting the Whole Temperature Response of a Heat-pulse Calorimeter. *Review of Scientific Instruments* **1997**, *68*, 94-101.
(31) Shepherd, J. P. Analysis of the Lumped $\tau_2$ Effect in Relaxation Calorimetry. *Review of Scientific Instruments* **1985**, *56*, 273-277.
(32) Hardy, V.; Bréard, Y.; Martin, C. Derivation of The Heat Capacity Anomaly at a First-order Transition by Using a Semi-adiabaticrelaxation Technique. *Journal of Physics: Condensed Matter* **2009**, *21*, 075403.
(33) Scheie, A. LongHCPulse: Code for processing long pulse heat capacity data taken on a Quantum Design PPMS. **2018**. DOI: https://github.com/asche1/LongHCPulse.

(34) Scheie, A. LongHCPulse: Long-pulse heat capacity on a Quantum Design PPMS. *Journal of Low Temperature Physics* **2018**, *193* (1), 60-73.
(35) Schnelle, W.; Engelhardt, J.; Gmelin, E. Specific heat capacity of Apiezon N high vacuum grease and of Duran borosilicate glass. *Cryogenics* **1999**, *39* (3), 271-275.
(36) Matsumoto, Y.; Nakatsuji, S. Relaxation calorimetry at very low temperatures for systems with internal relaxation. *Review of Scientific Instruments* **2018**, *89* (3).
(37) Ramesh Kumar, K.; Nair, H. S.; Christian, R.; Thamizhavel, A.; Strydom, A. M. Magnetic, specific heat and electrical transport properties of Frank–Kasper cage compounds RTM2Al20 [R= Eu, Gd and La; TM= V, Ti]. *Journal of Physics: Condensed Matter* **2016**, *28* (43), 436002.
(38) Stewart, G. R. Measurement of low-temperature specific heat. *Review of Scientific Instruments* **1983**, *54*, 1-11.
(39) Gmelin, E. Classical temperature-modulated calorimetry: A review. *Thermochimica Acta* **1997**, *304*, 1-26.
(40) Tari, A. Lattice Specific Heat. In *The Specific Heat of Matter at Low Temperatures*, Tari, A. Ed.; Imperial College Press, 2003; pp 19-58.
(41) Litzbarski, L.; Klimczuk, T.; Winiarski, M. J. Synthesis, Structure and Physical Properties of New Intermetallic Spin Glass-like Compounds $RE_2PdGe_3$ (RE = Tb and Dy). *Journal of Physics: Condensed Matter* **2020**, *32*, 225706.
(42) Gopal, E. S. R. Lattice Heat Capacity. In *Specific Heats at Low Temperatures*, Plenum Press, 1966; pp 19-54.
(43) Sinha, M.; Vivanco, H. K.; Wan, C.; Siegler, M. A.; Stewart, V. J.; Pogue, E. A.; Pressley, L. A.; Berry, T.; Wang, Z.; Johnson, I. Twisting of 2D Kagomé Sheets in Layered Intermetallics. *ACS Central Science* **2021**, *7*, 1381-1390.
(44) Trump, B. A.; Tutmaher, J. A.; McQueen, T. M. Anion–anion Bonding and Topology in Ternary Iridium Seleno–stannides. *Inorganic Chemistry* **2015**, *54*, 11993-12001.
(45) Pressley, L. A.; Torrejon, A.; Phelan, W. A.; McQueen, T. M. Discovery and single crystal growth of high entropy pyrochlores. *Inorganic Chemistry* **2020**, *59* (23), 17251-17258.
(46) *MaterialsAutomated2-CpPhononAnalysis*; 2020. https://github.com/materialsautomated/materialsautomated.github.io/tree/master/MaterialsAutomated2-CpPhononAnalysis (accessed.
(47) Do, S.-H.; Park, S.-Y.; Yoshitake, J.; Nasu, J.; Motome, Y.; Kwon, Y. S.; Adroja, D.; Voneshen, D.; Kim, K.; Jang, T.-H. Majorana Fermions in the Kitaev Quantum Spin System α-$RuCl_3$. *Nature Physics* **2017**, *13*, 1079-1084.
(48) Gopal, E. S. R. Electronic Specific Heat. In *Specific Heats at Low Temperatures*, Plenum Press, 1966; pp 55-83.
(49) Scheie, A.; Liu, Y.; Ghioldi, E.; Fender, S.; Rosa, P. F.; Bauer, E. D.; Zhu, J.-X.; Ronning, F. Excess heat capacity in magnetically ordered Ce heavy-fermion metals. *Physical Review B* **2024**, *110* (8), 085123.
(50) Tari, A. Electronic Specific Heat. In *The Specific Heat of Matter at Low Temperatures*, Tari, A. Ed.; Imperial College Press, 2003; pp 59-136.
(51) Hallas, A. M.; Gaudet, J.; Gaulin, B. D. Experimental Insights into Ground-state Selection of Quantum XY Pyrochlores. *Annual Review of Condensed Matter Physics* **2018**, *9*, 105-124.
(52) Winiarski, M. J.; Wiendlocha, B.; Sternik, M.; Wiśniewski, P.; O'Brien, J. R.; Kaczorowski, D.; Klimczuk, T. Rattling-enhanced Superconductivity in $MV_2Al_{20}$ (M = Sc, Lu, Y) Intermetallic Cage Compounds. *Physical Review B* **2016**, *93*, 134507.
(53) Ryżyńska, Z.; Marshall, M.; Xie, W.; Klimczuk, T.; Winiarski, M. J. Evolution of Physical Properties of $RE_3Ni_5Al_{19}$ Family (RE= Y, Nd, Sm, Gd, Tb, Dy, Ho, and Er). *Crystal Research and Technology* **2023**, *58*, 2200170.

(54) Klimczuk, T.; Xie, W.; Winiarski, M. J.; Kozioł, R.; Litzbarski, L.; Luo, H.; Cava, R. J. Crystal Structure and Physical Properties of New $Ca_2TGe_3$ (T = Pd and Pt) Germanides. *Journal of Solid State Chemistry* **2016**, *243*, 95-100.
(55) Ortiz, B. R.; Gomes, L. C.; Morey, J. R.; Winiarski, M.; Bordelon, M.; Mangum, J. S.; Oswald, I. W.; Rodriguez-Rivera, J. A.; Neilson, J. R.; Wilson, S. D. New Kagome Prototype Materials: Discovery of $KV_3Sb_5$, $RbV_3Sb_5$, and $CsV_3Sb_5$. *Physical Review Materials* **2019**, *3*, 094407.
(56) Elmslie, T. A.; Startt, J.; Soto-Medina, S.; Yang, Y.; Feng, K.; Baumbach, R. E.; Zappala, E.; Morris, G. D.; Frandsen, B. A.; Meisel, M. W. Magnetic Properties of Equiatomic CrMnFeCoNi. *Physical Review B* **2022**, *106*, 014418.
(57) Pokharel, G.; Teicher, S. M.; Ortiz, B. R.; Sarte, P. M.; Wu, G.; Peng, S.; He, J.; Seshadri, R.; Wilson, S. D. Electronic Properties of The Topological Kagome Metals $YV_6Sn_6$ and $GdV_6Sn_6$. *Physical Review B* **2021**, *104*, 235139.
(58) Pokharel, G.; Ortiz, B.; Chamorro, J.; Sarte, P.; Kautzsch, L.; Wu, G.; Ruff, J.; Wilson, S. D. Highly Anisotropic Magnetism in the Vanadium-based Kagome Metal $TbV_6Sn_6$. *Physical Review Materials* **2022**, *6*, 104202.
(59) Wiendlocha, B.; Winiarski, M.; Muras, M.; Zvoriste-Walters, C.; Griveau, J.-C.; Heathman, S.; Gazda, M.; Klimczuk, T. Pressure effects on the superconductivity of the Hf Pd 2 Al Heusler compound: experimental and theoretical study. *Physical Review B* **2015**, *91* (2), 024509.
(60) Królak, S.; Winiarski, M. J.; Yazici, D.; Shin, S.; Klimczuk, T. Kondo-like Behavior in a Mixed Valent Oxypnictide $La_3Cu_4P_4O_2$. *Scientific Reports* **2025**, *15*, 7019.
(61) Bernier, S.; Sinha, M.; Pearson, T. J.; Sushko, P. V.; Oyala, P. H.; Siegler, M. A.; Adam Phelan, W.; Neill, A. N.; Freedman, D. E.; McQueen, T. M. Symmetry-mediated Quantum Coherence of $W^{5+}$ Spins in an Oxygen-deficient Double Perovskite. *npj Quantum Materials* **2025**, *10*, 62.
(62) Ferrenti, A. M.; Meschke, V.; Ghosh, S.; Davis, J.; Drichko, N.; Toberer, E. S.; McQueen, T. M. Hydrothermal Synthesis of Ordered Corkite, $PbFe_3(PO_4)(SO_4)(OH)_6$, a $S$ = 5/2 Kagomé Antiferromagnet. *Journal of Solid State Chemistry* **2023**, *317*, 123620.
(63) Kamadurai, R. K.; Fobasso, R. D.; Maurya, A.; Ramankutty, P. K.; Strydom, A. M. Ferromagnetic Ordering and Heavy Fermion Behaviour in $Ce_2Ru3Ge_5$. *Journal of the Physical Society of Japan* **2020**, *89*, 064705.
(64) Gopal, E. S. R. Magnetic Contribution to Specific Heats. In *Specific Heats at Low Temperatures*, Plenum Press, 1966; pp 84-112.
(65) Tari, A. Magnetic Specific Heat. In *The Specific Heat of Matter at Low Temperatures*, Tari, A. Ed.; Imperial College Press, 2003; pp 137-180.
(66) Rodríguez, J. F.; Blanco, J. Temperature Dependence of the Molar Heat Capacity for Ferromagnets Within the Mean Field Theory. *Physica Scripta* **2005**, *71*, CC19.
(67) Howard, L. N.; Smart, J. S. The Specific Heat Discontinuity in Antiferromagnets and Ferrites. *Physical Review* **1953**, *91*, 17.
(68) Onsager, L. Crystal statistics. I. A two-dimensional model with an order-disorder transition. *Physical review* **1944**, *65* (3-4), 117.
(69) De Jongh, L. J.; Miedema, A. R. Experiments on simple magnetic model systems. *Advances in Physics* **2001**, *50* (8), 947-1170.
(70) Xing, J.; Taddei, K. M.; Sanjeewa, L. D.; Fishman, R. S.; Daum, M.; Mourigal, M.; dela Cruz, C.; Sefat, A. S. Stripe Antiferromagnetic Ground State of the Ideal Triangular Lattice Compound $KErSe_2$. *Physical Review B* **2021**, *103*, 144413.
(71) Bordelon, M. M.; Bocarsly, J. D.; Posthuma, L.; Banerjee, A.; Zhang, Q.; Wilson, S. D. Antiferromagnetism and Crystalline Electric Field Excitations in Tetragonal $NaCeO_2$. *Physical Review B* **2021**, *103*, 024430.

(72) Marjerrison, C. A.; Thompson, C. M.; Sala, G.; Maharaj, D. D.; Kermarrec, E.; Cai, Y.; Hallas, A. M.; Wilson, M. N.; Munsie, T. J.; Granroth, G. E. Cubic $Re^{6+}$ ($5d^1$) Double Perovskites, $Ba_2MgReO_6$, $Ba_2ZnReO_6$, and $Ba_2Y_{2/3}ReO_6$: Magnetism, Heat Capacity, $\mu$SR, and Neutron Scattering Studies and Comparison with Theory. *Inorganic chemistry* **2016**, *55*, 10701-10713.
(73) Ryżyńska, Z.; Błachowski, A.; Żukrowski, J.; Huai, X.; Tran, T. T.; Klimczuk, T.; Winiarski, M. J. Magnetic Properties of Orthorhombic $EuFe_2Al_8$ and $EuRh_2Al_8$ Intermetallic Compounds. *Journal of Alloys and Compounds* **2025**, 183671.
(74) Oyeka, E. E.; Winiarski, M. J.; Świątek, H.; Balliew, W.; McMillen, C. D.; Liang, M.; Sorolla, M.; Tran, T. T. $Ln_2(SeO_3)_2(SO_4)(H_2O)_2$ (Ln= Sm, Dy, Yb): A Mixed-Ligand Pathway to New Lanthanide (III) Multifunctional Materials Featuring Nonlinear Optical and Magnetic Anisotropy Properties. *Angewandte Chemie International Edition* **2022**, *61*, e202213499.
(75) Suchomel, M. R.; Shoemaker, D. P.; Ribaud, L.; Kemei, M. C.; Seshadri, R. Spin-induced Symmetry Breaking in Orbitally Ordered $NiCr_2O_4$ and $CuCr_2O_4$. *Physical Review B* **2012**, *86*, 054406.
(76) Bhutani, A.; Zuo, J. L.; McAuliffe, R. D.; dela Cruz, C. R.; Shoemaker, D. P. Strong Anisotropy in the Mixed Antiferromagnetic System $Mn_{1-x}Fe_xPSe_3$. *Physical Review Materials* **2020**, *4*, 034411.
(77) Yang, K.; Kang, K.; Diao, Z.; Ramanathan, A.; Karigerasi, M. H.; Shoemaker, D. P.; Schleife, A.; Cahill, D. G. Magneto-optic Response of the Metallic Antiferromagnet $Fe_2As$ to Ultrafast Temperature Excursions. *Physical Review Materials* **2019**, *3*, 124408.
(78) Kemei, M. C.; Moffitt, S. L.; Shoemaker, D. P.; Seshadri, R. Evolution of Magnetic Properties in The Normal Spinel Solid Solution $Mg_{1-x}Cu_xCr_2O_4$. *Journal of Physics: Condensed Matter* **2012**, *24*, 046003.
(79) Riedel, Z. W.; Jiang, Z.; Avdeev, M.; Schleife, A.; Shoemaker, D. P. Zero-field Magnetic Structure and Metamagnetic Phase Transitions of the Cobalt Chain Compound $Li_2CoCl_4$. *Physical Review Materials* **2023**, *7*, 104405.
(80) Morey, J. R.; Scheie, A.; Sheckelton, J. P.; Brown, C. M.; McQueen, T. M. $Ni_2Mo_3O_8$: Complex Aiferromagnetic Oder on a Hneycomb Lttice. *Physical Review Materials* **2019**, *3*, 014410.
(81) Berry, T.; Stewart, V. J.; Redemann, B. W.; Lygouras, C.; Varnava, N.; Vanderbilt, D.; McQueen, T. M. A-type Antiferromagnetic Order in the Zintl-phase Insulator $EuZn_2P_2$. *Physical Review B* **2022**, *106*, 054420.
(82) Berry, T.; Parkin, S. R.; McQueen, T. M. Antiferro-and Metamagnetism in the $S$ = 7/2 Hollandite Analog $EuGa_2Sb_2$. *Physical Review Materials* **2021**, *5*, 114401.
(83) Kelly, Z. A.; Tran, T. T.; McQueen, T. M. Nonpolar-to-polar Trimerization Transitions in The $S$ = 1 Kagomé Magnet $Na_2Ti_3Cl_8$. *Inorganic Chemistry* **2019**, *58*, 11941-11948.
(84) Paddison, J. A.; Ong, H. S.; Hamp, J. O.; Mukherjee, P.; Bai, X.; Tucker, M. G.; Butch, N. P.; Castelnovo, C.; Mourigal, M.; Dutton, S. Emergent Order in The Kagome Ising Magnet $Dy_3Mg_2Sb_3O_{14}$. *Nature Communications* **2016**, *7*, 13842.
(85) Dun, Z.; Bai, X.; Paddison, J. A.; Hollingworth, E.; Butch, N. P.; Cruz, C. D.; Stone, M. B.; Hong, T.; Demmel, F.; Mourigal, M. Quantum Versus Classical Spin Fragmentation in Dipolar Kagome Ice $Ho_3Mg_2Sb_3O_{14}$. *Physical Review X* **2020**, *10*, 031069.
(86) Gui, X.; Chang, T.-R.; Wei, K.; Daum, M. J.; Graf, D. E.; Baumbach, R. E.; Mourigal, M.; Xie, W. A Novel Magnetic Material by Design: Observation of $Yb^{3+}$ with Spin-1/2 in $Yb_xPt_5P$. *ACS Central Science* **2020**, *6*, 2023-2030.
(87) Ruiz, A.; Nagarajan, V.; Vranas, M.; Lopez, G.; McCandless, G. T.; Kimchi, I.; Chan, J. Y.; Breznay, N. P.; Frañó, A.; Frandsen, B. A. High-temperature Magnetic Anomaly in the Kitaev Hyperhoneycomb Compound β-$Li_2IrO_3$. *Physical Review B* **2020**, *101*, 075112.
(88) Bordelon, M. M.; Liu, C.; Posthuma, L.; Kenney, E.; Graf, M.; Butch, N.; Banerjee, A.; Calder, S.; Balents, L.; Wilson, S. D. Frustrated Heisenberg $J_1$-$J_2$ Model Within the Stretched Diamond Lattice of $LiYbO_2$. *Physical Review B* **2021**, *103*, 014420.

(89) Oyeka, E. E.; Huai, X.; Marshall, M.; Winiarski, M. J.; Błachowski, A.; Cao, H.; Tran, T. T. Noncollinear Polar magnet $Fe_2(SeO_3)_3(H_2O)_3$ with Inequivalent $Fe^{3+}$ sites. *APL Materials* **2024**, *12*.
(90) Huai, X.; Cairns, L. P.; Delles, B.; Winiarski, M. J.; Sorolla II, M.; Zhang, X.; Chen, Y.; Calder, S.; Kanyowa, T.; Kogar, A. Jeff= 1/2 Diamond Magnet $CaCo_2TeO_6$: A Pathway toward New Spin Physics and Quantum Functions. *arXiv preprint arXiv:2503.18977* **2025**.
(91) Chamorro, J. R.; Jackson, A. R.; Watkins, A. K.; Seshadri, R.; Wilson, S. D. Magnetic Order in the S eff= 1/2 Triangular-lattice Compound $NdCd_3P_3$. *Physical Review Materials* **2023**, *7*, 094402.
(92) Ortiz, B. R.; Sarte, P. M.; Avidor, A. H.; Hay, A.; Kenney, E.; Kolesnikov, A. I.; Pajerowski, D. M.; Aczel, A. A.; Taddei, K. M.; Brown, C. M. Quantum Disordered Ground State in the Triangular-lattice Magnet $NaRuO_2$. *Nature Physics* **2023**, *19*, 943-949.
(93) Amusia, M. Y.; Popov, K. G.; Shaginyan, V. R.; Stephanovich, V. A. Theory of heavy-fermion compounds. *Springer series in solid-state sciences* **2014**, *182*, 33.
(94) Ramirez, A. P.; Hayashi, A.; Cava, R. J.; Siddharthan, R.; Shastry, B. Zero-point entropy in 'spin ice'. *Nature* **1999**, *399* (6734), 333-335.
(95) Binder, K.; Young, A. P. Spin glasses: Experimental facts, theoretical concepts, and open questions. *Reviews of Modern Physics* **1986**, *58* (4), 801.
(96) Phillips, T.; Rosenberg, H. Spin Waves in Ferromagnets. *Reports on Progress in Physics* **1966**, *29*, 285.
(97) Gmelin, E. Modern Low-temperature Calorimetry. *Thermochimica Acta* **1979**, *29*, 1-39.
(98) Ashcroft, N. W.; Mermin, N. D. Solid state. *Physics (New York: Holt, Rinehart and Winston) Appendix C* **1976**, *1*.
(99) Pan, B.; Guan, T.; Hong, X.; Zhou, S.; Qiu, X.; Zhang, H.; Li, S. Specific Heat and Thermal Conductivity of Ferromagnetic Magnons in Yttrium Iron Garnet. *Europhysics Letters* **2013**, *103*, 37005.
(100) Anand, V.; Fraile, A.; Adroja, D.; Sharma, S.; Tripathi, R.; Ritter, C.; De La Fuente, C.; Biswas, P.; Sakai, V. G.; Del Moral, A. Crystal Electric Field and Possible Coupling with Phonons in Kondo Lattice $CeCuGa_3$. *Physical Review B* **2021**, *104* (17), 174438.
(101) Windsor, C.; Stevenson, R. Spin waves in RbMnF3. *Proceedings of the Physical Society* **1966**, *87* (2), 501.
(102) Scheie, A.; Sanders, M.; Gui, X.; Qiu, Y.; Prisk, T. R.; Cava, R. J.; Broholm, C. Beyond magnons in Nd 2 ScNbO 7: An Ising pyrochlore antiferromagnet with all-in–all-out order and random fields. *Physical Review B* **2021**, *104* (13), 134418.
(103) Quilliam, J.; Ross, K.; Del Maestro, A.; Gingras, M.; Corruccini, L.; Kycia, J. Evidence for gapped spin-wave excitations in the frustrated Gd 2 Sn 2 O 7 pyrochlore antiferromagnet from low-temperature specific heat measurements. *Physical Review Letters* **2007**, *99* (9), 097201.
(104) Tari, A. Specific-Heat Anomalies. In *The Specific Heat of Matter at Low Temperatures*, Tari, A. Ed.; Imperial College Press, 2003; pp 211-276.
(105) Gopal, E. S. R. Specific-Heat Anomalies. In *Specific Heats at Low Temperatures*, Plenum Press, 1966; pp 158-180.
(106) Huai, X.; Oyeka, E.; Chinaegbomkpa, U.; Winiarski, M. J.; Sanabria, H.; Tran, T. T. Inductive-Effect-Driven Tunability of Magnetism and Luminescence in Triangular Layers $ANd(SO_4)_2$ (A= Rb, Cs). *Inorganic Chemistry* **2025**, *64*, 17301-17312.
(107) Pressley, L. A.; Torrejon, A.; Phelan, W. A.; McQueen, T. M. Discovery and Single Crystal Growth of High Entropy Pyrochlores. *Inorganic Chemistry* **2020**, *59*, 17251-17258.
(108) McQueen, T. M.; Regulacio, M.; Williams, A. J.; Huang, Q.; Lynn, J. W.; Hor, Y. S.; West, D. V.; Green, M. A.; Cava, R. J. Intrinsic Properties of Stoichiometric LaFePO. *Physical Review B—Condensed Matter and Materials Physics* **2008**, *78*, 024521.

(109) Tran, T. T.; Pocs, C. A.; Zhang, Y.; Winiarski, M. J.; Sun, J.; Lee, M.; McQueen, T. M. Spinon Excitations in the Quasi-one-dimensional *S* = 1/2 Chain Compound $Cs_4CuSb_2Cl_{12}$. *Physical Review B* **2020**, *101*, 235107.
(110) Sheckelton, J. P.; Plumb, K. W.; Trump, B. A.; Broholm, C. L.; McQueen, T. M. Rearrangement of Van der Waals Stacking and Formation of a Singlet State at *T* = 90 K in a Cluster Magnet. *Inorganic Chemistry Frontiers* **2017**, *4*, 481-490.
(111) Daum, M. J.; Ramanathan, A.; Kolesnikov, A. I.; Calder, S.; Mourigal, M.; La Pierre, H. S. Collective Excitations in the Tetravalent Lanthanide Honeycomb Antiferromagnet $Na_2PrO_3$. *Physical Review B* **2021**, *103*, L121109.
(112) Ding, Z.-F.; Yang, Y.-X.; Zhang, J.; Tan, C.; Zhu, Z.-H.; Chen, G.; Shu, L. Possible Gapless Spin Liquid in the Rare-earth Kagome Lattice Magnet $Tm_3Sb_3Zn_2O_{14}$. *Physical Review B* **2018**, *98* (17), 174404.
(113) Moler, K. A.; Baar, D.; Urbach, J.; Liang, R.; Hardy, W.; Kapitulnik, A. Magnetic Field Dependence of the Density of States of $YBa_2Cu_3O_{6.95}$ as Determined from the Specific Heat. *Physical Review Letters* **1994**, *73*, 2744.
(114) Konic, A. M.; Adhikari, R.; Kunwar, D.; Kirmani, A. A.; Breindel, A.; Sheng, R.; Maple, M.; Dzero, M.; Almasan, C. C. Evolution of Non-Kramers Doublets in Magnetic Field in $PrNi_2Cd_{20}$ and $PrPd_2Cd_{20}$. *Physical Review B* **2021**, *104*, 205139.
(115) Early, S. R.; Hellman, F.; Marshall, J.; Geballe, T. H. A Silicon on Sapphire Thermometer for Small Sample Low Temperature Calorimetry. *Physica B+C* **1981**, *107*, 327-328.
(116) Mahoney, J.; Wallace, W.; Craig, R. Influence of The Crystalline Electric Field on the Heat Capacity and Resistivity of $PrAl_3$. *Journal of Applied Physics* **1974**, *45*, 2733-2738.
(117) Paddison, J. A.; Daum, M.; Dun, Z.; Ehlers, G.; Liu, Y.; Stone, M. B.; Zhou, H.; Mourigal, M. Continuous Excitations of the Triangular-lattice Quantum Spin Liquid $YbMgGaO_4$. *Nature Physics* **2017**, *13*, 117-122.
(118) Souza, M. d.; Paupitz, R.; Seridonio, A.; Lagos, R. E. Specific Heat Anomalies in Solids Described by a Multilevel Model. *Brazilian Journal of Physics* **2016**, *46*, 206-212.
(119) Westrum Jr, E. F. Lattice and Schottky Contributions to the Morphology of Lanthanide Heat Capacities. **1983**.
(120) Kim, S.-Y. Generalized Schottky Anomaly. *Journal of the Korean Physical Society* **2014**, *65*, 970-972.
(121) Joshi, D. A.; Burger, P.; Adelmann, P.; Ernst, D.; Wolf, T.; Sparta, K.; Roth, G.; Grube, K.; Meingast, C.; Löhneysen, H. v. Magnetic Properties of Single-crystalline $CeCuGa_3$. *Physical Review B* **2012**, *86*, 035144.
(122) Bleaney, B. Hyperfine Interactions in Rare‐Earth Metals. *Journal of Applied Physics* **1963**, *34* (4), 1024-1031.
(123) Winiarski, M. J.; Wiendlocha, B.; Gołąb, S.; Kushwaha, S. K.; Wiśniewski, P.; Kaczorowski, D.; Thompson, J. D.; Cava, R. J.; Klimczuk, T. Superconductivity in CaBi 2. *Physical Chemistry Chemical Physics* **2016**, *18* (31), 21737-21745.
(124) Kimchi, I.; Sheckelton, J. P.; McQueen, T. M.; Lee, P. A. Scaling and Data Collapse From Local Moments in Frustrated Disordered Quantum Spin Systems. *Nature Communications* **2018**, *9*, 4367.
(125) Ortiz, B. R.; Sarte, P. M.; Kenney, E. M.; Graf, M. J.; Teicher, S. M.; Seshadri, R.; Wilson, S. D. Superconductivity in the $Z_2$ Kagome Metal $KV_3Sb_5$. *Physical Review Materials* **2021**, *5*, 034801.
(126) Ortiz, B. R.; Pokharel, G.; Gundayao, M.; Li, H.; Kaboudvand, F.; Kautzsch, L.; Sarker, S.; Ruff, J. P.; Hogan, T.; Alvarado, S. J. G. $YbV_3Sb_4$ and $EuV_3Sb_4$ Vanadium-based Kagome Metals with $Yb_{2+}$ and $Eu_{2+}$ Zigzag Chains. *Physical Review Materials* **2023**, *7*, 064201.

(127) Bissengaliyeva, M. R.; Bespyatov, M. A.; Gogol, D. B.; Sadyrbekov, D. T.; Taimassova, S. T. Measurements of the Heat Capacity of Erbium Titanate $Er_2Ti_2O_7$. *Journal of Chemical & Engineering Data* **2022**, *67*, 2059-2066.
(128) Huai, X.; Acheampong, E.; Delles, E.; Winiarski, M. J.; Sorolla, M.; Nassar, L.; Liang, M.; Ramette, C.; Ji, H.; Scheie, A. Noncentrosymmetric Triangular Magnet $CaMnTeO_6$: Strong Quantum Fluctuations and Role of $s^0$ Versus $s^2$ Electronic States in Competing Exchange Interactions. *Advanced Materials* **2024**, *36*, 2313763.
(129) Scheie, A.; Sanders, M.; Gui, X.; Qiu, Y.; Prisk, T. R.; Cava, R. J.; Broholm, C. Beyond Magnons in $Nd_2ScNbO_7$: An Ising Pyrochlore Antiferromagnet with All-in–all-out Order and Random Fields. *Physical Review B* **2021**, *104*, 134418.
(130) Radousky, H.; Dunlap, B.; Knapp, G.; Niarchos, D. Heat-capacity studies of crystal-field effects in dilute R Rh 4 B 4 compounds. *Physical Review B* **1983**, *27* (9), 5526.
(131) CATALANO, E.; Phillips, N. The low-temperature heat capacities of antiferromagnetic MnF2 and CoF2. *Journal of the Physical Society of Japan* **1962**, *17*, 527-&.
(132) Chinaegbomkpa, U. V.; Huai, X.; Winiarski, M. J.; Sanabria, H.; Tran, T. T. Chemical Origins of Optically Addressable Spin States in $Eu_2(P_2S_6)$ and $Eu_2(P_2Se_6)$. *ACS Materials Au* **2024**, *5*, 182-190.
(133) Scheie, A. Exotic Magnetism in Frustrated Pyrochlore-Based Magnets. Ph.D., The Johns Hopkins University, United States -- Maryland, 2019.
(134) Tinkham, M. *Introduction to Superconductivity*; Courier Corporation, 2004.
(135) Simon, A. Superconductivity and Chemistry. *Angewandte Chemie International Edition in English* **1997**, *36*, 1788-1806.
(136) Wen, H.-H. Specific Heat in Superconductors. *Chinese Physics B* **2020**, *29*, 017401.
(137) Zhou, X.; Lee, W.-S.; Imada, M.; Trivedi, N.; Phillips, P.; Kee, H.-Y.; Törmä, P.; Eremets, M. High-Temperature Superconductivity. *Nature Reviews Physics* **2021**, *3*, 462-465.
(138) Stewart, G. Superconductivity in Iron Compounds. *Reviews of Modern Physics* **2011**, *83*, 1589-1652.
(139) Ryżyńska, Z.; Chamorro, J. R.; McQueen, T. M.; Wiśniewski, P.; Kaczorowski, D.; Xie, W.; Cava, R. J.; Klimczuk, T.; Winiarski, M. J. $RuAl_6$—An Endohedral Aluminide Superconductor. *Chemistry of Materials* **2020**, *32*, 3805-3812.
(140) Winiarski, M. J.; Wiendlocha, B.; Gołąb, S.; Kushwaha, S. K.; Wiśniewski, P.; Kaczorowski, D.; Thompson, J. D.; Cava, R. J.; Klimczuk, T. Superconductivity in $CaBi_2$. *Physical Chemistry Chemical Physics* **2016**, *18*, 21737-21745.
(141) Shoemaker, D. P.; Chung, D. Y.; Claus, H.; Francisco, M. C.; Avci, S.; Llobet, A.; Kanatzidis, M. G. Phase Relations in $K_xFe_{2-y}Se_2$ and the Structure of Superconducting $K_xFe_2Se_2$ via High-resolution Synchrotron Diffraction. *Physical Review B* **2012**, *86*, 184511.
(142) Lygouras, C. J.; Zhang, J.; Gautreau, J.; Pula, M.; Sharma, S.; Gao, S.; Berry, T.; Halloran, T.; Orban, P.; Grissonnanche, G. Type I and type II Superconductivity in a Quasi-2D Dirac Metal. *Materials Advances* **2025**, *6*, 1685-1694.
(143) Wagner, K. E.; Morosan, E.; Hor, Y.; Tao, J.; Zhu, Y.; Sanders, T.; McQueen, T. M.; Zandbergen, H. W.; Williams, A. J.; West, D. V. Tuning the Charge Density Wave and Superconductivity in $Cu_xTaS_2$. *Physical Review B* **2008**, *78*, 104520.
(144) Ortiz, B. R.; Teicher, S. M.; Hu, Y.; Zuo, J. L.; Sarte, P. M.; Schueller, E. C.; Abeykoon, A. M.; Krogstad, M. J.; Rosenkranz, S.; Osborn, R. $CsV_3Sb_5$: AZ 2 Topological Kagome Metal with a Superconducting Ground State. *Physical Review Letters* **2020**, *125*, 247002.
(145) Hirshfeld, A. T.; Leupold, H.; Boorse, H. Superconducting and Normal Specific Heats of Niobium. *Physical Review* **1962**, *127*, 1501.
(146) Phillips, N. E.; Fisher, R.; Gordon, J. The Specific Heat of $YBa_2Cu_3O_7$. *Chinese Journal of Physics* **1992**, *30*, 799-807.

(147) Bud'ko, S. L.; Canfield, P. C. Superconductivity of Magnesium Diboride. *Physica C: Superconductivity and its Applications* **2015**, *514*, 142-151.
(148) Oudah, M.; Kung, H.-H.; Sahu, S.; Heinsdorf, N.; Schulz, A.; Philippi, K.; De Toro Sanchez, M.-V.; Cai, Y.; Kojima, K.; Schnyder, A. P. Discovery of Superconductivity and Electron-phonon Drag in the Non-centrosymmetric Weyl Semimetal $LaRhGe_3$. *npj Quantum Materials* **2024**, *9*, 88.
(149) Cheng, P.-Y.; Oudah, M.; Hung, T.-L.; Hsu, C.-E.; Chang, C.-C.; Haung, J.-Y.; Liu, T.-C.; Cheng, C.-M.; Ou, M.-N.; Chen, W.-T. Physical Properties and Electronic Structure of the Two-gap Superconductor $V_2Ga_5$. *Physical Review Research* **2024**, *6*, 033253.
(150) Klimczuk, T.; Królak, S.; Cava, R. J. Superconductivity of Ta-Hf and Ta-Zr alloys: Potential alloys for use in superconducting devices. *Physical Review Materials* **2023**, *7* (6), 064802.
(151) Allenspach, P. M.; Maple, M. B. Heat capacity. *Handbook on The Physics and Chemistry of Rare Earths* **2001**, *31*, 351-390.
(152) Yang, H.; Lin, J.-Y.; Lin, J.; Ho, J. Low-temperature Specific Heat of Superconductors II: Progress in The New Era. *Chinese Journal of Physics* **2019**, *61*, 212-226.
(153) Imfeld, N.; Bestgen, W.; Rinderer, L. Specific Heat and Magnetization of Superconducting Niobium in the Mixed State. *Journal of Low Temperature Physics* **1985**, *60*, 223-238.
(154) Prozorov, R.; Zarea, M.; Sauls, J. A. Niobium in the Clean Limit: An Intrinsic Type-I Superconductor. *Physical Review B* **2022**, *106*, L180505.
(155) Tanatar, M. A.; Torsello, D.; Joshi, K. R.; Ghimire, S.; Kopas, C. J.; Marshall, J.; Mutus, J. Y.; Ghigo, G.; Zarea, M.; Sauls, J. A. Anisotropic Superconductivity of Niobium Based on Its Response to Nonmagnetic Disorder. *Physical Review B* **2022**, *106*, 224511.
(156) Pronin, A. V.; Dressel, M.; Pimenov, A.; Loidl, A.; Roshchin, I. V.; Greene, L. Direct observation of the superconducting energy gap developing in the conductivity spectra of niobium. *Physical Review B* **1998**, *57* (22), 14416.
(157) Hussein, A. A.; Hussein, A. A.; Hasan, N. A. Study of The Properties of YBCO Superconductor Compound in Various Preparation Methods: A Short Review. *Journal of Applied Sciences and Nanotechnology* **2023**, *3*, 65-79.
(158) Shen, K. M.; Davis, J. S. Cuprate High-$T_c$ Superconductors. *Materials Today* **2008**, *11*, 14-21.
(159) Phillips, N. E.; Emerson, J. P.; Fisher, R. A.; Gordon, J. E.; Woodfield, B. F.; Wright, D. A. Specific Heat of $YBa_2Cu_3O_7$. *Journal of Superconductivity* **1994**, *7*, 251-255. DOI: 10.1007/BF00730406.
(160) Nagamatsu, J.; Nakagawa, N.; Muranaka, T.; Zenitani, Y.; Akimitsu, J. Superconductivity at 39 K in Magnesium Diboride. *Nature* **2001**, *410*, 63-64.
(161) Buzea, C.; Yamashita, T. Review of the Superconducting Properties of $MgB_2$. *Superconductor Science and Technology* **2001**, *14*, R115.
(162) Canfield, P. C.; Crabtree, G. W. Magnesium Diboride: Better Late Than Never. *Physics Today* **2003**, *56*, 34-40.
(163) Bouquet, F.; Wang, Y.; Fisher, R.; Hinks, D.; Jorgensen, J.; Junod, A.; Phillips, N. Phenomenological Two-gap Model for The Specific Heat of $MgB_2$. *Europhysics Letters* **2001**, *56*, 856.
(164) Finocchio, G.; Büttner, F.; Tomasello, R.; Carpentieri, M.; Kläui, M. Magnetic Skyrmions: From Fundamental to Applications. *Journal of Physics D: Applied Physics* **2016**, *49*, 423001.
(165) Kanazawa, N.; Seki, S.; Tokura, Y. Noncentrosymmetric Magnets Hosting Magnetic Skyrmions. *Advanced Materials* **2017**, *29* (25), 1603227.
(166) Fert, A.; Reyren, N.; Cros, V. Magnetic Skyrmions: Advances in Physics and Potential Applications. *Nature Reviews Materials* **2017**, *2*, 1-15.
(167) Muhlbauer, S.; Binz, B.; Jonietz, F.; Pfleiderer, C.; Rosch, A.; Neubauer, A.; Georgii, R.; Boni, P. Skyrmion lattice in a chiral magnet. *Science* **2009**, *323* (5916), 915-919.

(168) Pappas, C.; Lelievre-Berna, E.; Falus, P.; Bentley, P.; Moskvin, E.; Grigoriev, S.; Fouquet, P.; Farago, B. Chiral Paramagnetic Skyrmion-like Phase in MnSi. *Physical Review Letters* **2009**, *102*, 197202.
(169) Yu, X.; Kanazawa, N.; Onose, Y.; Kimoto, K.; Zhang, W.; Ishiwata, S.; Matsui, Y.; Tokura, Y. Near room-temperature formation of a skyrmion crystal in thin-films of the helimagnet FeGe. *Nature materials* **2011**, *10* (2), 106-109.
(170) Bocarsly, J. D.; Need, R. F.; Seshadri, R.; Wilson, S. D. Magnetoentropic Signatures of Skyrmionic Phase Behavior in FeGe. *Physical Review B* **2018**, *97*, 100404.
(171) Kézsmárki, I.; Bordács, S.; Milde, P.; Neuber, E.; Eng, L. M.; White, J. S.; Rønnow, H. M.; Dewhurst, C. D.; Mochizuki, M.; Yanai, K. Néel-type skyrmion lattice with confined orientation in the polar magnetic semiconductor GaV4S8. *Nature materials* **2015**, *14* (11), 1116-1122.
(172) Tokunaga, Y.; Yu, X.; White, J.; Rønnow, H. M.; Morikawa, D.; Taguchi, Y.; Tokura, Y. A new class of chiral materials hosting magnetic skyrmions beyond room temperature. *Nature communications* **2015**, *6* (1), 7638.
(173) Kakihana, M.; Aoki, D.; Nakamura, A.; Honda, F.; Nakashima, M.; Amako, Y.; Takeuchi, T.; Harima, H.; Hedo, M.; Nakama, T. Unique magnetic phases in the skyrmion lattice and Fermi surface properties in cubic chiral antiferromagnet EuPtSi. *Journal of the Physical Society of Japan* **2019**, *88* (9), 094705.
(174) Hirschberger, M.; Nakajima, T.; Gao, S.; Peng, L.; Kikkawa, A.; Kurumaji, T.; Kriener, M.; Yamasaki, Y.; Sagayama, H.; Nakao, H. Skyrmion phase and competing magnetic orders on a breathing kagomé lattice. *Nature communications* **2019**, *10* (1), 5831.
(175) Paddison, J. A.; Rai, B. K.; May, A. F.; Calder, S.; Stone, M. B.; Frontzek, M. D.; Christianson, A. D. Magnetic Interactions of the Centrosymmetric Skyrmion Material $Gd_2PdSi_3$. *Physical Review Letters* **2022**, *129*, 137202.
(176) Hayami, S.; Motome, Y. Square Skyrmion Crystal in Centrosymmetric Itinerant Magnets. *Physical Review B* **2021**, *103*, 024439.
(177) Rathnaweera, D. N.; Huai, X.; Kumar, K. R.; Winiarski, M. J.; Klimczuk, T.; Tran, T. T. Atomically Modulating Competing Exchange Interactions in Centrosymmetric Skyrmion Hosts $GdRu_2X_2$ (X= Si, Ge). *arXiv preprint arXiv:2502.21169* **2025**.
(178) Savary, L.; Balents, L. Quantum Spin Liquids: A Review. *Reports on Progress in Physics* **2016**, *80* (1), 016502.
(179) Chamorro, J. R.; McQueen, T. M.; Tran, T. T. Chemistry of quantum spin liquids. *Chemical Reviews* **2020**, *121*, 2898-2934.
(180) Clark, L.; Abdeldaim, A. H. Quantum Spin Liquids From a Materials Perspective. *Annual Review of Materials Research* **2021**, *51*, 495-519.
(181) Sheckelton, J. P.; Neilson, J. R.; Soltan, D. G.; McQueen, T. M. Possible Valence-bond Condensation in the Frustrated Cluster Magnet $LiZn_2Mo_3O_8$. *Nature Materials* **2012**, *11*, 493-496.
(182) Hasan, Z.; Zoghlin, E.; Winiarski, M.; Arpino, K. E.; Halloran, T.; Tran, T. T.; McQueen, T. M. Quasi-one-dimensional Exchange Interactions and Short-range Magnetic Correlations in $CuTeO_4$. *Physical Review B* **2024**, *109*, 195168.
(183) Ramanathan, A.; Mourigal, M.; La Pierre, H. S. Frustrated Magnetism and Spin Anisotropy in a Buckled Square Net $YbTaO_4$. *Inorganic Chemistry* **2024**, *64*, 158-165.
(184) Bordelon, M. M.; Wang, X.; Pajerowski, D. M.; Banerjee, A.; Sherwin, M.; Brown, C. M.; Eldeeb, M.; Petersen, T.; Hozoi, L.; Rößler, U. Magnetic properties and signatures of moment ordering in the triangular lattice antiferromagnet KCeO 2. *Physical Review B* **2021**, *104*, 094421.
(185) Frandsen, B. A.; Fischer, H. E. A New Spin on Material Properties: Local Magnetic Structure in Functional and Quantum Materials. *Chemistry of Materials* **2024**, *36*, 9089-9106.

(186) Nuttall, K. M.; Suggs, C. Z.; Fischer, H. E.; Bordelon, M. M.; Wilson, S. D.; Frandsen, B. A. Quantitative Investigation of the Short-range Magnetic Correlations in the Candidate Quantum Spin Liquid $NaYbO_2$. *Physical Review B* **2023**, *108*, L140411.
(187) Furukawa, T.; Kobashi, K.; Kurosaki, Y.; Miyagawa, K.; Kanoda, K. Quasi-continuous transition From a Fermi Liquid to a Spin Liquid in $\kappa$-$(ET)_2Cu_2(CN)_3$. *Nature communications* **2018**, *9*, 307.
(188) Powell, B.; McKenzie, R. H. Quantum Frustration in Organic Mott Insulators: From Spin Liquids to Unconventional Superconductors. *Reports on Progress in Physics* **2011**, *74*, 056501.
(189) Itou, T.; Oyamada, A.; Maegawa, S.; Kato, R. Instability of a Quantum Spin Liquid in an Organic Triangular-lattice Antiferromagnet. *Nature Physics* **2010**, *6*, 673-676.
(190) Norman, M. R. Colloquium: Herbertsmithite and the Search for the Quantum Spin Liquid. *Reviews of Modern Physics* **2016**, *88*, 041002.
(191) Wang, J.; Yuan, W.; Singer, P. M.; Smaha, R. W.; He, W.; Wen, J.; Lee, Y. S.; Imai, T. Freezing of the Lattice in the Kagome Lattice Heisenberg Antiferromagnet Zn-barlowite $ZnCu_3(OD)_6FBr$. *Physical Review Letters* **2022**, *128*, 157202.
(192) Breidenbach, A. T.; Campello, A. C.; Wen, J.; Jiang, H.-C.; Pajerowski, D. M.; Smaha, R. W.; Lee, Y. S. Identifying universal spin excitations in candidate spin-1/2 kagome quantum spin liquid materials. *Nature Physics* **2025**, 1-8.
(193) Huai, X.; Tran, T. T. Universal behaviour: Quantum spin liquids. *Nature Physics* **2025**, 1-2.
(194) Banerjee, A.; Yan, J.; Knolle, J.; Bridges, C. A.; Stone, M. B.; Lumsden, M. D.; Mandrus, D. G.; Tennant, D. A.; Moessner, R.; Nagler, S. E. Neutron Scattering in the Proximate Quantum Spin Liquid $\alpha$-$RuCl_3$. *Science* **2017**, *356*, 1055-1059.
(195) Chaloupka, J.; Jackeli, G.; Khaliullin, G. Zigzag Magnetic Order in the Iridium Oxide $Na_2IrO_3$. *Physical Review Letters* **2013**, *110*, 097204.
(196) Takagi, H.; Takayama, T.; Jackeli, G.; Khaliullin, G.; Nagler, S. E. Concept and Realization of Kitaev Quantum Spin Liquids. *Nature Reviews Physics* **2019**, *1*, 264-280.
(197) Widmann, S.; Tsurkan, V.; Prishchenko, D. A.; Mazurenko, V. G.; Tsirlin, A. A.; Loidl, A. Thermodynamic Evidence of Fractionalized Excitations in $\alpha$-$RuCl_3$. *Physical Review B* **2019**, *99*, 094415.
(198) Banerjee, A.; Lampen-Kelley, P.; Knolle, J.; Balz, C.; Aczel, A. A.; Winn, B.; Liu, Y.; Pajerowski, D.; Yan, J.; Bridges, C. A. Excitations in the Field-induced Quantum Spin Liquid State of $\alpha$-$RuCl_3$. *npj Quantum Materials* **2018**, *3*, 8.
(199) Kubota, Y.; Tanaka, H.; Ono, T.; Narumi, Y.; Kindo, K. Successive Magnetic Phase Transitions in $\alpha$-$RuCl_3$: XY-like Frustrated Magnet on the Honeycomb Lattice. *Physical Review B* **2015**, *91*, 094422.
(200) Imamura, K.; Suetsugu, S.; Mizukami, Y.; Yoshida, Y.; Hashimoto, K.; Ohtsuka, K.; Kasahara, Y.; Kurita, N.; Tanaka, H.; Noh, P. Majorana-fermion Origin of the Planar Thermal Hall Effect in the Kitaev Magnet $\alpha$-$RuCl_3$. *Science advances* **2024**, *10*, eadk3539.
(201) Kitaev, A. Anyons in an exactly solved model and beyond. *Annals of Physics* **2006**, *321* (1), 2-111.
(202) Fisk, Z.; Sarrao, J. L.; Smith, J.; Thompson, J. The Physics and Chemistry of Heavy Fermions. *Proceedings of the National Academy of Sciences* **1995**, *92*, 6663-6667.
(203) Stewart, S. G. Heavy-Fermion Systems. *Reviews of Modern Physics* **1984**, *56*, 755.
(204) Coleman, P. Heavy fermions: electrons at the edge of magnetism Handbook of Magnetism and Advanced Magnetic Materials ed H Kronmuller and S Parkin. New York: Wiley: 2007.
(205) Yoshimochi, H.; Takagi, R.; Ju, J.; Khanh, N.; Saito, H.; Sagayama, H.; Nakao, H.; Itoh, S.; Tokura, Y.; Arima, T.-h. Multistep topological transitions among meron and skyrmion crystals in a centrosymmetric magnet. *Nature Physics* **2024**, *20* (6), 1001-1008.

(206) Imamura, K.; Namba, R.; Ishioka, R.; Ishihara, K.; Matsuda, Y.; Lee, S.; Moon, E.-G.; Hashimoto, K.; Shibauchi, T. Bulk excitations in ultraclean α-RuCl 3: Quantitative evidence for Majorana dispersions in a Kitaev quantum spin liquid. *Physical Review B* **2025**, *111* (18), 184440.
(207) Kadowaki, K.; Woods, S. Universal Relationship of the Resistivity and Specific Heat in Heavy-fermion Compounds. *Solid State Communications* **1986**, *58*, 507-509.
(208) Tsujii, N.; Kontani, H.; Yoshimura, K. Universality in Heavy Fermion Systems With General Degeneracy. *Physical Review Letters* **2005**, *94*, 057201.
(209) Jacko, A.; Fjærestad, J.; Powell, B. A Unified Explanation of the Kadowaki–Woods Ratio in Strongly Correlated Metals. *Nature Physics* **2009**, *5*, 422-425.
(210) Okabe, T. Kadowaki-Woods Ratio for Strongly Coupled Fermi Liquids. *Physical Review B* **2007**, *76*, 193109.
(211) Gschneidner Jr, K.; Tang, J.; Dhar, S.; Goldman, A. False Heavy Fermions. *Physica B: Condensed Matter* **1990**, *163*, 507-510.
(212) Yuan, H.; Grosche, F.; Deppe, M.; Geibel, C.; Sparn, G.; Steglich, F. Observation of Two Distinct Superconducting Phases in $CeCu_2Si_2$. *Science* **2003**, *302*, 2104-2107.
(213) Andraka, B.; Fraunberger, G.; Kim, J.; Quitmann, C.; Stewart, G. High-field Specific Heat of $CeCu_2Si_2$ and $CeAl_3$. *Physical Review B* **1989**, *39*, 6420.
(214) Flouquet, J.; Lasjaunias, J.; Peyrard, J.; Ribault, M. Low-temperature Properties of $CeAl_3$. *Journal of Applied Physics* **1982**, *53*, 2127-2130.
(215) Stewart, G.; Fisk, Z.; Willis, J.; Smith, J. Possibility of Coexistence of Bulk Superconductivity and Spin Fluctuations in $UPt_3$. *Physical Review Letters* **1984**, *52*, 679.
(216) Bulka, B. R. Electron Correlations in Molecular Systems. *arXiv preprint cond-mat/9812392* **1998**.
(217) Stewart, G.; Fisk, Z.; Willis, J. Characterization of single crystals of Ce Cu 2 Si 2. A source of new perspectives. *Physical Review B* **1983**, *28* (1), 172.
(218) Blatt, J. *Thermoelectric power of metals*; Springer Science & Business Media, 2012.
(219) Wang, Y.; Rogado, N. S.; Cava, R. J.; Ong, N. P. Spin entropy as the likely source of enhanced thermopower in Na x Co2O4. *Nature* **2003**, *423*, 425-428.
(220) Polash, M. M. H.; Moseley, D.; Zhang, J.; Hermann, R. P.; Vashaee, D. Understanding and design of spin-driven thermoelectrics. *Cell Reports Physical Science* **2021**, *2*.
(221) Sun, P.; Kumar, K. R.; Lyu, M.; Wang, Z.; Xiang, J.; Zhang, W. Generic Seebeck effect from spin entropy. *The Innovation* **2021**, *2*.
(222) Tang, S.; Craig, P.; Kitchens, T. Seebeck coefficient at the Curie temperature: specific heat of charge carriers in ferromagnets. *Physical Review Letters* **1971**, *27*, 593.
(223) Matusiak, M.; Plackowski, T.; Bukowski, Z.; Zhigadlo, N.; Karpinski, J. Evidence of Spin-density-wave Order in $RFeAsO_{1-x}F_x$ From Measurements of Thermoelectric Power. *Physical Review B—Condensed Matter and Materials Physics* **2009**, *79*, 212502.

**Supplemental Information**

# Heat Capacity−A Powerful Tool for Studying Exotic States of Matter

K. Ramesh Kumar,[1,⊥] Xudong Huai,[1,⊥] Michał J. Winiarski,[2]
Allen O. Scheie,[3,†] and Thao T. Tran[1,4]*

[1] Department of Chemistry, Clemson University, Clemson, SC, 29634, USA

[2] Faculty of Applied Physics and Mathematics and Advanced Materials Center, Gdansk University of Technology, ul. Narutowicza 11/12, 80-233 Gdansk, Poland

[3] MPA-Q, Los Alamos National Laboratory, Los Alamos, NM 87545, USA

[4] Department of Physics and Astronomy, Clemson University, Clemson, SC, 29634, USA

[⊥]These authors contributed equally

[†] scheie@lanl.gov

* thao@clemson.edu

**Additional Measurement Guidelines**

For metallic samples, heat capacity measurements are straightforward due to their high thermal conductivity. If we work with metallic samples that have at least one flat surface, heat capacity measurements can be carried out with less than 10 mg of material, since the sample coupling and thermal conductance are usually sufficient even for smaller masses. To improve accuracy and signal-to-noise ratio, the sample mass can be slightly increased. This will not cause any internal thermal gradient, but the total measurement time will increase because the time constant is directly proportional to the sample heat capacity.

For insulating samples, the measurement poses additional challenges due to their low thermal conductivity. Because the sample's thermal conductance is poor, it may not reach thermal equilibrium during heating and cooling. As a result of internal thermal gradients, there can be a temperature lag between the sample and the platform. If we use a small sample to reduce internal thermal gradients, then the sample's heat capacity may be less than 50 percent of the total heat capacity. To compensate and improve the signal-to-noise ratio, one might consider using a larger sample. However, doing so can compromise both accuracy and precision, as heat flows across a larger volume becomes non-uniform, leading to temperature gradients within the sample. In this case, two independent measurements may not reproduce the same value. Although a larger sample increases the relative contribution of the sample heat capacity, it does not eliminate the underlying systematic error. The relaxation curve may still appear well-behaved, but the heat-flow conditions are unlikely to be identical between measurements, leading to inconsistencies in the extracted values.

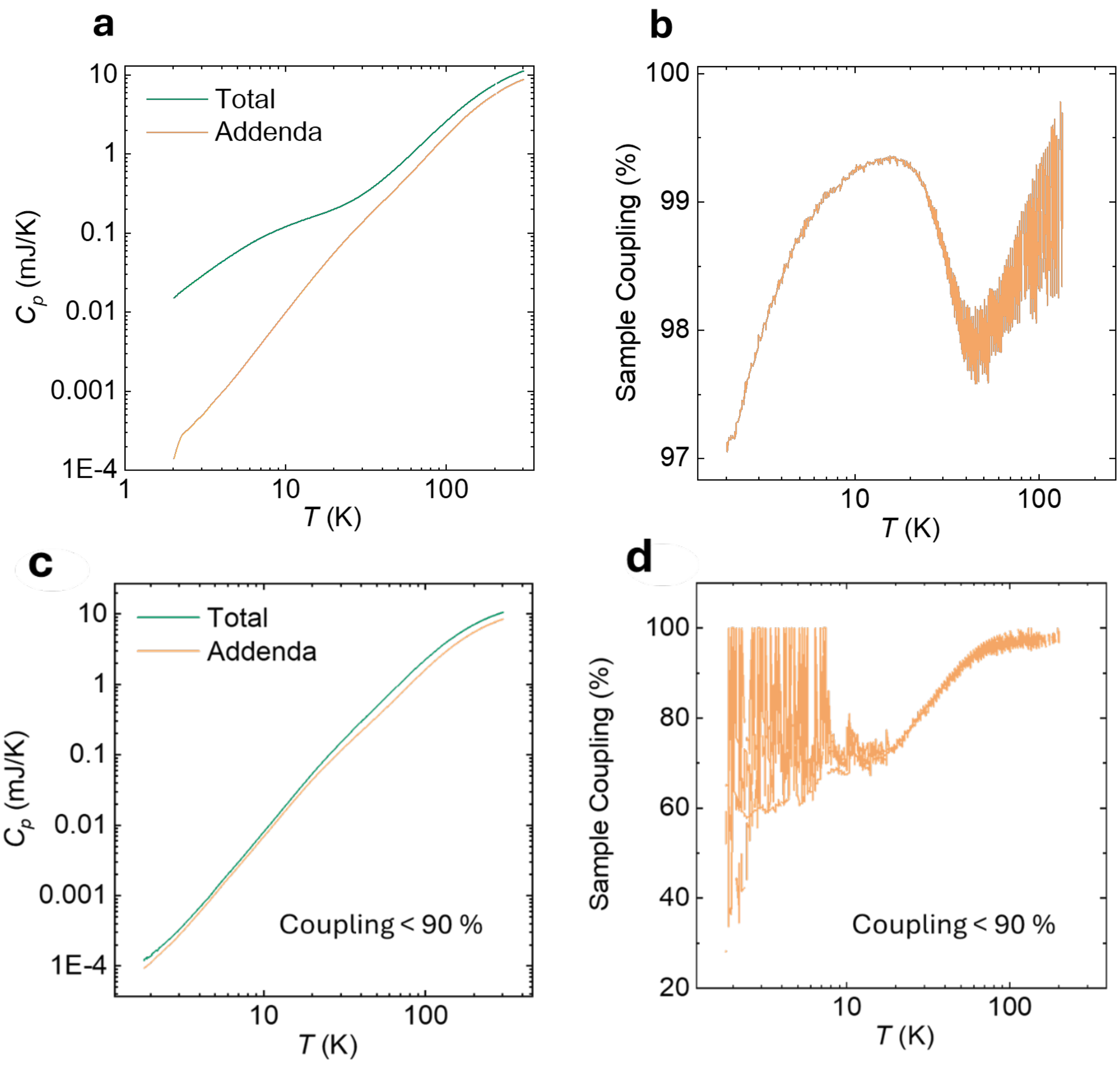

**Figure S1**. (a) Total and addenda heat capacity as function of temperature for a metallic sample (b) Temperature variation of sample coupling for a metal. (c) Total and addenda heat capacity as a function of temperature for an insulator (d) Temperature variation of sample coupling for an insulator.

For powdered samples, several methodologies can be used. For example, the powder can be pressed into a pellet, and a cut section of the pellet can be used for measurement. The sample coupling will be reduced in such cases, but as long as the sample heat capacity is 50–60% of the total heat capacity, as suggested by Lashley et al., it is still possible to collect high-quality data for powdered samples [1]. The powder sample must be pressed into a pellet using a die and plunger in a pelletizer. It is better to use a 10–12 mm die in order to apply higher pressure, up to 4 tons. The pellet can then be easily cut with a surgical knife into a parallelepiped shape of the desired mass. If the available sample quantity is limited, a die with a smaller diameter can be used, but it is recommended to sinter the sample to reduce porosity. There are other methods that have been suggested for powder samples. The pan with grease method, in which the sample is sealed in an aluminum DSC pan with Apiezon N grease, can yield accuracy within ±1–5%, but this approach is time-consuming. Another method is the flat pan technique, where the powder is pressed into an aluminum pan without grease, which reduces grease-related errors but still requires careful subtraction of the pan's heat capacity contribution.

**Triple Relaxation Method**

Another relaxation calorimetry is useful at very low temperature (~mK) proposed for samples that exhibit internal relaxation or inhomogeneous heat flow. In the millikelvin range several systems show nuclear Schottky contribution (section3.4) .[2] Typically, the electronic and phonon contribution is negligible in this range hence this may not possess any difficulty for the nuclear Schottky analysis. Heavy fermion systems show exception here because of mixed nuclear and electronic contributions. In the conventional PPMS measurement the heat flow is governed by two coupled differential equations as discussed in section 2.3. The two characteristics of time constants ($\tau_1$, $\tau_2$) and coefficients ($A_1$ and $A_2$) is linked to the $K_w$, $K_g$ and $C_s$ and $C_{addenda}$. In the low temperatures, heavy fermion system possesses not only two contributions such as nuclear and electronic, but their thermal relaxation is typically not of the same order, and it introduces additional relaxation in between the sample and grease. Now the heat flow is governed by three differential equations instead of two, consequently the platform temperature relaxation shows triple exponential decay and corresponding to each thermal bath the equation (5) of the main text can be modified

$$T_p(t) = \bar{T} + A_1{}^{e^{-t/\tau_s}} + A_2{}^{e^{-t/\tau_i}} + A_3{}^{e^{-t/\tau_f}} \qquad (S1)$$

It is not practical to extract heat capacity and thermal conductance from full solution of the coupled equation. But all three-time constants are well separated and follow the constrain equation $\tau_s > \tau_i > \tau_f$. The $\tau_f$ is extremely small hence it can be omitted by masking the initial regime. The remaining two time constants are related to the electronic ($C_e$) and nuclear ($C_n$) heat capacity. The total heat capacity in triple exponential case is $C_e + C_n$ whereas in the two-exponential relaxation method $C_{total} = C_s + C_{addenta}$ hence It is emphasized here, $\tau_s$ and $\tau_i$ , in the physical sense, are not related to the $\tau_1$ and $\tau_2$ introduced in section 2.3.

**Extreme condition for $\frac{C_p}{T^3}$ function**

It is customary to plot $\frac{C_p}{T^3}$ vs. *T* in literature to observe the non-Debye type signatures in the heat capacity. The plot $\frac{C_p}{T^3}$ vs. *T* suppresses the Debye contribution and becomes more pronounced if the system possesses low frequency optical modes of vibration or rattling phenomena in the form of a peak at a particular temperature *T**. The peak position is directly correlated with Einstein temperature $\Theta_E$ by the expression $T^* = \Theta_E/5$. The coefficient (1/5) is not an arbitrary number, but it can be derived from the first derivative of the $\frac{C_p}{T^3}$ function. As we discussed in the main text, the Einstein contribution to the heat capacity is written as:

$$C_p = 3rR\frac{x^2e^x}{(e^x-1)^2} \qquad \text{(S2)}$$

where $x = \frac{\Theta_E}{T}$ and *r* is the oscillator strength. By dividing $T^3$ both sides and omitting the constant term for simplicity, we arrive at:

$$\frac{C_p}{T^3} \propto \frac{x^5e^x}{(e^x-1)^2} \qquad \text{(S3)}$$

By setting the extreme condition $\frac{d}{dx}\left(\frac{x^5e^x}{(e^x-1)^2}\right) = 0$, we get an expression $5 + x = \frac{2xe^x}{e^x-1}$. This equation is a transcendental equation that cannot be solved algebraically. But we can employ either numerical or graphical solutions. Figure S2 displays both functions with respect to x. It is clear from the plot that the solution to the equation is 4.91, which confirms the Einstein oscillator peak position in $\frac{C_p}{T^3}$ plot.

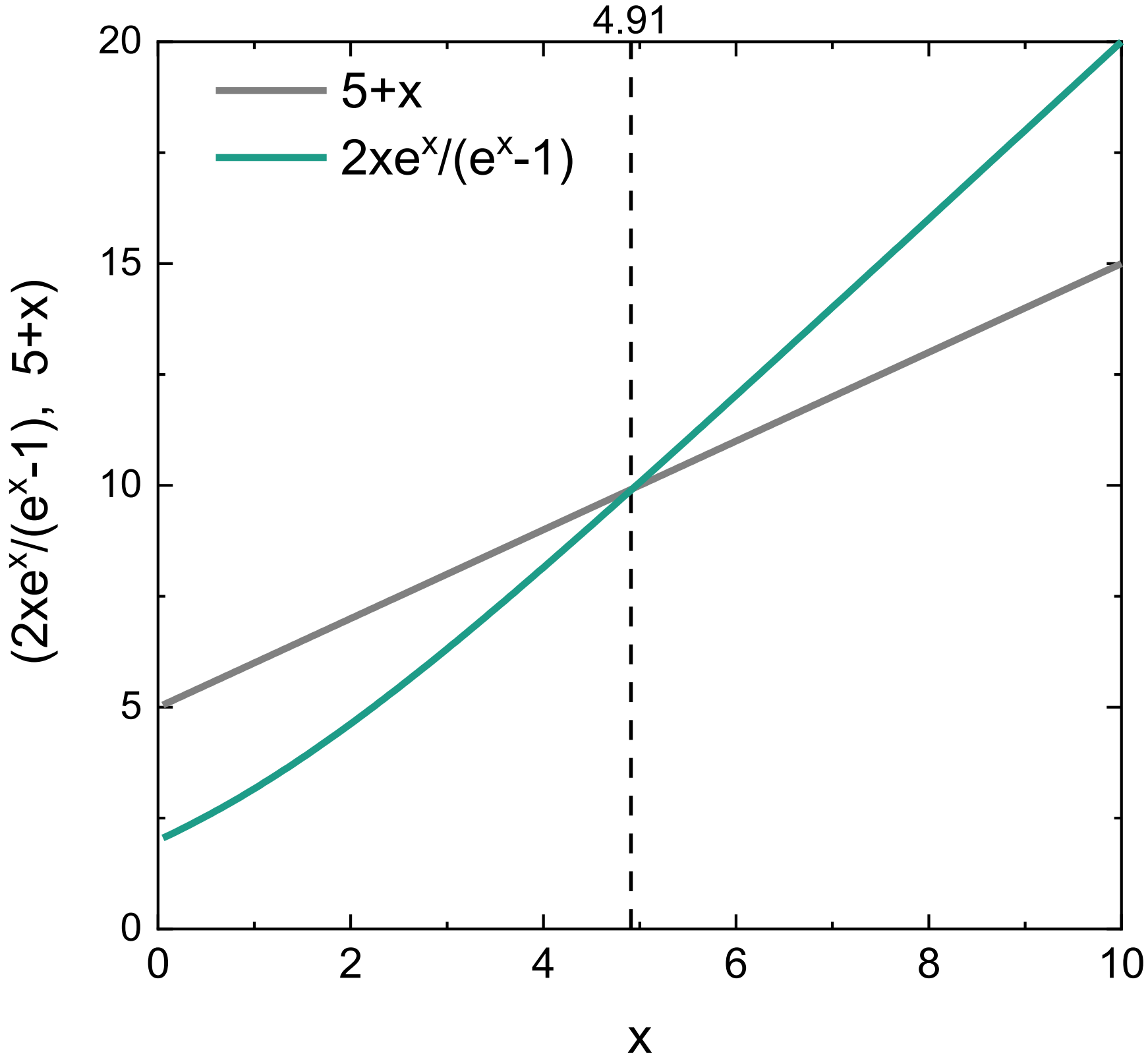

**Figure S2.** Graphical representation for the solution to the transcendental equation for $\frac{C_p}{T^3}$ . The vertical line represents the intersection of the two curves at x = 4.91.

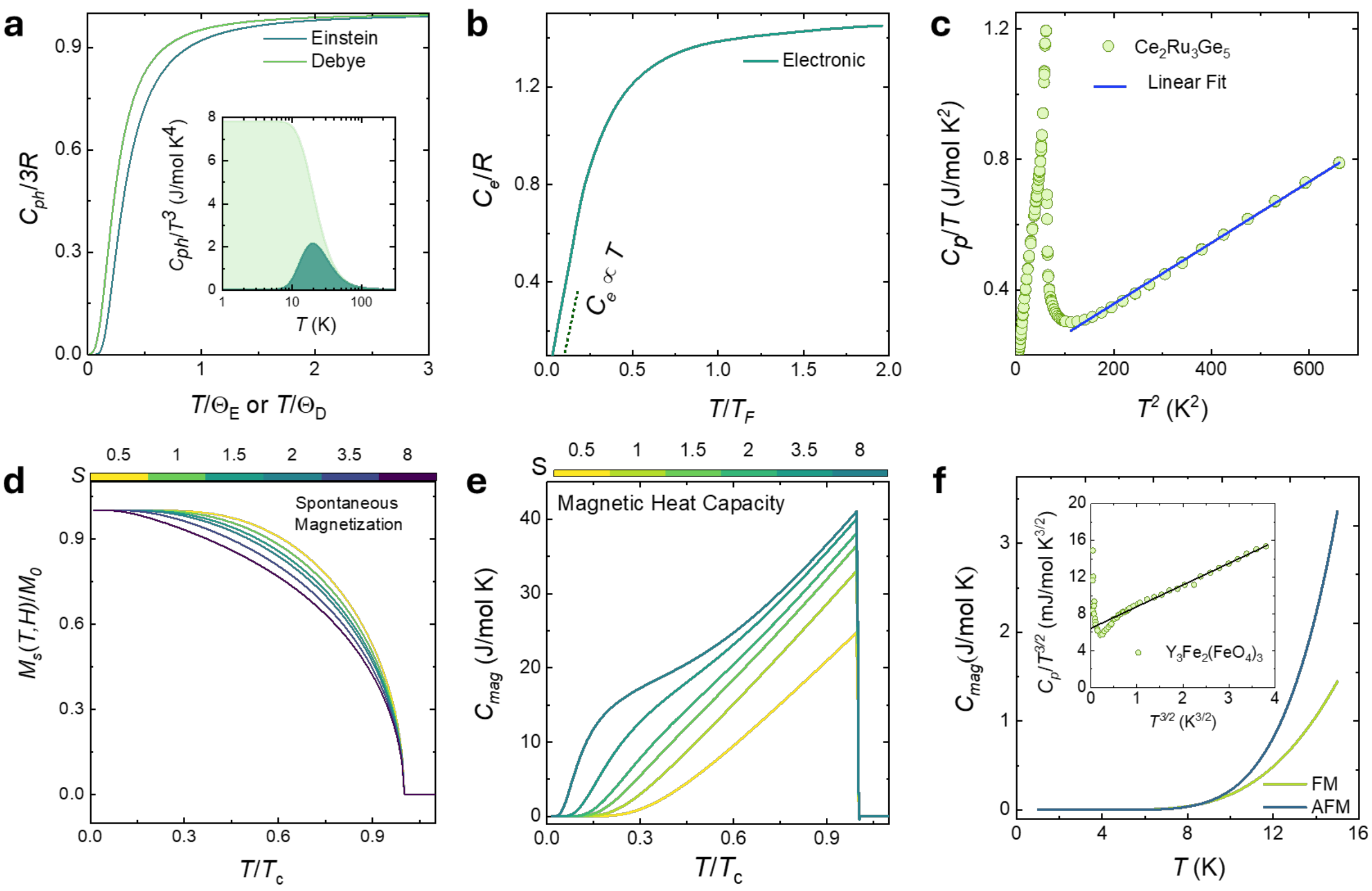

**Figure S3.** (a) Temperature variation of Einstein and Debye model heat capacity, with the inset showing $C_p/T^3$ vs $T$ plot showing signature peak at $\Theta_E/5$. (b) Electronic heat capacity as a function of temperature. The dotted line represents a region $T << T_F$ where linear fit is effective. (c). $C_p/T$ vs $T^2$ plot for $Ce_2Ru_3Ge_5$ .The blue solid line represents a linear fit using the expression. Simulated spontaneous magnetization (d) and magnetic heat capacity (e) using the mean field theory expressions. (f) Simulated spin wave heat capacity for isotropic ferromagnetic and antiferromagnetic cases, with the inset showing: $C_p/T^{3/2}$as function of $T^{3/2}$ for yttrium iron garnet (Adapted with permission from Ref[3] © 2013 EDP Sciences, IOP Publishing and the European Physical Society).

**Functional form of gapped magnon specific heat**

As noted in the main text, magnon specific heat takes on an activated character when the magnons have a gap $\Delta$, for example from single-ion anisotropy or a magnetic field. This can be modeled phenomenologically with $C_{\text{mag}} = \delta T^3 e^{-\Delta/k_B T}$ , but this form of the equation yields inaccurate values for the gap. This is shown with Figure S4, which compares the numerically exact specific heat (Figure S4a) computed with main text equation . (30) to the exponential form above Figure S4b. The same gap $\Delta$ yields wildly different curves, and for small $\Delta$ the difference is more than an order of magnitude in certain temperature ranges.

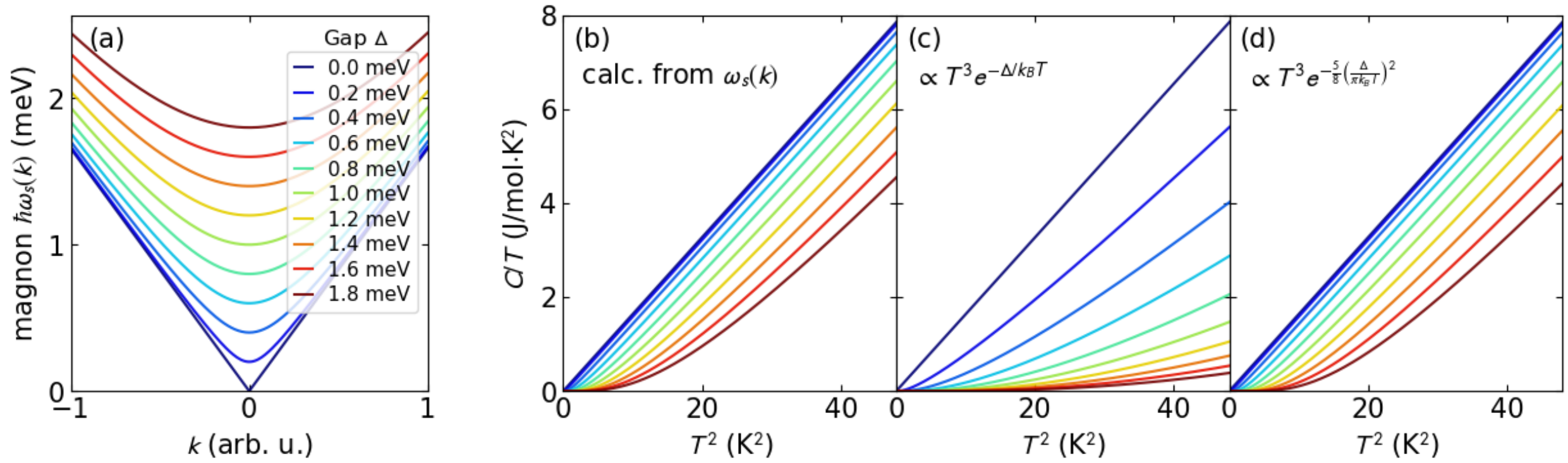

**Figure S4:** Simulated specific heat for three-dimensional antiferromagnetic magnon modes as a function of gap. (a) shows the theoretical dispersions. (b) shows the numerically exact calculation of specific heat from the dispersions using main text Eq. (30). (c) shows the specific heat of a magnon model with the same velocity but a phenomenological exponential. The effect of the gap is dramatically overestimated. (d) shows a different approximation with an exponential in temperature squared, that comes much closer to the true theoretical curve and is more reliable for fitting a gap from specific heat.

A more accurate (though still approximate) equation is main text equation. (32), which is proportional to an exponential in inverse temperature squared. As shown in Figure. S4(c), this matches the numerical curves much more closely (becoming exact in the high $T$ limit).[4] Thus, when fitting gapped heat capacity to estimate the quasiparticle gap, it is important to use main text equation (32) rather than a simple $T^3$ times an exponential. Note that the exponent for gapped magnon specific heat for ferromagnets or lower-dimensional spin systems is different than three-dimensional antiferromagnets, and the correct equation needs to be chosen to correctly estimate $\Delta$.[4]

### Multilevel versus two-level Schottky anomalies

The qualitative shape of the Schottky anomaly does not change with the number of levels: it remains a broad hump. However, the precise form of the anomaly does depend on the number of excited levels, even when the spacing between those levels is uniform. In Figure S5 we show the calculated Schottky anomaly from main text Eq. (34) with different number of levels of equal spacing $\Delta_i$, such that the energy between the highest and lowest level ($\Delta_{tot} = \sum_i \Delta_i$) is fixed to be 1 meV. In the high-temperature limit the curves all overlap, and as the number of levels grows large the heat capacity curve begins to converge. However, for few-level systems at lower temperatures ($k_B T < 2\Delta_{tot}$) the difference between curves is dramatic. Thus, for determining the energy-level splitting for an experimental Schottky anomalies, it is crucial to use the correct number of levels that describe the system at hand.

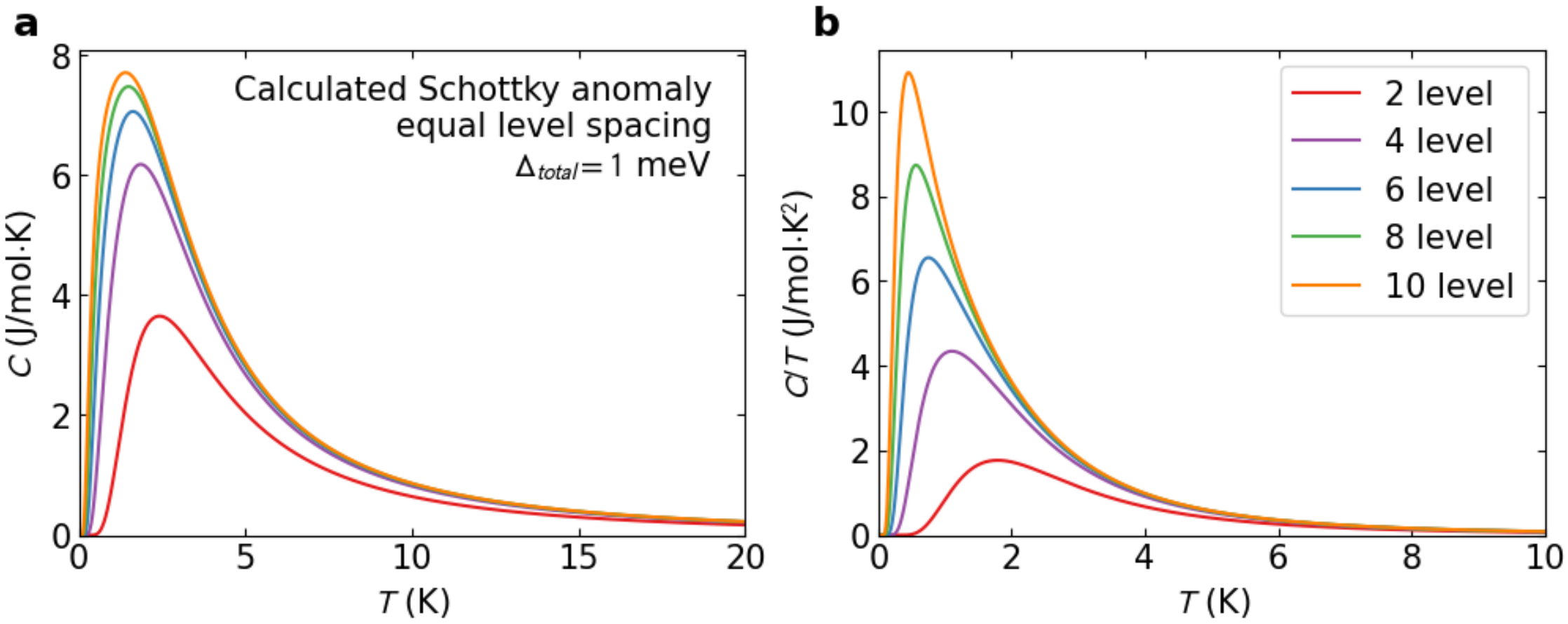

**Figure S5.** Calculated Schottky anomaly heat capacity (main text Eq. 33) for different numbers of energy levels such that the difference between the highest and lowest level $\Delta_{tot}$= 1 meV and all other levels are equally spaced between. Panel (a) shows heat capacity $C$, panel (b) shows the same data plotted as $C/T$. Although in the high-temperature limit the curves converge, at temperature scales up to twice $\Delta_{tot}$ there is a noticeable difference between the different numbers of energy levels, especially for small numbers of levels.

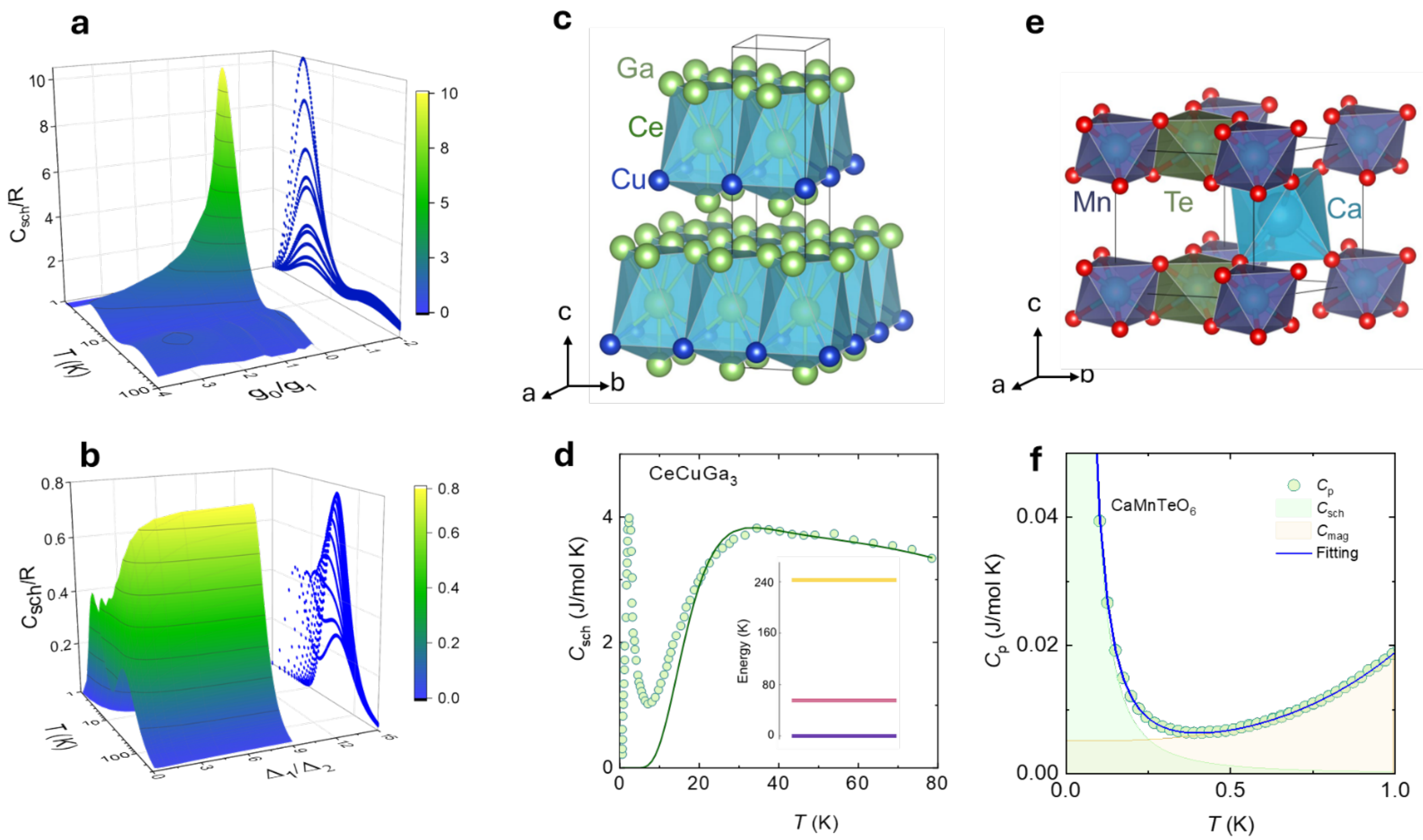

**Figure S6.** (a) Schottky heat capacity as a function of temperature and relative degeneracy. (b) Temperature and relative energy gap variation of Schottky heat capacity. (c) Non-centrosymmetric tetragonal crystal structure of $CeCuGa_3$. (d) $C_{mag}$ as a function of temperature for $CeCuGa_3$ (Adapted from Ref [5] © 2021 American Physical Society. Solid line represents the three-level Schottky model for the $Ce^{3+}$ CEF ground state. (e) Crystal structure of Non-centrosymmetric $CaMnTeO_6$ in space group *P*312. (f) The nuclear Schottky contribution from the Mn nuclear spin of $I = 5/2$. The experimental heat capacity data are presented as symbols, and the electronic and Schottky modelling curves are shown as lines.

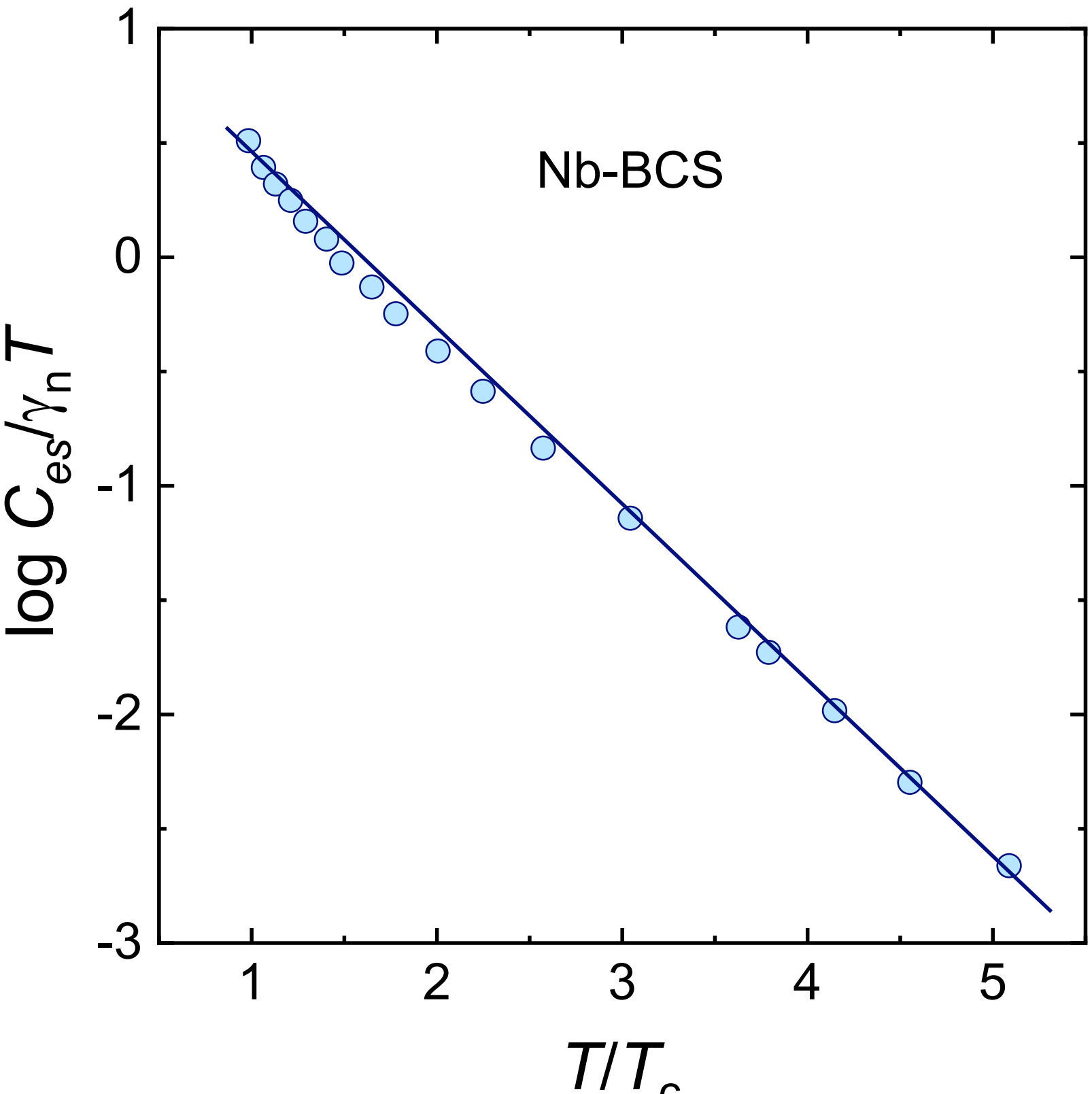

**Figure S7**: Log $\frac{C_{es}}{\gamma_n T}$ vs $\frac{T}{T_c}$ plot for Niobium. The linear fit indicates the exponential suppression of the superconducting gap. (Adapted with permission from Ref[6] © 1962 American Physical Society)

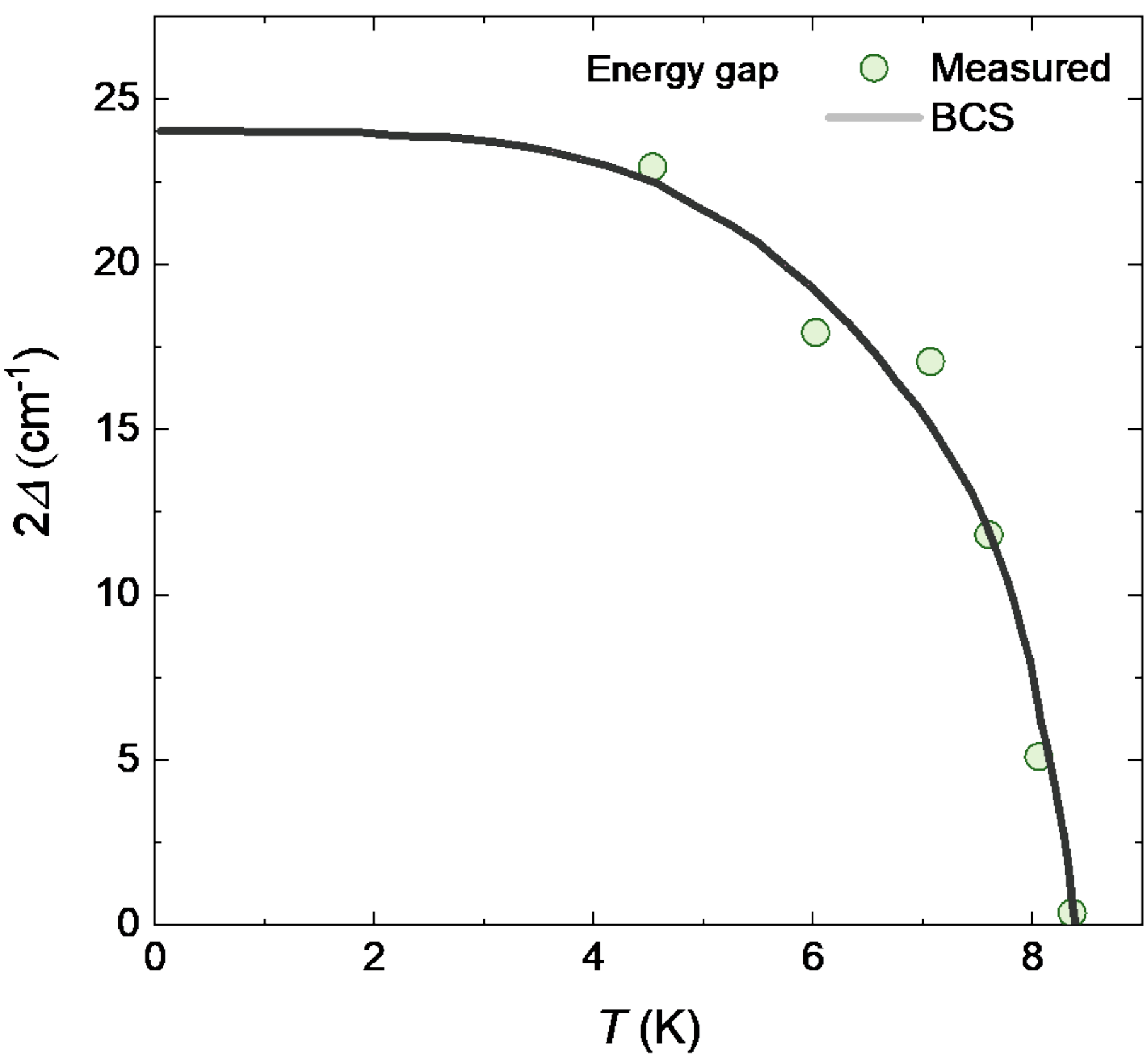

Figure S8. Temperature dependence of the superconducting energy gap 2 $\Delta$(T) for niobium. The solid line represents the prediction from the BCS theory. (Adapted with permission from Ref[7] © 1998 American Physical Society)

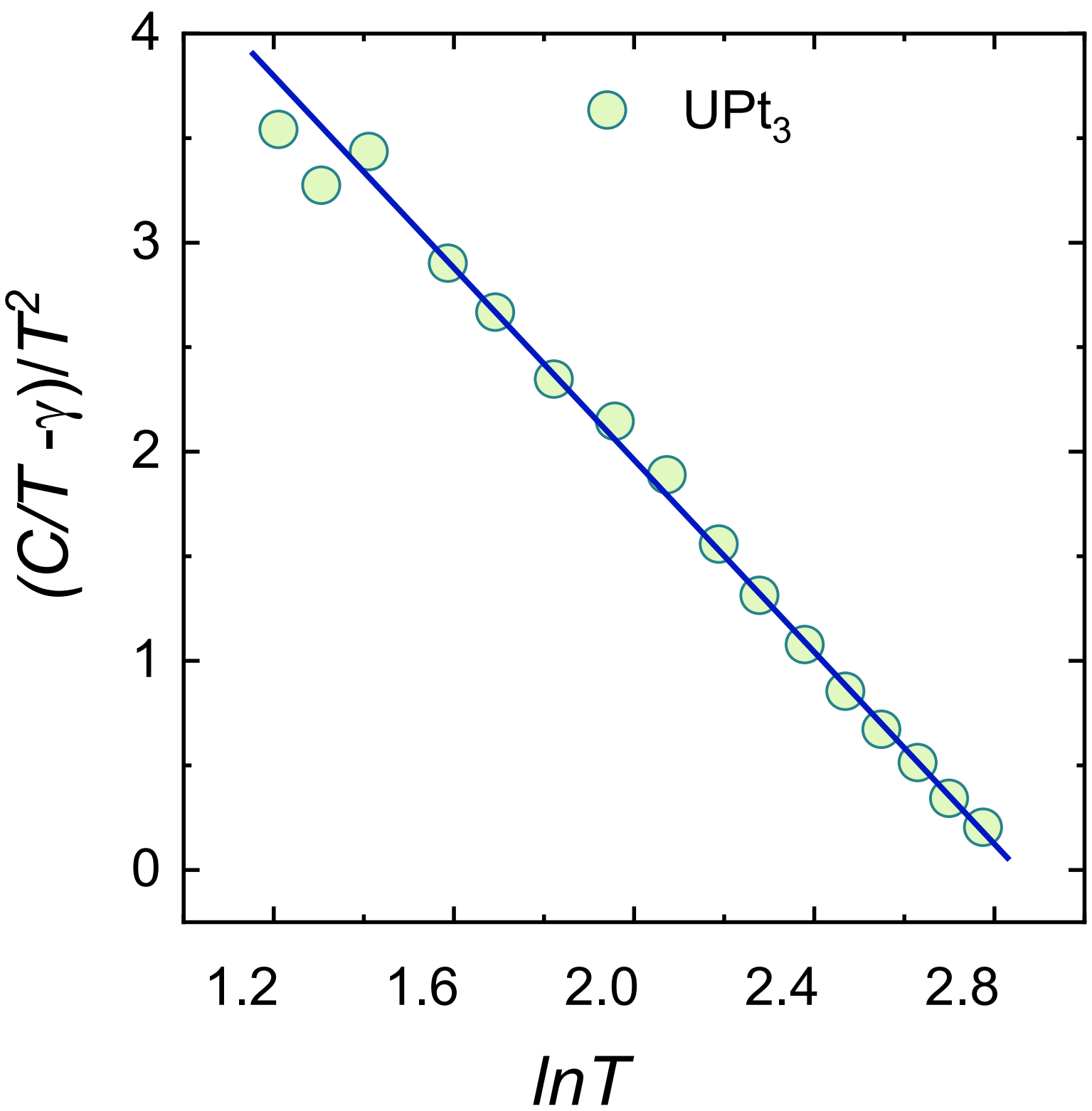

**Figure S9** : $C_p/T - \gamma/T^2$ vs. $\ln T$ plot indicating non-Fermi liquid behavior in $UPt_3$. (Adapted with permission from Ref[8] © 1984 American Physical Society).

**Internal Relaxation and Non-Fermi Liquid Phenomena**

α-YbAlB4 is a heavy fermion compound with a Fermi liquid ground state in zero field with $\gamma$ = 130 mJ/mol.K$^2$ and a large Wilson ratio of 7 puts this system in proximity to the quantum critical point. [9, 10] The field suppresses the heavy fermion behavior, and the system enters the non-Fermi liquid ground state .[9, 10] The total heat capacity shows smooth variation, which, in hindsight, resembles either nuclear Schottky or heavy fermion behavior. However, only if the internal relaxation is taken into account and the system is treated as two separate subsystems, the crossover behavior can be explained. The $C_{total}$ below 200 mK shows a large upturn and $1/T^2$ behavior, expected for a nuclear contribution, and quasi-linear behavior above 200 mK, expected for Fermi-liquid type behavior. The interesting aspect of triple exponential relaxation is visible at 80 mK and 64 mK where $C_e$ shows two distinct kinks due to the crossover regime where the nuclear-electronic thermal conductance is comparable. Physically, this indicates a change in the effective thermal description of the system. Above 80 mK nuclear degrees of freedom act as one reservoir weakly coupled to the electronic degrees of freedom. As decreases, due to the strong coupling between the Yb nuclear moment and the 4f electronic degrees of freedom, the system behaves thermally as a single system. Further details about the importance of the estimation $K_n$ values and how it correlates with NMR (Korringa) time scale are seen here. [2]

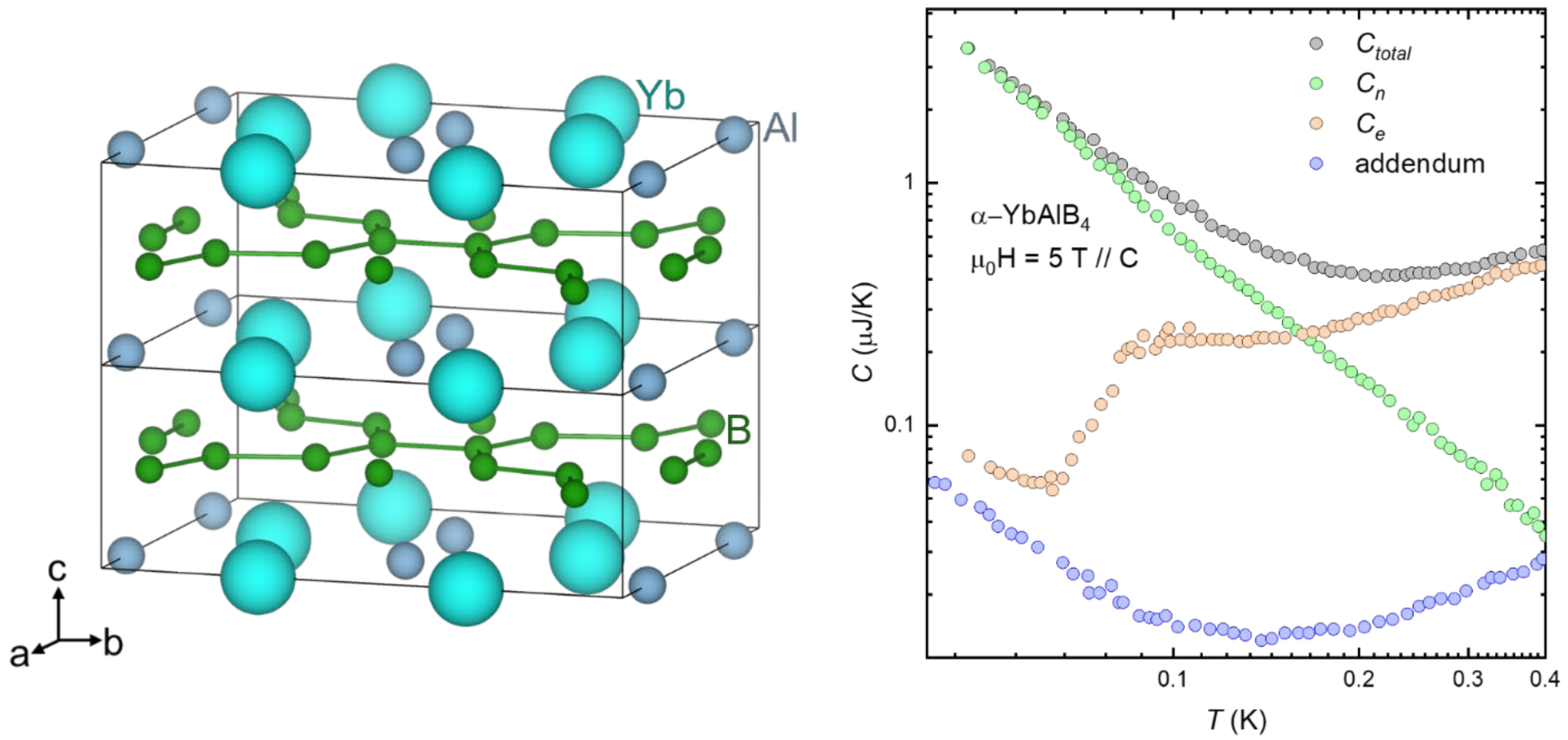

**Figure S10.** (a) Layer boron network and the orthorhombic crystal structure of α-$YbAlB_4$. (b) Total, nuclear, electronic and addendum heat capacity as function of temperature employing triple relaxation method. (Adapted with permission from Ref[2] © 2018 AIP Publishing LLC).

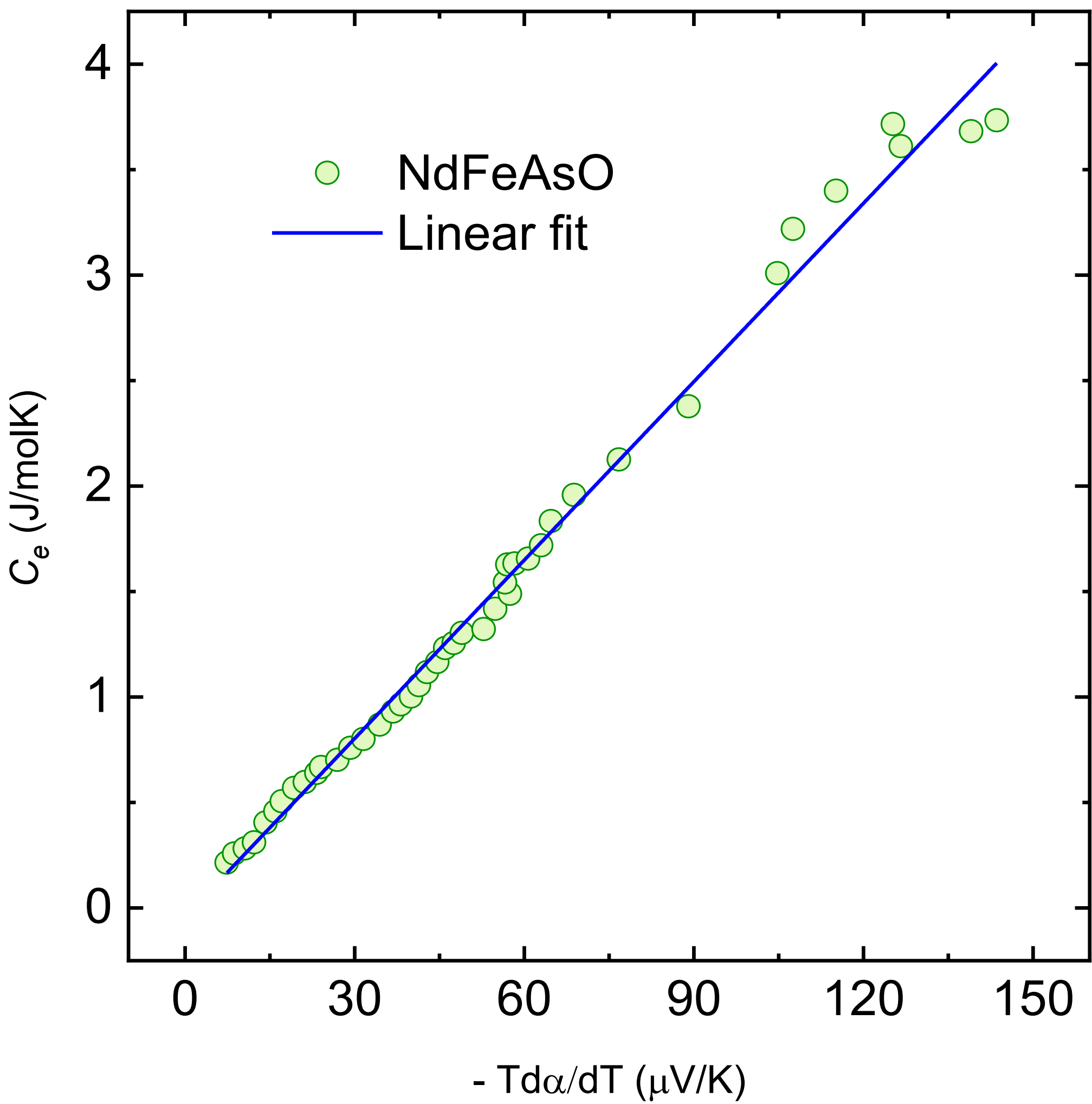

**Figure S11**: Scaling plot between -$Td\alpha/dT$ and $C_e$ indicating the correlation between spin entropy and thermopower. (Adapted with permission from Ref[11] © 2009 American Physical Society).

**References**

(1) Lashley, J.; Hundley, M.; Migliori, A.; Sarrao, J.; Pagliuso, P.; Darling, T.; Jaime, M.; Cooley, J.; Hults, W.; Morales, L. Critical Examination of Heat Capacity Measurements Made on a Quantum Design Physical Property Measurement System. *Cryogenics* **2003**, *43*, 369-378.
(2) Matsumoto, Y.; Nakatsuji, S. Relaxation calorimetry at very low temperatures for systems with internal relaxation. *Review of Scientific Instruments* **2018**, *89* (3).
(3) Pan, B.; Guan, T.; Hong, X.; Zhou, S.; Qiu, X.; Zhang, H.; Li, S. Specific Heat and Thermal Conductivity of Ferromagnetic Magnons in Yttrium Iron Garnet. *Europhysics Letters* **2013**, *103*, 37005.
(4) Straub, A.; Scheie, A. O. Manuscript in Preparation. **2026**.

(5) Anand, V.; Fraile, A.; Adroja, D.; Sharma, S.; Tripathi, R.; Ritter, C.; De La Fuente, C.; Biswas, P.; Sakai, V. G.; Del Moral, A. Crystal Electric Field and Possible Coupling with Phonons in Kondo Lattice $CeCuGa_3$. *Physical Review B* **2021**, *104* (17), 174438.
(6) Hirshfeld, A. T.; Leupold, H.; Boorse, H. Superconducting and Normal Specific Heats of Niobium. *Physical Review* **1962**, *127*, 1501.
(7) Pronin, A. V.; Dressel, M.; Pimenov, A.; Loidl, A.; Roshchin, I. V.; Greene, L. Direct observation of the superconducting energy gap developing in the conductivity spectra of niobium. *Physical Review B* **1998**, *57* (22), 14416.
(8) Stewart, S. G. Heavy-Fermion Systems. *Reviews of Modern Physics* **1984**, *56*, 755.
(9) Matsumoto, Y.; Hong, J.; Kuga, K.; Nakatsuji, S. Anisotropic transverse magnetoresistivity in α-YbAlB4. In *Journal of Physics: Conference Series*, 2015; IOP Publishing: Vol. 592, p 012086.
(10) Matsumoto, Y.; Kentaro, K.; Nakatsuji, S. Suppression of the Heavy Fermion State in Magnetic Fields in the Mixed Valent α-YbAIB4. In *Proceedings of the International Conference on Strongly Correlated Electron Systems (SCES2013)*, 2014; p 011076.
(11) Matusiak, M.; Plackowski, T.; Bukowski, Z.; Zhigadlo, N.; Karpinski, J. Evidence of Spin-density-wave Order in $RFeAsO_{1-x}F_x$ From Measurements of Thermoelectric Power. *Physical Review B—Condensed Matter and Materials Physics* **2009**, *79*, 212502.